\documentclass[journal]{IEEEtran}
\usepackage{bm}

\usepackage{cite}
\usepackage{amssymb}
\usepackage{setspace}

\ifCLASSINFOpdf
\usepackage[pdftex]{graphicx}
\else
\fi

\usepackage{caption}
\usepackage{amsmath,amsthm,bm,amssymb}
\newtheorem{defi}{Definition}
\newtheorem{prop}{Proposition}

\usepackage{amsfonts} 
\usepackage{algorithm}  
\usepackage{algorithmic}
\usepackage{color}
\usepackage{tabularx}
\newcolumntype{C}{>{\centering\arraybackslash}X} 
\usepackage[dvipsnames]{xcolor}

\usepackage{array}
\usepackage{subfigure}
\usepackage{url}
\begin{document}
\title{Market Mechanisms for Low-Carbon Electricity Investments: A Game-Theoretical Analysis}
\author{Dongwei~Zhao,
		Sarah~Coyle,
		Apurba~Sakti,
		and 	Audun~Botterud

  \thanks{This work is supported by Equinor ASA.}
		\thanks{Dongwei Zhao is with the MIT Energy Initiative and the  Laboratory for Information and Decision Systems, Massachusetts Institute of Technology, Cambridge, MA 02139 (e-mail: zhaodw@mit.edu).

    Sarah Coyle is with the Department of System Design \& Management, Massachusetts Institute of Technology, Cambridge, MA 02139 (e-mail: sarahcoyle@chevron.com).

  		 Apurba Sakti is with the MIT Energy Initiative, Massachusetts Institute of Technology, Cambridge, MA 02139 (e-mail:  sakti@mit.edu).

  Audun Botterud is with the  Laboratory for Information and Decision Systems, Massachusetts Institute of Technology, Cambridge, MA 02139 (e-mail: audunb@mit.edu).
		}
	
	}
	\maketitle

\begin{abstract} 
Electricity markets are transforming from the dominance of conventional energy resources (CERs), e.g., fossil fuels, to low-carbon energy resources (LERs), e.g., renewables and energy storage. This work examines market mechanisms to incentivize LER investments, while ensuring adequate market revenues for investors, guiding investors' strategic investments towards social optimum, and protecting consumers from scarcity prices. To reduce the impact of excessive scarcity prices, we present a new market mechanism, which consists of a \textit{P}enalty payment for lost load, a supply \textit{I}ncentive, and an energy price \textit{U}plift (PIU). We establish a game-theoretical framework to analyze market equilibrium. We prove that one Nash equilibrium under the penalty payment and supply incentive can reach the social optimum  given quadratic supply costs of CERs. Although the price uplift can ensure adequate revenues, the resulting system cost deviates from the social optimum while the gap decreases as more CERs retire. Furthermore,  under the traditional marginal-cost pricing (MCP) mechanism, investors may withhold investments to cause scarcity prices, but  such behavior is absent under the PIU mechanism.  Simulation results show that the PIU mechanism can reduce consumers' costs by over 30\% compared with the MCP mechanism by reducing excessive revenues of low-cost CERs from scarcity prices.
\end{abstract}
\vspace{-1mm}

\IEEEpeerreviewmaketitle

\section*{ {Nomenclature}}
\addcontentsline{toc}{section}{Nomenclature}
\subsection{Sets and Indexes}

\begin{IEEEdescription}[\IEEEusemathlabelsep\IEEEsetlabelwidth{$TT$}]

\item[$\mathcal{D}$]  Set of operational horizons (days) in the investment horizon with total $D_a$ days, indexed by $d$
\item[$\mathcal{I}$] Set of LER investors, indexed by $i$

\item[$\mathcal{I}^{\text{RE}}$] Set of VRE investors
\item[$\mathcal{I}^{\text{ES}}$] Set of ES investors
\item[$\mathcal{K}$] Set of types of LER investors, indexed by $k$
\item[$\mathcal{T}$]  Set of hours in an operational horizon, indexed by $t$
\item[$\mathcal{Z}_i^{o}$] Self-constraint set of LER $i$'s resource operations
\item[$\tilde{\mathcal{Z}}_i^{o}$] Self-constraint set of LER $i$'s resource operations including lost load
\item[$\mathcal{Z}_i^{c}$] Joint-constraint set of LER $i$'s resource operations associated with other investors
\item[$\tilde{\mathcal{Z}}_i^{c}$] Joint-constraint set of LER $i$'s resource operations associated with other investors including lost load

\item[$\Omega$] Set of scenarios, indexed by $\omega$
\end{IEEEdescription}

\subsection{Parameters}
\begin{IEEEdescription}[\IEEEusemathlabelsep\IEEEsetlabelwidth{$TTTT$}]

\item[$a^\omega{[t]}$ ]  Quadratic coefficient of CER supply curve 
\item[$b^\omega{[t]}$ ]  Linear coefficient of CER supply curve 
\item[$c^\omega{[t]}$ ]  No-load cost of CER supply curve 
\item[$c_i^S$ ]  Energy-capacity cost of ES
\item[$c_i^P$ ]  Power-capacity cost of ES
\item[$c_i^{X}$ ]  Power-capacity cost of VRE
\item[$c_i^{ch/dis}$ ]  Charge/Discharge cost of ES
\item[$D^\omega{[t]}$ ]  System demand
\item[$\bar{p}^{cv}$ ]  Initial capacity of CERs
\item[$\underline{U}_i/\overline{U}_i$ ]  Lower/Upper bound of duration of ES

\item[VOLL ]  Value of lost load
\item[$\rho^\omega$]  Probability of scenario $\omega$
\item[$\gamma$]  Ratio of remaining capacities of CERs to the initial capacities
\item[$\nu_i^\omega{[t]}$ ]  Generation output factor of VRE $i$ 
\item[$\eta_i^{c/d}$ ]  Charge/Discharge efficiency of ES
\item[$\Delta \pi^\omega{[t]}$ ]  Energy-price uplift 
\end{IEEEdescription}

\subsection{Decision Variables}
\begin{IEEEdescription}[\IEEEusemathlabelsep\IEEEsetlabelwidth{$TTTTTt$}]
\item[$A_{i}^{\omega}{[t]}$] Net output of LERs 
\item[$\tilde{A}_{i}^{\omega}{[t]}$] Net output of LERs including  lost load
\item[$e^{\omega}{[t]}$] State-of-charge of ES
\item[$f_i$] Profit of LERs
\item[$\tilde{f}_i$] Profit of LERs including lost-load cost
\item[$\tilde{\tilde{f}}_i$] Profit of LERs including lost-load cost and supply incentive
\item[$g^{\omega}{[t]}$] Supply cost of CERs
\item[$p^{ch/dis,\omega}{[t]}$] Charge/ Discharge of ES
\item[$p^{cv,\omega}{[t]}$] Output of CERs
\item[$p^{sh,\omega}{[t]}$] Load shedding in the market
\item[$P_{i}$] Power capacity of ES
\item[$S_{i}$] Energy capacity of ES
\item[$X_{i}$] Power Capacity of VRE
\item[$\bm{z}_i$] LER decision vector
\item[$\tilde{\bm{z}_i}$] LER   decision vectors including the  lost load
\item[${\pi}^\omega{[t]}$] General notation of market prices
\item[$\hat{\pi}^\omega{[t]}$] Market price in the PIU mechanism without price uplift
\item[$\tilde{\pi}^\omega{[t]}$] Market price  in the PIU mechanism with price uplift

\end{IEEEdescription}

\section{Introduction}
The threat of climate change is accelerating the global transition of electricity supply from conventional energy resources (CERs), e.g., fossil fuels, to low-carbon energy resources (LERs), e.g., variable renewable energy (VRE) and energy storage (ES). Many countries and states have set zero-carbon targets for electricity supplies. For example,  California aims to achieve the target of 100\% zero-carbon electricity by 2045  \cite{de2018sb}. From 2010 to 2020, the globally installed solar PV capacity increased from about 40 GW to 760 GW, while wind energy grew from about 200 GW to 740 GWh \cite{renreport2021}.  Due to the variable and uncertain nature of VRE, ES is becoming a crucial flexible resource to support VRE integration. The global stationary ES market is estimated to grow from $15.2$ GWh in 2019 to $222.7$ GWh in 2035 \cite{storagelux}. In this work, we will focus on VRE (wind and solar energy) and ES  for LERs.

The increasing integration of LERs is transforming the electricity market, which is possibly on a path towards a 100\%-LER market in the future. Such transition, however, is bringing new challenges, e.g., more frequent scarcity prices \cite{leslie2020designing} and inadequate revenues for  LER investments \cite{machado2022estimating}. Due to the uncertain and variable nature of VRE, it is economically difficult to guarantee adequate supply. This may lead to an increase in the frequency of scarcity price spikes and thus face consumers' backlash due to political and social reasons \cite{fundamentals}. Meanwhile, as more CER generators retire, ensuring adequate revenues for the LER investment may be more difficult. Since VRE has zero marginal cost, the marginal-cost-based market prices are expected to fall with increasing VRE\cite{ketterer2014impact,leslie2020designing}. Furthermore, the retirement of CERs may require much larger capacities of ES and VRE  to fill the gap due to supply variability. Thus, the high capital costs, e.g., of ES \cite{mongird2020}, combined with large capacity needs, will shift the cost structure in the system towards high investment and low operating costs, possibly reducing revenues under current market designs.

This prospect of volatile prices and diminishing revenues can influence the strategic behaviors of LER investors. In the deregulated market, where investors make independent investment and operation decisions, they will make decisions strategically to pursue profits. Since most electricity markets are imperfectly competitive \cite{fundamentals}, the diminishing market returns may discourage profit-seeking investors from investing in LERs, or encourage them to withhold the investment and supply to increase prices \cite{wolak2001impact}. Under the decarbonization target, the strategic behaviors of VRE and ES may incur a higher frequency of supply shortages and price spikes.  {Although most markets perform market power mitigation to reduce the strategic behavior of suppliers in the short term \cite{guo2019market}, it is hard to monitor and regulate investments in the long term.  Besides, compared with CERs' certain outputs and certain marginal costs, VRE generation is uncertain. ES marginal cost is not straightforward and can be dependent on the opportunity cost. It is still an open problem for the market monitor how to mitigate the market power of VRE and  ES.}

Market mechanism design under an increasing penetration of LERs still poses a number of open questions, e.g.,  how to set the price in zero-carbon markets \cite{zhou2022price}, and how to incentivize ES operations and investments into the market \cite{zheng2022arbitraging}. In this paper, we do not focus on a 100\%-LER market, but rather on the transition from a CER-dominated mix to 100\%-LER. We study ways to adjust the CER-based market pricing mechanism  to incentivize investors' strategic investments in LERs and to address the following key question: 

\begin{itemize}
\item \textit{Under the increasing percentage of LERs in electricity markets, how can we ensure adequate revenues for LER investors, guide their investments towards the social optimum, and protect consumers from excessive prices?}
\end{itemize}

 {Notably, our work focuses on the long-term planning of LERs in the energy-only electricity market. We do not consider the capacity market or capacity requirements for energy resources. Our model is based on joint optimization over the long-term investment horizon and short-term operations.}

\vspace{-1ex}
\subsection{Related work}
\vspace{-0.5ex}

 {There is extensive literature on  incentive mechanisms for encouraging investments in VRE and ES in electricity markets. In terms of market mechanisms, the approaches are centered around the energy market,  capacity market, and supportive financial tools. (i) For the energy market, besides the short-term marginal-cost pricing, extensive studies proposed long-term enhancement. For example, long-term bilateral contracts (e.g., power purchase agreements) \cite{joskow2022hierarchies}, long-term  auction-based energy markets \cite{pierpont2017markets}, and pricing mechanisms based on long-run marginal costs \cite{stevenson2018transitioning}.  (ii) Besides energy markets, most of the US electricity markets have established the long-term capacity market or  resource adequacy requirements \cite{fer2020energy}, which provides remuneration for resource investments. Approaches (e.g., effective load carrying capability) have been proposed to effectively characterize  capacity credits  for  VRE and ES \cite{garver1966effective,wang2021crediting,Chen2023Capacity}. (iii) Financial tools have been discussed to hedge the risk of lost load or high prices and meanwhile provide adequate incentives for the VRE and ES investors, e.g., through reliability options \cite{andreis2020pricing} \cite{vazquez2003security} and insurance contracts \cite{zhao2022insurance}\cite{billimoria2019market}. Finally, in addition to the market mechanisms,  government policies also play an important role in incentivizing the deployment of VRE and ES \cite{zhou2022price} \cite{levin2019long}. For example, a renewable portfolio standard (RPS)  is a regulatory mandate to increase electricity supply from renewable energy.  Eligible
resources that satisfy  specific
RPS obligations receive credits. Production and investment tax credits  incentivize the expansion of clean resources by offering tax credits to investors. Carbon pricing usually imposes a direct tax on  carbon emissions or implements a carbon cap-and-trade framework for resources, which benefits clean energy resources.}

In this work, we study how to adjust marginal-cost-based pricing  mechanisms to incentivize the investment of LERs in energy-only markets. The marginal-cost pricing (MCP) mechanism has been widely adopted in current electricity markets and extensively studied in the literature. Earlier works showed that the MCP mechanism can achieve short-term and long-term market efficiency under convex assumptions \cite{schweppe2013spot, Kazempour2018Market, korpaas2020optimality}. However, the MCP mechanism can cause high scarcity prices, which may create consumers' concerns \cite{fundamentals}. Some markets enforce a price cap, e.g., 2000\$/ MWh in CAISO, but such a price cap is still very high for consumers and additionally contributes to the missing money problem \cite{newbery2016missing}. In contrast, our work proposes adjustments to the traditional MCP mechanism by reducing the impact of excessive scarcity prices and meanwhile ensuring adequate revenues for investors. Furthermore, in the traditional MCP mechanism, consumers bear the cost of the lost load. Our work lets investors pay this cost by designing a penalty payment.  Such a penalty payment works similarly to the contract for price difference \cite{newbery2020club}\cite{joskow2022hierarchies} or insurance \cite{fumagalli2004quality} by reducing the risk for consumers. We analyze the market equilibrium under imperfect competition under such a penalty payment, which is not captured in \cite{newbery2020club, joskow2022hierarchies,fumagalli2004quality}. 

The market equilibrium under the MCP mechanism is mostly analyzed under perfect competition \cite{schweppe2013spot,Kazempour2018Market,korpaas2020optimality}, which assumes market participants as price-takers. However, considering imperfect competition, a finite number of market participants can behave strategically to influence the market price. It is highly challenging to analyze and compute the equilibrium under the MCP mechanism under imperfect competition, which usually requires solving equilibrium programs with equilibrium constraints  (EPEC) for multiple players \cite{pozo2011finding,conejo2016investment,devine2022strategic}. A pure-strategy Nash equilibrium of EPEC generally may not exist \cite{hobbs2003complementarity}. In this work, we characterize a special pure-strategy Nash equilibrium for the traditional MCP mechanism to demonstrate that strategic LER investors may withhold investments to obtain more frequent scarcity prices. 

We establish an imperfect competition model of LERs based on the Cournot model where firms compete in quantity \cite{novshek1985existence,allaz1993cournot,mas1995microeconomic}.  The classic Cournot model \cite{novshek1985existence,allaz1993cournot,mas1995microeconomic} focused on theoretical analysis with simplified assumptions on firms. The works \cite{hobbs2007nash, ruiz2008some,Huang2021storage,cruise2018impact,kramer2021strictly,zhang2015competition,zhang2018competition,zhao2022storagecaiso} applied Cournot competition  in electricity markets to study suppliers' strategic behaviors.  However, none of the prior studies include market competition in the joint investment of VRE and ES, while our work will capture it. The studies \cite{hobbs2007nash, ruiz2008some,Huang2021storage,cruise2018impact,kramer2021strictly} were established based on assumed demand curves, which cannot reflect the impact of zero-marginal-cost VRE. Furthermore, electricity demand is usually not elastic whose demand curve is difficult to quantify. In contrast, our work focuses on inelastic demand. We utilize real market data to construct the supply curve, in an attempt to reflect the impact of VRE on market revenues and costs. 

\vspace{-2ex}
\subsection{Our contributions}

Expanding on the literature above, we summarize the key contributions of this work in the following.

 \textit{Strategic LER investments under imperfect competition:} 
To the best of our knowledge, our work is the first to study strategic investments of LERs, including VRE and ES, under imperfect competition through a game-theoretical framework. We use market data from the California Independent System Operator (CAISO) to approximate the system marginal cost curve. We analyze the pure-strategy Nash equilibrium of VRE and ES investors' investment and operation under the conventional MCP mechanism and the proposed PIU mechanism. 
	
\textit{PIU mechanism:} We present a novel mechanism consisting of a \textbf{P}enalty payment for lost load, a supply \textbf{I}ncentive, and a price \textbf{U}plift (PIU), to LER investors. Under the assumption of quadratic supply cost of CERs, we prove that one Nash equilibrium coincides with the social optimum under perfect competition in the P-mechanism, and under imperfect competition in the PI-mechanism. To solve the resulting profit loss under the retirement of CERs, we introduce a price uplift to ensure adequate revenues for LER investors. The PIU mechanism will incentivize more LER investment than the social optimum. However,  the system-cost gap between the PIU mechanism and the social optimum decreases as more CERs retire.
	
     \textit{Comparison between the PIU and MCP mechanisms:} The traditional MCP mechanism can induce the social optimum under perfect competition. However, under imperfect competition, we prove that market participants may withhold investments and intentionally cause scarcity prices for consumers, which will not occur in the PIU mechanism. Simulation results show that the PIU mechanism can reduce consumers' cost by over 30\% than the traditional MCP mechanism even under perfect competition, by reducing the excessive profit of low-cost CERs under scarcity prices.

The rest of this paper is organized as follows. In Section \ref{sec:system}, we present the models of the LER investors and system operator. In Section \ref{section:game}, we formulate a game-theoretical framework for market competition. Then, we analyze the Nash equilibrium for the traditional MCP mechanism in Section \ref{section:benchmark_scarce}. We introduce a new PIU mechanism and analyze the Nash equilibrium in Section \ref{section:design}. In Section \ref{section:sim}, we conduct simulations to demonstrate the equilibrium results under the PIU mechanism. Finally, we conclude in Section \ref{sec:con}.

\vspace{-2ex}

\section{System model}\label{sec:system}

We consider a set of LER investors in the market denoted by $\mathcal{I}=\mathcal{I}^{\text{RE}}\bigcup \mathcal{I}^{\text{ES}}$, which contains two subsets $\mathcal{I}^{\text{RE}}$ and $\mathcal{I}^{\text{ES}}$ corresponding to two \textit{classes} of LERs: VRE and ES, respectively. The VRE can be solar energy or wind energy. Without loss of generality, We assume that each investor has only one \textit{type} of resources. One type means the same class of energy resources and the same technology parameters.  

Besides the new additions of VRE and ES, there are also existing CERs, mainly fossil thermal generations, in the market.\footnote{For conventional generators, first, we assume that in the future they will gradually retire in the market and there will be no new investment due to  decarbonization targets. Therefore, we do not consider the strategic long-term investment behavior of CERs. Second, CER generators are dispatchable with controllable outputs. Current electricity markets have sophisticated mechanisms to mitigate the market power of CER short-term supply. Thus, we assume no strategic short-term withholding from CERs.  However, CERs may strategically decide how much capacity they want to retire, which may impact the market equilibrium outcome. Our current model does not capture retirement decisions, which we leave as future work.} We assume that the system operator will schedule these existing CERs, and ensure the demand and supply balance. During periods of scarcity, the system operator can also leave some demand unserved, as we assume that the demand is inelastic without demand response. Besides, our work assumes no congestion in the network and captures key insights into market mechanisms. We  partially generalize our model to incorporate
networks using DC power flow, where some key results will
 hold as shown in Appendix.VIII \cite{marketappendix}.

Next, we will introduce the timescale of the investors' decision-making. 

\vspace{-1ex}
\subsection{Timescale of decision-making}
\begin{figure}[t]
	\centering
	\includegraphics[width=2.3in]{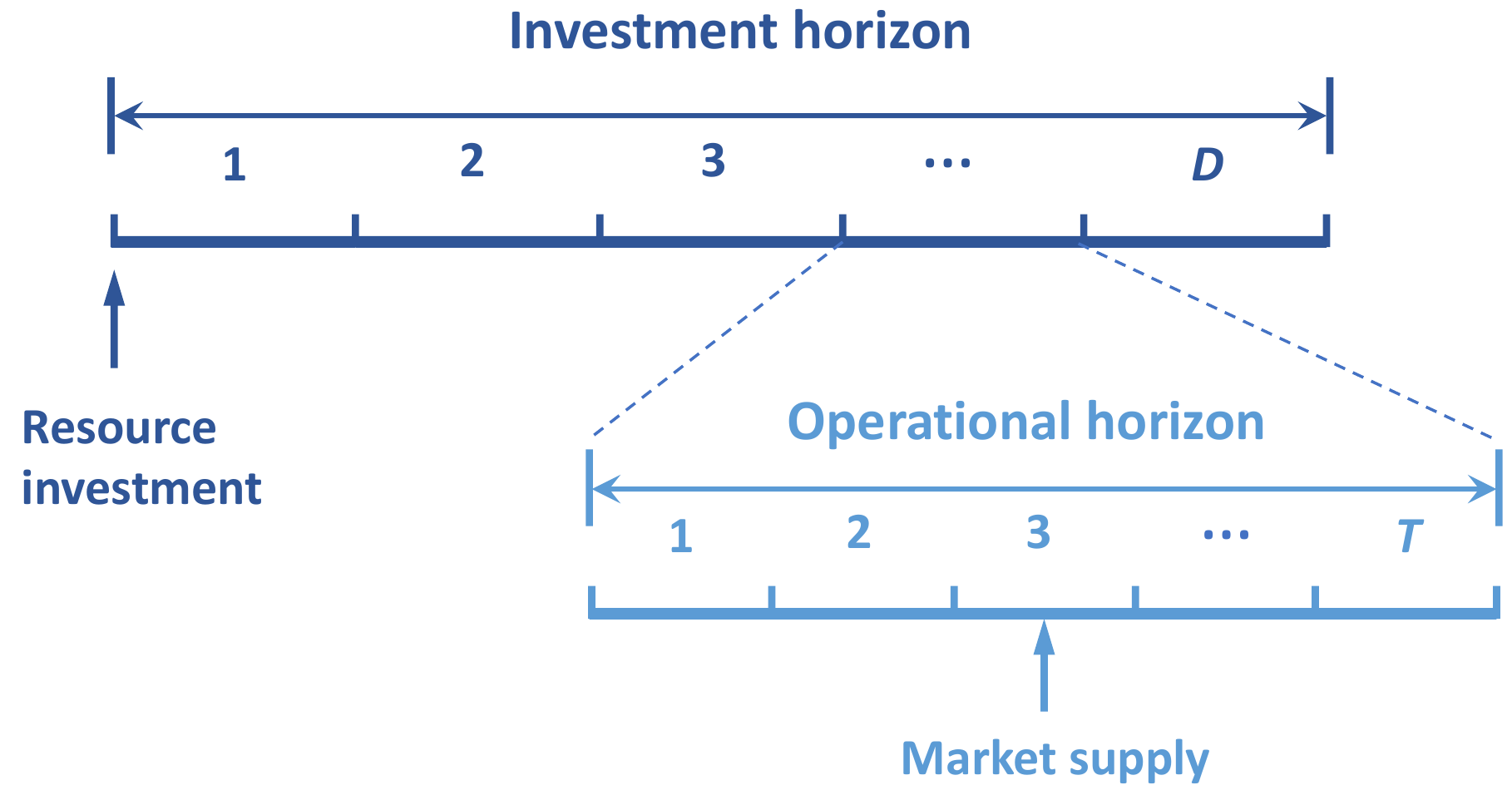}
	\vspace{-1ex}
	\caption{Timescales.}
	\label{fig:time0}
	\vspace{-2ex}
\end{figure}

Figure \ref{fig:time0} illustrates two timescales of decision making. At the beginning of an investment horizon $\mathcal{D}\hspace{-1mm}=\hspace{-1mm}\{1,2,...,D_a\}$ of $D_a$ days (e.g., 10 years), each investor $i\in \mathcal{I}$ decides the investment decision $\bm{X}_i$ of VRE or ES.\footnote{ {We focus on the one-time investment decisions and capital costs of LERs to capture key insights from long-term planning.  We do not consider the multiple sequential investment decisions during the investment horizon and we leave it to future work. }}

The investment horizon is divided into a number of operational horizons, i.e., each day $d\in \mathcal{D}$ corresponds to one operational horizon, which is further divided into multiple time slots $\mathcal{T}=\{1,2,...,T\}$ (e.g., 24 hours). The system demand is denoted by $D^d[t]$ for hour $t$ of day $d$.  For day $d \in \mathcal{D}$, each investor $i$ schedules the operation decision over time slots $\mathcal{T}$, i.e., $\bm{p}_i^{d}=({p}_i^{d}[t], t\in \mathcal{T})$. We denote by a vector $\bm{z}_i=\Big(\bm{X}_i, ( \bm{p}_i^{d}, d\in  \mathcal{D})\Big)$ all the decision variables of investor $i$ over the two timescales,  which we will elaborate later. The system operator will  schedule the CER operation $\bm{p}^{cv,d}=({p}_i^{cv,d}[t], t\in \mathcal{T})$ and the shed load $\bm{p}^{sh,d}=({p}^{sh,d}[t], t\in \mathcal{T})$ to maintain the power balance.

 To facilitate the formulation using stochastic programming, we use historical data to construct a  scenario set $\Omega$ that represents all the days in the investment horizon. Specifically, for any day $d\in \mathcal{D}$ in the operation,  scenario $\omega \in \Omega$ occurs with a probability $\rho^\omega$. We use scenarios to model the discrete distributions of daily demand, operational cost, and renewable energy outputs. We calculate the expected revenues and operation costs for investors on a daily basis and scale the long-term capital cost into one day. For example, we use three-year data from CAISO to construct over 1000 scenarios for daily market operation profiles. This can effectively capture multiple periods of operations within the investment horizon. Since we model the operation by scenarios, we substitute $d\in \mathcal{D}$ by  $\omega\in  \Omega$ for the related decisions and parameters.

\vspace{-2ex}
\subsection{Model of VRE and ES investors, and system operator}

We will introduce the decision variables, constraints, and costs for the VRE and ES investors, and the system operator. 

\subsubsection{VRE investor}

For the VRE investor $i\in \mathcal{I}^{\text{RE}}$, the investment decision $\bm{X}_i$ is the invested capacity $X_i$. The operational decision $\bm{p}_i$ includes the market supply $\bm{p}_i^{mk}=({p}_i^{mk, \omega}[t],\forall t\in \mathcal{T}, \forall \omega \in \Omega)$  and the curtailment  $\bm{p}_i^{cur}=({p}_i^{cur, \omega}[t],\forall t\in \mathcal{T}, \forall \omega \in \Omega)$ for each hour $t$ of  scenario $\omega$.

The operational constraints for investor $i\in \mathcal{I}^{\text{RE}}$ satisfy 
\begingroup
     \begin{subequations} \label{eq:const_ren}
     \allowdisplaybreaks
		\begin{align}
			& p_i^{mk,\omega}[t]+p_i^{cur,\omega}[t]=\nu_i^{\omega}[t] \cdot X_i,\forall t\in\mathcal{T},\forall \omega \in {\Omega},\label{eq:const_ren_a}\\
			& X_i\geq 0,  p_i^{mk,\omega}[t]\geq 0, p_i^{cur,\omega}[t]\geq 0, \forall t\in\mathcal{T},\forall \omega \in {\Omega}.\label{eq:const_ren_b}
		\end{align}
     \end{subequations}
\endgroup
Here  the parameter $\nu_i^{\omega}[t] $ denotes the capacity factor for time $t\in \mathcal{T}$ of scenario $ \omega \in \Omega$. Note that such capacity factors can be quite different between solar and wind energy. Equation \eqref{eq:const_ren_a} ensures the power balance of VRE. Constraint \eqref{eq:const_ren_b} gives non-negative variables. We denote the constraint set constructed by \eqref{eq:const_ren_a}-\eqref{eq:const_ren_b} by $\mathcal{Z}_i^o$, $\forall  i \in \mathcal{I}^{\text{RE}}$.

In terms of the costs, the variable operation cost of VRE is zero $C_i^{opr}(\bm{z}_i)=0$.  The capital cost is $C_i^{inv}(\bm{z}_i)=\kappa_i \cdot  c_i^{X}X_i$, where $c_i^{X}$ denotes the unit capacity cost of the wind or solar energy. The factor $\kappa_i$ scales the capital cost over the investment horizon into one day for stochastic programming \cite{zhao2022storagecaiso}.

\subsubsection{Storage investor} For the ES investor $i\in \mathcal{I}^{\text{ES}}$,  the investment decision $\bm{X}_i$ includes the invested energy capacity and power capacity $(S_i, P_i)$. The operation decision $\bm{p}_i$ includes  the charge and discharge profiles for each hour $t$ of scenario $\omega$, i.e.,  $\bm{p}_i^{ch}=({p}_i^{ch,\omega}[t],\forall t\in \mathcal{T}, \forall \omega \in \Omega)$ and $\bm{p}_i^{dis}=({p}_i^{dis,\omega}[t], t\in \mathcal{T},\forall \omega \in \Omega)$, as well as  the state-of-charge (SOC) in  storage by $\bm{e}_i^\omega=\{e_i^\omega[t],~ \forall t \in 0\bigcup  \mathcal{T},\forall \omega \in \Omega\}$. 

The operation constraints for investor $i\in \mathcal{I}^{\text{RE}}$ satisfy 
    \begin{subequations}\label{eq:const_storage}
    \begin{align}
	&0\leq 	p_i^{dis,\omega}[t]\leq P_i,\forall t\in\mathcal{T},\forall \omega\in\Omega, \label{eq:ch}\\
	&0\leq 	p_i^{ch,\omega}[t]\leq P_i,\forall t\in\mathcal{T},\forall \omega\in\Omega, \label{eq:dis}\\
	&e_i^\omega[t]\hspace{-0.3mm}=\hspace{-0.3mm}e_i^\omega[t-1]\hspace{-0.3mm}+\hspace{-0.3mm}\eta_i^cp_i^{ch,\omega}[t]\hspace{-0.3mm}-\hspace{-0.3mm}p_i^{dis,\omega}[t]/\eta_i^d, \forall t\in\mathcal{T},\forall \omega\in\Omega, \label{eq:dynamics1}\\
	&0\leq e_i^\omega[t]\leq S_i,\forall t\in\mathcal{T}',\forall \omega\in\Omega, \label{eq:dynamics2}\\
	&e_i^\omega[0]= e_i^\omega[T],\forall \omega\in\Omega,\label{eq:dynamics3}\\
	&\underline{U}_i\leq \frac{S_i}{P_i}\leq \overline{U}_i. \label{eq:duration}
\end{align}
\end{subequations}\label{eq:storage}\par{\vspace{-2ex}}

 We elaborate the constraints next. First, constraints \eqref{eq:ch}-\eqref{eq:dis} are the power-capacity limits for charge and discharge, respectively.
Second, constraints \eqref{eq:dynamics1}-\eqref{eq:dynamics3} are the constraints for the SOC in the storage. Constraint \eqref{eq:dynamics1} is the SOC change with time due to the charge and discharge operation, where $\eta_i^c$ and $\eta_i^d$ denote the charge and discharge efficiency, respectively. Equation \eqref{eq:dynamics2} ensures the SOC  bounded by the storage capacity.  Equation \eqref{eq:dynamics3} makes the initial SOC equal to the final level, which decouples the storage operation across different days. Finally, constraint \eqref{eq:duration} enforces a general duration bound on the investment alternatives of the energy capacity and power capacity, where $\underline{U}_i$/$\overline{U}_i$ are the lower/upper bounds.\footnote{We omit the constraint of non-simultaneous charge and discharge 	$p_i^{dis,\omega}[t]\cdot 	p_i^{ch,\omega}[t] =0$ to maintain the convexity of the model. In simulations, simultaneous charge and discharge does not occur.}  We denote the strategy set constructed by \eqref{eq:ch}-\eqref{eq:duration} by $\mathcal{Z}_i^o$, $\forall  i \in \mathcal{I}^{\text{ES}}$.

In terms of the costs, the  investment cost includes the capital costs for both energy and power capacity $ 
	C_i^{inv}(\bm{z}_i)=\kappa_i \left(c_i^{S}S_i+c_i^{P}P_i\right)$, 
where $c_i^{S}$ and $c_i^{P}$ denote the unit costs for energy and power capacities, respectively. For the operation cost incurred by charge and discharge, we adopt a  linear model and calculate the expected cost by $C_i^{opr}(\bm{z}_i)$ on a daily basis.
\begin{align}
	C_i^{opr}(\bm{z}_i)=\mathbb{E}_{\omega \in \Omega} \Big[\sum_{t\in \mathcal{T}}\left(c_i^{ch}{p}_i^{ch,\omega}[t]+c_i^{dis}{p}_i^{dis,\omega}[t]\right)\Big],
\end{align}
where $c_i^{ch}$ and $c_i^{dis}$ denote the unit costs for charge and discharge, respectively. This can model the O\&M cost as well as the potential degradation cost.
    
\subsubsection{System operator} The system operator schedules the generation of CERs $p^{cv,\omega}[t]$ and the lost load $p^{sh,\omega}[t]$.

To introduce the system constraints, we first let $A_i^\omega[t]$ be investor $i$'s net market supply.  For the VRE investor $i\in \mathcal{I}^{\text{RE}}$, $A_i^\omega[t]\triangleq p_i^{mk,\omega}[t]$. For the storage investor $i\in \mathcal{I}^{\text{ES}}$, $A_i^\omega[t]\triangleq-p_i^{ch,\omega}[t]+p_i^{dis,\omega}[t]$. We list the system constraints.
\begingroup
\allowdisplaybreaks
	\begin{align}
	& \hspace{-2mm}\sum_{i\in \mathcal{I}}A_i^{\omega}[t]\hspace{-0.5mm}+\hspace{-0.5mm}p^{cv,\omega}[t]\hspace{-0.5mm}+\hspace{-0.5mm}p^{sh,\omega}[t]\hspace{-0.5mm}=\hspace{-0.5mm}D^\omega[t],\forall t\in\mathcal{T},\forall \omega \in {\Omega},\label{eq:const_balance}\\
	& 0\leq p^{cv,\omega}[t]\leq \gamma \cdot \bar{p}^{cv},	\forall t\in\mathcal{T},\forall \omega \in {\Omega},\label{eq:const_convention}\\
	& 0\leq p^{sh,\omega}[t],	\forall t\in\mathcal{T},\forall \omega \in \Omega.\label{eq:const_loss}
	\end{align}
 \endgroup
Here constraint \eqref{eq:const_balance} ensures power balance. Constraint \eqref{eq:const_convention} gives the bound on the capacity of available CERs.  The parameter $\gamma \in [0,1]$ adjusts the retirement amount of the current CERs. For example, $\gamma=0.4$ means the retirement of 60\% CERs.

In terms of the costs,  we calculate the expected cost of lost load based on  the value of lost load (VOLL):
\vspace{-0.5ex}
\begin{align}
G^{sh}(\bm{p}^{sh})=\mathbb{E}_\omega \sum_{t\in \mathcal{T}} \text{VOLL} \cdot p^{sh,\omega}[t].
\end{align}
For the operation cost of CERs, we denote it as a function $g^{cv,\omega}[t](\cdot)$ in $p^{cv,\omega}[t]$ for  hour $t$ of scenario $\omega$. The overall expected cost is $G^{cv}(\bm{p}^{cv})$.
\vspace{-0.5ex}
\begin{align}
&G^{cv}(\bm{p}^{cv})=\mathbb{E}_\omega \sum_{t\in \mathcal{T}} g^{cv,\omega}[t]({p}^{cv,\omega}[t]).
\end{align}
 We assume that the function $g^{cv,\omega}[t](\cdot)$ is convex, i.e., the marginal cost $\frac{d g^{cv,\omega}[t]}{d p^{cv,\omega}[t]}$ is non-decreasing,  and $\frac{d g^{cv,\omega}[t]}{d p^{cv,\omega}[t]}<\text{VOLL}$.  Next, we will introduce a social-optimum benchmark.

\vspace{-1ex}
\subsection{Social-optimum benchmark: System cost minimization}\label{section:benchmark_system}

We consider a social-optimum benchmark where the system operator decides the investment and operation of all the LER investors.  The system operator aims to  minimize the system cost, which includes the investment cost and operation cost of new LERs, the operation cost of the existing CERs, and the cost of system lost load. Since we consider inelastic demand, the social optimum is equivalent to the least system cost.

\noindent \textbf{Problem SO: System Cost Minimization}
\begingroup
\allowdisplaybreaks
\begin{align*}
\min	&\sum_{i\in \mathcal{I}} C_i^{inv}(\bm{X}_i)+\sum_{i\in \mathcal{I}} C_i^{opr}(\bm{p}_i)+ G^{cv}(\bm{p}^{cv})+G^{sh}(\bm{p}^{sh})\\
\text{s.t.}~& \eqref{eq:const_ren_a}-\eqref{eq:const_ren_b}, \eqref{eq:ch}-\eqref{eq:duration}, \forall i \in \mathcal{I},\\
~&\eqref{eq:const_balance},~~~~~~~~~~~~~~~~~~~~~\text{dual}:~\lambda^{b,\omega}[t]\\
~&\eqref{eq:const_convention},~~~~~~~~~~~~~~~~~~~~~\text{dual}:~\underline{\mu}^{cv,\omega}[t], ~\overline{\mu}^{cv,\omega}[t]\\
~&\eqref{eq:const_loss},~~~~~~~~~~~~~~~~~~~~~\text{dual}:~\underline{\mu}^{sh,\omega}[t]\\
\text{var:}~& (\bm{X}_i, \bm{p}_i, \forall i \in \mathcal{I}), \bm{p}^{cv},\bm{p}^{sh}.
\end{align*}
\endgroup

Problem \textbf{SO} is convex. We denote its optimal solution as $\bm{z}^\dagger=(\bm{X}_i^\dagger, \bm{p}_i^\dagger, \forall i \in \mathcal{I}), ~\bm{p}^{cv\dagger}$, and $\bm{p}^{sh\dagger}$. We denote the dual variables corresponding to the system constraints \eqref{eq:const_balance}-\eqref{eq:const_loss} as $\lambda^{b,\omega}[t]$, $(\underline{\mu}^{cv,\omega}[t], ~\overline{\mu}^{cv,\omega}[t])$, and $\underline{\mu}^{sh,\omega}[t]$, respectively.

\vspace{-0.5ex}
\section{Deregulated market: Game theory model}\label{section:game}

We now focus on a deregulated market where each new LER investor can decide the invested capacity and market supply by themselves. Each LER investor aims to maximize its profit. The system operator will schedule CERs and shed load over the operation horizons to ensure power balance.

The market energy price is the key to investors' market revenues.  Note that the market price is affected by all the investors' decisions. Thus, we model the market price of time $t$ of scenario $\omega$ as a function of all the investors' decision variables, i.e.,  ${\pi}^\omega[t](\bm{z}_1, \bm{z}_2\ldots \bm{z}_I)$.  Next, we introduce the revenue, profit, and constraint formulations for the VRE and ES investors, based on which we establish a game-theoretical model among these competing investors.

\vspace{-2ex}
\subsection{Revenue and profits} 

For investor $i$, the market revenue $R_i(\bm{z}_i, \bm{z}_{-i})$ comes from the net supply ${A}_i^{\omega}[t]$ on the market.
\begin{align}
	&R_i(\bm{z}_i, \bm{z}_{-i})=\mathbb{E}_{\omega \in \Omega} \Big[\sum_{t\in \mathcal{T}} {A}_i^{\omega}[t] \cdot \pi^\omega[t] (\bm{z}_i, \bm{z}_{-i})\Big],
\end{align}
where $\bm{z}_{-i}$ denotes all the investors' decision variables other than investor $i$. The profit $f_i(\bm{z}_i, \bm{z}_{-i})$ is the revenue minus the investment and operation cost, which is coupled with others.
\begin{align}
	f_i(\bm{z}_i, \bm{z}_{-i})=R_i(\bm{z}_i, \bm{z}_{-i})-C_i^{inv}(\bm{z}_i)-C_i^{opr}(\bm{z}_i).
\end{align}

\vspace{-3ex}
\subsection{Strategy set: constraints} 

There are two parts of each investor's constraint. First, each investor's operation decision is constrained by its resource $\bm{z}_i \in \mathcal{Z}_i^o$.  Second, investors' decision sets are coupled due to the  system constraints \eqref{eq:const_balance}-\eqref{eq:const_loss}, which leads to the following constraint on the total supply $\sum_{i\in \mathcal{I}}A_i^{\omega}[t]$ of all the investors.
    	\begin{align}
	& D^\omega[t]-\gamma \cdot \bar{p}^{cv} \leq \sum_{i\in \mathcal{I}}A_i^{\omega}[t]\leq D^\omega[t],\forall t\in\mathcal{T},\forall \omega \in {\Omega}.\label{eq:const_balancemodify}
	\end{align}
Given other investors' decisions $\bm{z}_{-i}$, we denote the coupled strategy set of investor $i$ from \eqref{eq:const_balancemodify} by  $\mathcal{Z}_i^c(\bm{z}_{-i})$. Overall,  the constraint set of investor $i$ is $\mathcal{Z}_i(\bm{z}_{-i})\triangleq \mathcal{Z}_i^o\bigcup \mathcal{Z}_i^c(\bm{z}_{-i})$. 
    
\vspace{-2ex}
\subsection{Game-theoretical model} 
Now we formulate a competition game $G$ among investors to model investors' strategic behaviors.

\begin{itemize}
	\item \textit{Players}: Investor $i\in \mathcal{I}$
	\item \textit{Strategy}: $\bm{z}_i\in \mathcal{Z}_i (\bm{z}_{-i})$ of investor $i$ 
	\item \textit{Profit}: $f_i(\bm{z}_i ,\bm{z}_{-i})$ of investor $i$ 	
\end{itemize}

We aim to find the Nash equilibrium of the game. If the players' strategy sets are also coupled, the Nash equilibrium is usually called a generalized Nash equilibrium, which we define for the competition game $G$ in Definition \ref{def:equilibrium} \cite{facchinei2010generalized}.

\vspace{-0.5ex}
\begin{defi}[Nash equilibrium]\label{def:equilibrium}
	In the  game $G =\langle \mathcal{I},(\mathcal{Z}_i),(f_i)\rangle$, the strategy profile $\bm{z}^*\in \Pi_{i\in \mathcal{I}} \mathcal{Z}_i(\bm{z}_{-i}^*)$ is  a generalized pure strategy Nash equilibrium if for every investor $i\in \mathcal{I}$, $f_i(\bm{z}_i^*,\bm{z}_{-i}^*)\geq f_i(\bm{z}_i,\bm{z}_{-i}^*)$ for any $\bm{z}_i\in \mathcal{Z}_i(\bm{z}_{-i}^*)$.
\end{defi}
The above definition states that the Nash equilibrium is a state where every investor's decision maximizes its profit given other investors' decisions. The Nash equilibrium will serve as the market equilibrium for analysis in this paper.

We make the following assumptions for this game-theoretical model. First, investors make decisions simultaneously, and they do not observe others' decisions when making their own. Second, we consider a complete-information setting, i.e., each investor knows the market price function and the technology parameters of all the other investors. Third,  every investor is rational and will make decisions following the Nash equilibrium. Otherwise, some investors can always change their strategies in an attempt to pursue higher profits.

Different market mechanisms lead to different market price functions ${\pi}^\omega[t](\bm{z})$. Next, in Section \ref{section:benchmark_scarce}, we discuss the traditional MCP mechanism,  which can lead to scarcity prices during supply shortages. In Section \ref{section:design}, we propose a new mechanism without scarcity prices and analyze  Nash equilibrium. We present mathematical proofs in appendices \cite{marketappendix}.
 
\vspace{-2ex}

\section{MCP mechanism}\label{section:benchmark_scarce}
The market price is set based on the shadow price $\lambda^{b,\omega}[t]$ corresponding to the power balance constraint \eqref{eq:const_balance} in Problem \textbf{SO} given investors' decisions of investment and operation, i.e., 
\vspace{-1ex}
\begin{align}
\pi^{\lambda,\omega}[t] (\bm{z})=\frac{\lambda^{b,\omega}[t]}{\rho^\omega}, \label{eq:shadowprice}
\end{align}
where we modify the shadow price by the  probability $\rho^\omega$ due to the probability coefficient for revenues.\footnote{ The reason for this price modification  is that in Problem \textbf{SO},  we optimize the expected system cost based on scenario probabilities.
According to the KKT conditions of Problem \textbf{SO}, we show the following stationary condition associated with the CER operation variable $p^{cv,\omega}[t]$:   $\frac{d g^{cv,\omega}[t]}{d p^{cv,\omega}[t]}-\frac{\lambda^{b,\omega}[t]}{\rho^\omega}+\frac{\overline{\mu}^{cv,\omega}[t]}{\rho^\omega}-\frac{\underline{\mu}^{cv,\omega}[t]}{\rho^\omega}=0.$
Given  each hour $t$ of each scenario $\omega$, the CER actual marginal cost (or roughly system marginal cost) is set up based on $\frac{\lambda^{b,\omega}[t]}{\rho^\omega}$. Therefore, we use $\frac{\lambda^{b,\omega}[t]}{\rho^\omega}$ as the market-clearing price.}

Problem \textbf{SO}  is a joint optimization problem for the investment and operation of energy resources. The shadow price \eqref{eq:shadowprice} can reflect investment capital costs.  In Proposition \ref{prop:shadowperfect} later, we prove that the shadow price \eqref{eq:shadowprice} can incentivize investors to invest and operate their resources reaching the long-term social-optimum outcome.

 However, in the energy-only market, if we fix the invested capacity for energy resources, conventional short-term shadow prices from the economics dispatch problem can only reflect short-term operation costs and thus  may not provide proper market signals for LER investment \cite{wogrin2022impact}.\footnote{If we fix investors' investment decisions and re-run  Problem \textbf{SO}, the  dual variable $\lambda^{b,\omega}[t]$ can have multiple solutions, one of which coincides with the dual variable computed when investors' investment decisions are variables \cite{wogrin2022impact}. Thus, the market signal is missing when we only focus on short-term operations under given LER capacities.} Short-term shadow prices may also take account of the long-term investment
cost if new mechanisms are introduced. The obtained shadow price \eqref{eq:shadowprice} actually can elicit this special shadow price to reach the long-term social optimum.

 In this work, we focus on the long-term planning of LERs. We will directly utilize the market price  \eqref{eq:shadowprice} for the MCP mechanism as it gives nice market signals to the LER investment. It is not this paper’s focus on how to practically incorporate investment factors into short-term shadow prices, which we leave as future work. Next,  we analyze the Nash equilibrium under  perfect and imperfect competition based on the price \eqref{eq:shadowprice}, respectively.

 \vspace{-2ex}
\subsection{Market equilibrium under perfect competition}

In Proposition \ref{prop:shadowperfect}, we prove that the social-optimum result is one Nash equilibrium under perfect competition (i.e., investors are price takers), which achieves zero surplus for investors.

\vspace{-1ex}
\begin{prop}[Perfect competition under the shadow price]\label{prop:shadowperfect}
Given the market price  \eqref{eq:shadowprice} , under perfect competition, the social-optimum solution $\bm{z}^\dagger$ is one Nash equilibrium, at which each investor's profit is zero, i.e., $f_i(z^\dagger)=0,\forall i \in \mathcal{I}$.
\end{prop}

Although the MCP mechanism can achieve the social optimum under perfect competition, the resulting high scarcity prices typically face criticism from political and consumer groups \cite{fundamentals}, and the social optimum may not be achieved under imperfect competition.

\vspace{-1ex}

\subsection{Market equilibrium under imperfect competition}

With a finite number of investors, the perfect competition assumption is not realistic as investors' decisions will impact the market price.  It is challenging to analyze the pure-strategy Nash equilibrium under the MCP mechanism.

We focus on a case of homogeneous VRE investors and characterize  a Nash equilibrium that can always achieve a market price at  VOLL  in Proposition \ref{prop:shadowimperfect}. The impact of storage charge and discharge can be complicated, but the VRE supply will simply reduce the market price. When the supply from LERs satisfies $\sum_{i\in \mathcal{I}}A_i^\omega[t] <D^\omega[t]-\gamma \cdot \bar{p}^{cv}$,  some demand is not served after utilizing the maximum capacity of CERs, which leads to the price at VOLL. When  $D^\omega[t]-\gamma\cdot \bar{p}^{cv}<\sum_{i\in \mathcal{I}}A_i^\omega[t]$, the market price is bounded by the marginal cost of the CERs. The market price is not unique at $\sum_{i\in \mathcal{I}}A_i^\omega[t]=D^\omega[t]-\gamma\cdot \bar{p}^{cv}$, which we will avoid in the analysis.

\vspace{-0.5ex}
\begin{prop}\label{prop:shadowimperfect}
\allowdisplaybreaks
Assume $N$ homogeneous VRE investors and $D^\omega[t]- \gamma \bar{p}^{cv}>0$ for any time $t$ of scenario  $\omega$. Given an $\epsilon^\omega[t]>0$, we consider the following optimization problem.
\begin{align*}
\max~	&\sum_{i\in \mathcal{I}} f_i(\bm{z}_i ,\bm{z}_{-i})\\
\text{s.t.}~&\pi^\omega[t] (\bm{z})=VOLL,\\
	& \sum_{i\in \mathcal{I}}A_i^{\omega}[t]\leq D^\omega[t]-\gamma \cdot \bar{p}^{cv}-\epsilon^\omega[t],\forall t\in\mathcal{T},\forall \omega \in {\Omega},\\
	&\eqref{eq:const_ren_a}-\eqref{eq:const_ren_b}.
\end{align*}
If VOLL  satisfies \eqref{eq:condition} for any time $t$ of scenario  $\omega$
\begin{align}
VOLL \geq \left(1+ \frac{N\cdot \gamma \bar{p}^{cv}}{D^\omega[t]-\gamma \bar{p}^{cv}}\right)\cdot \frac{d g^{cv,\omega}[t]}{d p^{cv,\omega}[t]}\Big|_{\gamma \cdot \bar{p}^{cv}},  \label{eq:condition}
\end{align}
then the solution to the above optimization problem approaches a Nash equilibrium as $\epsilon^\omega[t]\rightarrow 0$.
\end{prop}

If all the CERs retire, i.e., $\gamma\rightarrow 0$,  \eqref{eq:condition} will be $\text{VOLL}>\frac{d g^{cv,\omega}[t]}{d p^{cv,\omega}[t]}\Big|_{\gamma \cdot \bar{p}^{cv}}$, which is easily satisfied. This implies that more retirement of the CERs makes it easier for the investors to exercise market power. If $N\rightarrow +\infty$, the condition \eqref{eq:condition} will require  $\text{VOLL}\rightarrow +\infty$, which implies that exercising market power is harder with more investors in the market. Next, we propose a new mechanism in an attempt to address the shortcomings of high scarcity prices.

\vspace{-1ex}
\section{PIU mechanism}\label{section:design}
\vspace{-0.5ex}

In the new mechanism, we let the market price be bounded by the marginal cost of the  CERs.  Even if there is unserved load in the market, the market price is still set by the marginal cost of the CERs at the maximum output. Specifically,  we have the price with LER supply as follows.

\vspace{-1.5ex}
\begingroup
\allowdisplaybreaks
\begin{align}
\hat{\pi}^\omega[t] (\bm{z})=\left \{
\begin{aligned}
&\frac{d g^{cv,\omega}[t]}{d p^{cv,\omega}[t]}\Bigg|_{\gamma\cdot \bar{p}^{cv}},  ~\text{if}~ \sum_{i\in \mathcal{I}}A_i^\omega[t]< D^\omega[t]-\gamma\cdot \bar{p}^{cv};\\
&\frac{d g^{cv,\omega}[t]}{d p^{cv,\omega}[t]},~~~~~\text{if}~  D^\omega[t]-\gamma \cdot \bar{p}^{cv} \\
&\hspace{22ex}\leq \sum_{i\in \mathcal{I}}A_i^\omega[t]\leq D^\omega[t].
\end{aligned}
\right. \label{eq:ourprice}
\end{align}
\endgroup

\begin{figure}[t]
	\centering
	\includegraphics[width=2.4in]{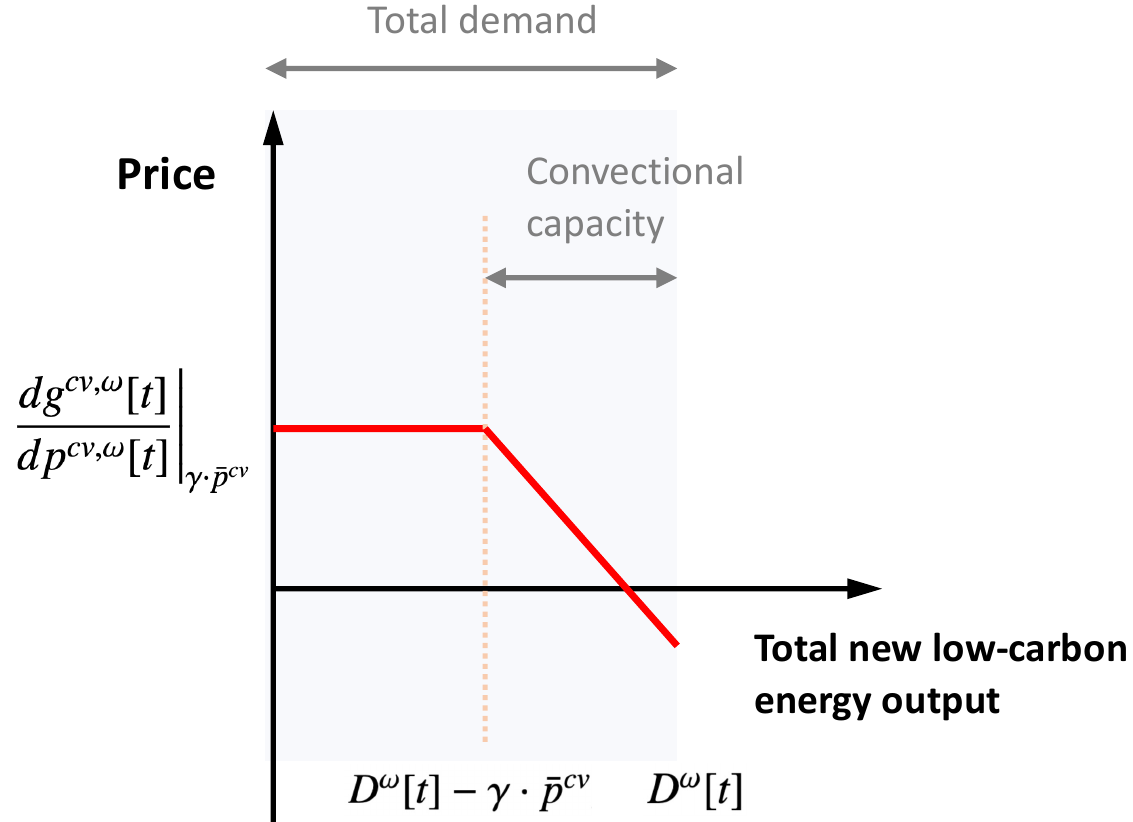}
	\caption{Market price with LER supply without scarce prices.}
	\label{fig:price_supply_cap}
	\vspace{-2ex}
\end{figure}

We illustrate the price $\hat{\pi}^\omega[t] (\bm{z})$ in Fig.\ref{fig:price_supply_cap}, which decreases with the supply from LERs.  The price $\hat{\pi}^\omega[t] (\bm{z})$ is bounded by the marginal cost of CERs.  This price bound shares some similarities with the existing practice that ISOs set a fixed value of price
cap. We clarify some key
differences in the following. First, the existing price cap is still very high for consumers, e.g., 5001\$/MWh in ERCOT and   2000\$/MWh in CAISO.  This price cap can  be much higher than the maximum marginal cost of CERs, especially when we see retirements of high-cost CERs under decarbonization. Such a high price cap may lead to strategic behaviors of investors,  as we show in Proposition \ref{prop:shadowimperfect}. Second, the existing price cap is fixed in the long term. It can be challenging to choose a proper cap to reflect the value of lost load while protecting consumers. In contrast, our work sets the price cap at the maximum marginal cost of CERs, which will vary when more CERs retire. Finally, the ISOs usually set a price cap lower than the actual values of lost load. Although such price caps may protect consumers, they can cause missing money problems for investors to recover investment costs. Our pricing restraints will work together with supply incentives and energy price lift to ensure revenues for suppliers as shown later.

However, such a price bound cannot reflect the value of lost load in the system in the resulting market prices. To solve this problem, we design a penalty payment for investors based on the lost load.

\subsection{Penalty payment}
 {The system cost includes the cost of lost load. After we remove the scarcity price, if investors' profits can not reflect the cost of lost load, the market equilibrium cannot eliminate the gap to the social-optimum benchmark. Imposing penalty payments for the lost load on investors can well bridge this gap and motivate investors to invest in more capacity.}  Next, we introduce the model of the penalty payment (P-mechanism). We discuss the Nash equilibrium under the penalty payment and the price setting $\hat{\pi}^\omega[t]$.

\subsubsection{Model of penalty payment}
Under the penalty payment, each investor $i$ decides an additional contribution item in the lost load  $p_i^{sh,\omega}[t]\geq 0$ at each hour $t$ of scenario $\omega$. The system operator ensures that all the investors' total lost load satisfies the demand balance in the market as in the following.
	\begin{align}
	& \sum_{i\in \mathcal{I}}A_i^{\omega}[t]+p^{cv,\omega}[t]+\sum_{i\in \mathcal{I}} p_i^{sh,\omega}[t]=D^\omega[t],\notag\\
	&\hspace{30ex}\forall t\in\mathcal{T},\forall \omega \in {\Omega},\label{eq:const_balancenew}\\
	&p_i^{sh,\omega}[t]\geq 0, \forall i \in \mathcal{I},\forall t\in\mathcal{T},\forall \omega \in \Omega.\label{eq:const_positive_loss}
	\end{align}
We let the new market supply of investor $i$ be $\tilde{A}_i^{\omega}[t]=A_i^{\omega}[t]+ p_i^{sh,\omega}[t]$ since the lost load also balances the demand. We assume that the system operator will also pay for $p_i^{sh,\omega}[t]$ at the market price to each investor $i$.  We denote $\Tilde{\bm{z}}_i=\left(\bm{z}_i,\bm{p}_i^{sh}\right)$, where $\bm{p}_i^{sh}=(p_i^{sh,\omega}[t], \forall t\in \mathcal{T}, \forall \omega \in \Omega)$.

We update the strategy set, market price function, and profit formulation based on the new variables $\tilde{A}_i^{\omega}[t]$ and $\Tilde{\bm{z}}_i$. First, the constraint \eqref{eq:const_balancemodify} is changed to 
    	\begin{align}
	& D^\omega[t]-\gamma \cdot \bar{p}^{cv} \leq \sum_{i\in \mathcal{I}}\tilde{A}_i^{\omega}[t]\leq D^\omega[t],\forall t\in\mathcal{T},\forall \omega \in {\Omega}.\label{eq:const_balancemodify_new}
	\end{align}
Based on the new constraint \eqref{eq:const_balancemodify_new}, we update the coupled strategy set  $\mathcal{Z}_i^c(\bm{z}_{-i})$ to $\tilde{\mathcal{Z}}_i^c(\tilde{\bm{z}}_{-i})$. We add a new constraint $p_i^{sh,\omega}[t]\geq 0$ in the strategy set $\mathcal{Z}_i^o$, which is now changed to $\tilde{\mathcal{Z}}_i^o$. Overall, we denote the strategy set for investor $i$ as  $\Tilde{\bm{z}}_i \in \Tilde{\mathcal{Z}}_i (\Tilde{\bm{z}}_{-i})\triangleq \tilde{\mathcal{Z}}_i^o \bigcup \tilde{\mathcal{Z}}_i^c(\tilde{\bm{z}}_{-i}) $. Second, we update the market price function in \eqref{eq:ourprice} by directly replacing ${A}_i^{\omega}[t]$  with $\tilde{A}_i^{\omega}[t]$ as the lost load will not impact the pricing. Finally,  the investor pays the additional penalty payment for the lost load $p_i^{sh,\omega}[t]$, which leads to the following new profit $\tilde{f}_i(\tilde{\bm{z}}_i, \tilde{\bm{z}}_{-i})$.
	\begin{align}
\tilde{f}_i(\tilde{\bm{z}}_i, \tilde{\bm{z}}_{-i})=f_i (\tilde{\bm{z}}_i, \tilde{\bm{z}}_{-i})- \mathbb{E}_\omega \sum_{t\in \mathcal{T}} \text{VOLL} \cdot p_i^{sh,\omega}[t]. \label{mech:penalty}
\end{align}
Accordingly, we update the game-theoretical model $G$.

\subsubsection{Analysis of the Nash equilibrium}
To demonstrate the key insights, we analyze the Nash equilibrium under the assumption of a quadratic supply cost of CERs, i.e., at hour $t$ of scenario $\omega$, 
\begin{align}
g^{cv,\omega}[t](p)=\frac{1}{2} a^\omega[t] \cdot  {p}^2+ b^\omega[t]\cdot  {p}+c^\omega[t],~a^\omega[t]>0.\label{eq:cercost}
\end{align}
The analytical results in the rest of Section \ref{section:design} are based on the quadratic-cost assumption. In Section \ref{section:sim}, we use   CAISO data to characterize such quadratic supply-cost functions. We leave a more general cost function as future work.

\textit{Computing Nash equilibrium:} We characterize one Nash equilibrium for the updated competition game $G$ considering the penalty payment, which is computed in Proposition \ref{prop:penaltyfinite}.

\begin{prop}\label{prop:penaltyfinite}
One Nash equilibrium coincides with the optimal solution to the following problem, where $\Phi^\omega(\tilde{\bm{z}}_i,\tilde{\bm{z}}_j)=\tilde{A}_i^\omega \cdot \tilde{A}_j^\omega.$
	\begin{align}	\hspace{-0.9ex}\max_{\bm{z}}~\hspace{-1.2mm}	&\sum_{i\in \mathcal{I}} \tilde{f}_i(\tilde{\bm{z}})+\mathbb{E}_{\omega \in \Omega}\Big[\sum_{t\in \mathcal{T}}\sum_{1\leq i<j\leq I} \hspace{-0.8mm}a^\omega[t] \hspace{-0.5mm}\cdot \hspace{-0.5mm}\Phi^\omega(\tilde{\bm{z}_i},\tilde{\bm{z}}_j)\Big],\label{eq:nashobj}\\
		\text{s.t.}~	&\tilde{\bm{z}}_i\in \tilde{\mathcal{Z}}_i^o,\forall i\in \mathcal{I},~\eqref{eq:const_balancemodify_new}.\notag
	\end{align}
\end{prop}
In the objective \eqref{eq:nashobj}, the first part $\sum_{i\in \mathcal{I}} \tilde{f}_i(\tilde{\bm{z}})$ is the total profit of all the investors. The
second part with $\Phi^\omega(\tilde{\bm{z}}_i,\tilde{\bm{z}}_j)$ is an additional item indicating the impact of strategic interactions among investors. 

The objective function \eqref{eq:nashobj} does not have a direct physical or economic interpretation. We characterize this objective function based on mathematical results to compute Nash equilibrium.   {One approach to prove Proposition \ref{prop:penaltyfinite} is based on the potential game theory \cite{monderer1996potential} \cite{la2016potential}, which states that if there exists a potential function for this game, the optimum of the potential function is one Nash equilibrium. Our game-theoretical model is in nature a Cournot competition model. Based on the linear supply function, we can characterize a potential function, i.e.,  the objective \eqref{eq:nashobj}. A stylized Cournot model with a characterized potential function is given in  \cite{monderer1996potential}. Our previous work applies the result in \cite{monderer1996potential} to the storage competition problem \cite{zhao2022storagecaiso}, and this work applies it to the mechanism design for the joint investment of VRE and ES. We show the details of the potential game theory in Appendix IV \cite{marketappendix}.  Another perspective of proving Proposition \ref{prop:penaltyfinite}  is to establish the KKT condition equivalence between the game model and the centralized optimization problem in Proposition \ref{prop:penaltyfinite}.  In Appendix IV, we show that the KKT conditions of the centralized optimization problem in Proposition \ref{prop:penaltyfinite} suffice to satisfy the KKT conditions for all the investors at the Nash equilibrium. }

The optimization problem in Proposition \ref{prop:penaltyfinite} is quadratic programming, which can
be efficiently solved. Note that our work proposes a penalty payment for investors and analyzes the equilibrium result. There can be multiple equilibria of the lost load allocation among investors. How to design the rules  to distribute the lost load to the individual investors is beyond the scope of this work but  deserves future investigations.

\textit{Discussion on the Nash equilibrium set:} The Nash equilibrium computed in Proposition \ref{prop:penaltyfinite} may not give all the pure-strategy Nash equilibria. If we do not have the coupled strategy constraint \eqref{eq:const_balancemodify_new}, the set of  all the pure-strategy  Nash equilibria will exactly coincide with the solutions in Proposition \ref{prop:penaltyfinite} based on the equivalent KKT conditions. 
However, with the coupled strategy constraint \eqref{eq:const_balancemodify_new}, the solutions in Proposition \ref{prop:penaltyfinite} can only give a subset of all the 
 Nash equilibria. The reason is that
the optimization problem in Proposition \ref{prop:penaltyfinite} will generate a uniform dual variable for  \eqref{eq:const_balancemodify_new}. However, when each investor optimizes its profit under the  constraint \eqref{eq:const_balancemodify_new} in the game model, the dual variables wrt.  \eqref{eq:const_balancemodify_new} can be different among investors. Thus, to ensure that Proposition \ref{prop:penaltyfinite} gives all the equilibria, we may need additional constraints to make the dual variables of \eqref{eq:const_balancemodify_new} identical among investors. How to introduce and implement such additional constraints for investors is beyond this paper's scope. We will just focus on the Nash equilibrium from  Proposition \ref{prop:penaltyfinite} in the rest of the paper.

\textit{Perfect competition:} We prove in Proposition \ref{prop:penaltyinf} that under perfect competition, the Nash equilibrium will coincide with the social optimum. 
\begin{prop}\label{prop:penaltyinf}
Considering a set of $K$ types of investors $\mathcal{K}$. Each type $k$ has $N^k$ investors. As $N^k\rightarrow +\infty$ for any $k$, the Nash equilibrium approaches the social-optimum solution.
\end{prop}

 {\textit{Proof sketch:} The objective function in Proposition \ref{prop:penaltyfinite}  for computing Nash equilibrium  is equivalent to 
\begin{align}
\max ~C^0\hspace{-0.6mm}-\hspace{-0.6mm}\text{System cost}\hspace{-0.6mm}-\hspace{-0.6mm}\mathbb{E}_\omega \sum_{i\in \mathcal{I}}\sum_{t\in \mathcal{T}}\frac{1}{2}\hspace{-0.4mm}\cdot \hspace{-0.4mm}a_n^\omega[t]\hspace{-0.4mm}\cdot\hspace{-0.4mm} (\tilde{A}_i^{\omega}[t])^2, \label{eq:eqiv}
\end{align}
where $C^0$ is some constant value. Here we can see that the gap between the Nash equilibrium and  the social-optimum benchmark comes from the item $\mathbb{E}_\omega \sum_{i\in \mathcal{I}}\sum_{t\in \mathcal{T}}\frac{1}{2}\cdot a_n^\omega[t]\cdot (\tilde{A}_i^{\omega}[t])^2$. Note that the total invested capacity of energy resources or the total market supply is upper-bounded in the market, which implies $\sum_{i\in \mathcal{I}^k} \mid \tilde{A}_i^{\omega}[t]\mid \leq \textit{(Certain upper bound)}$ for any type $k\in \mathcal{K}$. As the number of each  investor type approaches infinity,  the item  $\sum_{i\in \mathcal{I}}\mathbb{E}_\omega \sum_{t\in \mathcal{T}}\frac{1}{2}\cdot a_n^\omega[t]\cdot (\tilde{A}_i^{\omega}[t])^2$ will approach zero due to the quadratic formulation.}

Although the Nash equilibrium coincides with the social optimum under perfect competition,  a finite number of investors  lead to imperfect competition, where investors can withhold capacities to increase market prices and  get higher profits.

\vspace{-1ex}
\subsection{Supply incentive}

 {Under the finite number of investors, as shown in \eqref{eq:eqiv},  the gap between the Nash equilibrium and the social-optimum benchmark comes from the item $\sum_{i\in \mathcal{I}}\sum_{t\in \mathcal{T}}\frac{1}{2}\cdot a_n^\omega[t]\cdot (\tilde{A}_i^{\omega}[t])^2$. To bridge the gap, we design a variable supply incentive for each hour of each scenario based on the  quadratic values of  supply $\tilde{A}_i^\omega[t]$ for each investor. We present the new profit $\tilde{\tilde{f}}_i$ of investor $i$ as follows.}\footnote{ {The design of penalty payments and supply incentives is to theoretically ensure that the Nash
equilibrium  approaches the social-optimum outcome under certain conditions. If we change the value of supply incentives, there is no such  theoretical guarantee. Besides, arbitrarily increasing the supply incentive can intuitively promote investment in LERs and mitigate investors' strategic behaviors. However, if we simply adjust the quadratic coefficient by a factor $\xi$, i.e., supply incentive  
$\frac{1}{2}\cdot \xi\cdot a^\omega[t] \cdot (\tilde{A}_i^\omega[t])^2$, we will have a non-convex structure of the investor's profit function when $\xi\geq 3$. The existence of the pure-strategy Nash equilibrium can not be ensured. The discussion on the design of other structures of supply incentives is  beyond the current game-theoretical framework. We leave it as future work.}}
\vspace{-1ex}
	\begin{align}
\tilde{\tilde{f}}_i (\tilde{\bm{z}}_i, \tilde{\bm{z}}_{-i})=\tilde{f}_i (\tilde{\bm{z}}_i, \tilde{\bm{z}}_{-i})+ \mathbb{E}_\omega \sum_{t\in \mathcal{T}} \frac{1}{2}a^\omega[t](\tilde{A}_i^\omega[t])^2. \label{mech:incentive}
\end{align}

 {With such a supply incentive, the objective function \eqref{eq:eqiv} for computing Nash equilibrium  is changed to
\begin{align}
\max ~C^0-\text{System cost},
\end{align}
which is exactly equivalent to the objective of the social-optimum benchmark. Therefore,  under the penalty payment and supply incentive (PI-mechanism), the Nash equilibrium can reach the social optimum, as concluded in Proposition \ref{prop:incentive}.}

\vspace{-0.5ex}
\begin{prop}\label{prop:incentive}
Under the PI mechanism, one Nash equilibrium coincides with the social-optimum solution.
\end{prop}

Although the penalty payment and supply incentive can make the  Nash equilibrium coincide with the social optimum, the investors may face profit loss. As more CERs retire in the market, i.e., $\gamma$ goes down, the maximum market price 
$\frac{d g^{cv,\omega}[t]}{d p^{cv,\omega}[t]}\Big|_{\gamma\cdot \bar{p}^{cv}}$ decreases, which can incur negative profits to investors due to the penalty payment. 
\begin{figure}[t]
	\centering
	\includegraphics[width=3.43in]{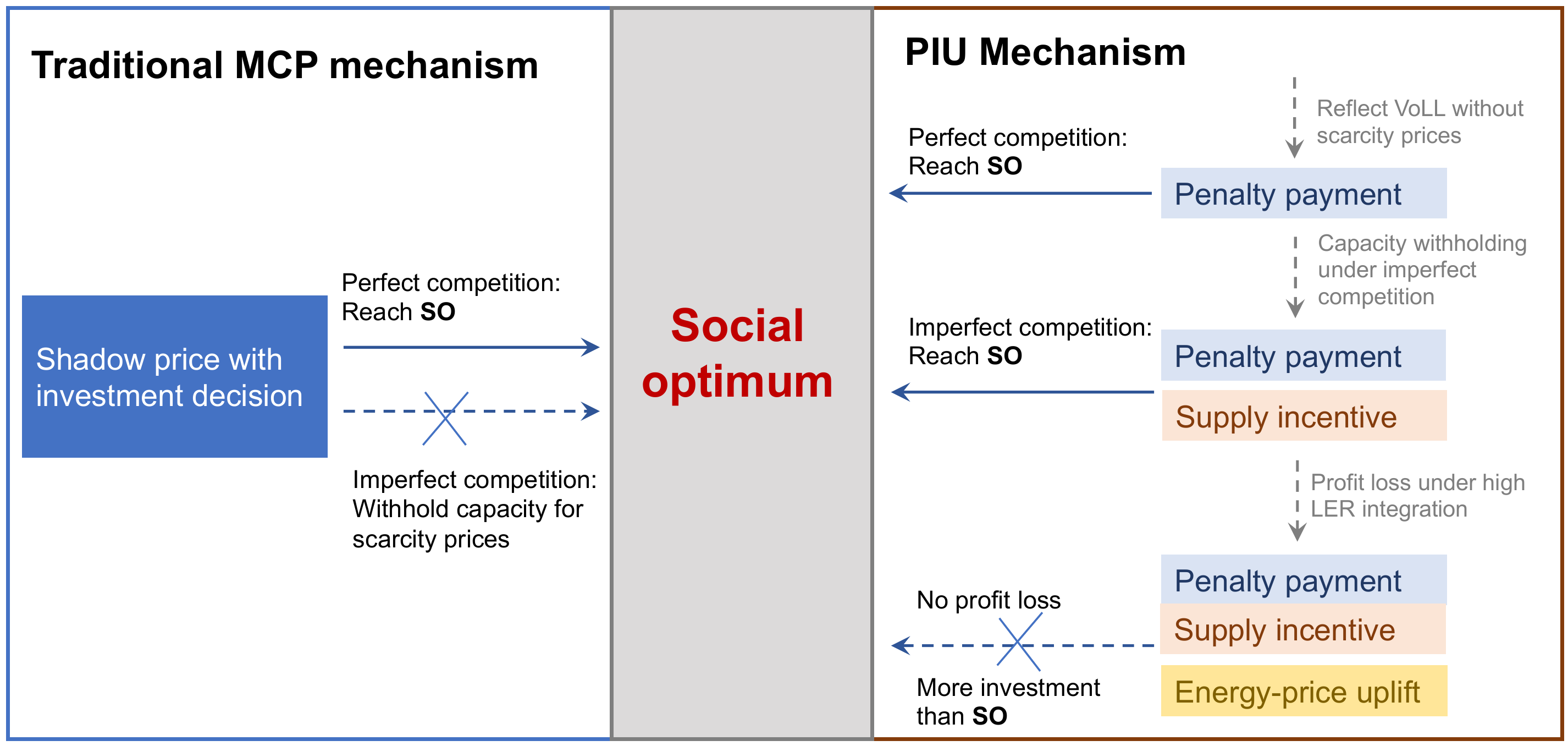}
	\vspace{-0.5ex}
	\caption{\small Equilibrium under the MCP and PIU mechanisms  vs. social-optimum benchmark.}
	\label{fig:comp}
	\vspace{-2ex}
\end{figure}
\vspace{-1ex}
\subsection{Energy-price uplift}

To ensure adequate revenue for investors, we propose an uplift of energy prices $\Delta \pi^\omega[t]\geq 0$:
\vspace{-0.5ex}
\begin{align}
\tilde{\pi}^\omega[t] (\bm{z})=\hat{\pi}^\omega[t] (\bm{z}) +\Delta \pi^\omega[t]. \label{mech:uplift}
\end{align}
which is added to the market price \eqref{eq:ourprice} as above. 

 {The energy-price uplift serves as the basic rate for market prices, which means suppliers will get additional  price uplift $\Delta \pi^\omega[t]$ besides the market rate $\hat{\pi}^\omega[t] (\bm{z}).$ This can provide additional revenues for suppliers. The uplift $\Delta \pi^\omega[t]$ is set as parameters in this work, which can be flexibly adjusted at any time $t$ and scenario $\omega$.}

Adding the price uplift increases investors' profits, but makes the Nash equilibrium deviate from the social optimum. Under the penalty payment, supply incentive, and price uplift (PIU-mechanism), Proposition \ref{prop:uplift} shows that the Nash equilibrium coincides with the solution to a modified Problem $\textbf{SO}$.

\begin{prop}\label{prop:uplift}
Under the PIU mechanism, one Nash equilibrium is  the optimum solution to Problem \textbf{SO} when we change the linear cost of the CERs from $b^\omega[t]$ in \eqref{eq:cercost} to $b^\omega[t]+\Delta \pi^\omega[t]$.
\end{prop}

 Proposition \ref{prop:uplift} shows that the price uplift will equivalently increase the cost of the CERs.  This will reduce the supply from the CERs and incentivize more investment and supply from LERs.  Furthermore, the increasing retirement of the CERs will lessen the impact of the CERs' cost, which implies the Nash equilibrium will be closer to the social optimum.   The uplift can be interpreted as  a long-term incentive in practice, such as premium fees of insurance. How to design and implement the uplift is an interesting problem for future research. We later show some simulation results on the uplift in Section \ref{section:sim}.
 
 In Fig.\ref{fig:comp}, we summarize the equilibrium under the  MCP and PIU mechanisms compared with the \textbf{SO} benchmark.  {Compared with MCP, the advantages of the PIU mechanism are further summarized below:  (i)  Under imperfect competition, the PIU mechanism can prevent the strategic behaviors of investors from incurring scarcity prices. As shown in Proposition \ref{prop:shadowimperfect}, when VOLL is higher than a threshold, the Nash equilibrium under MCP is that investors will withhold capacity and supply to incur scarcity prices. However, such strategic behaviors will not occur under the PIU mechanism. (ii) Under perfect competition, the proposed PIU mechanism can significantly reduce the energy cost of consumers and reduce excessive revenues of low-cost CERs compared with MCP. We demonstrate it in detail in Section \ref{section:sim}.E.}

\vspace{-2ex}
\section{Numerical studies}\label{section:sim}
\vspace{-1ex}

We conduct numerical studies using CAISO data. We first introduce the simulation setup, especially how we use the market data to approximate the supply cost of CERs. Then, we discuss the equilibrium results under the P, PI, and PIU mechanisms, respectively. In Proposition \ref{prop:shadowimperfect}, we have shown that the traditional MCP mechanism may cause manipulated scarcity prices and  zero surplus for consumers under imperfect competition. In simulations,  we will further compare the PIU mechanism with the  MCP mechanism under perfect competition, which shows that the PIU mechanism can still achieve a higher consumer surplus.

\vspace{-2ex}
\subsection{Simulation setup}
\vspace{-0.5ex}

\subsubsection{Data-driven system cost model}  We  assume a quadratic supply-cost function for CERs and introduce how to utilize the historical data from the CAISO to characterize it.

 We consider the historical market price as the system marginal cost, and use this to approximate the marginal cost of  CERs. Specifically, we examine the three-year market data from 2019 to  2021 in CAISO, where each day corresponds to one scenario $\omega$ and is divided into 24 hours $\mathcal{T}$.  We utilize the day-ahead prices $\pi_0^\omega[t]$, day-ahead forecast system demand $D_0^\omega[t]$, and system VRE of solar and wind energy $R_0^\omega[t]$ \cite{calidata}. Recall that the marginal cost of the CERs is represented as $a^\omega[t] \cdot  p^{cv,\omega}[t] + b^\omega[t]$ based on \eqref{eq:cercost}. Then, the net demand $D^\omega[t]=D_0^\omega[t]-R_0^\omega[t]$ is assumed to be served by the existing CERs only in the historical data, which leads to the following expression for the historical market price $\pi_0^\omega[t]$, i.e., 
\begin{align}
\pi_0^\omega[t]=a^\omega[t] \cdot  (D_0^\omega[t]-R_0^\omega[t])+ b^\omega[t]. \label{eq:con_cost_characterization}
\end{align}
Next, we will analyze the historical data to characterize the quadratic coefficient $a^\omega[t]$ based on linear regression. Once we obtain $a^\omega[t]$, we  compute $b^\omega[t]$ according to \eqref{eq:con_cost_characterization}. Note that the fixed cost $c^\omega[t]$ in (19) will not impact the social-optimum solution or Nash equilibrium result; so, we set it to zero.

\begin{figure*}[ht]
	\centering
	\hspace{-1ex}
	\subfigure[]{
		\raisebox{-2mm}{\includegraphics[width=3in]{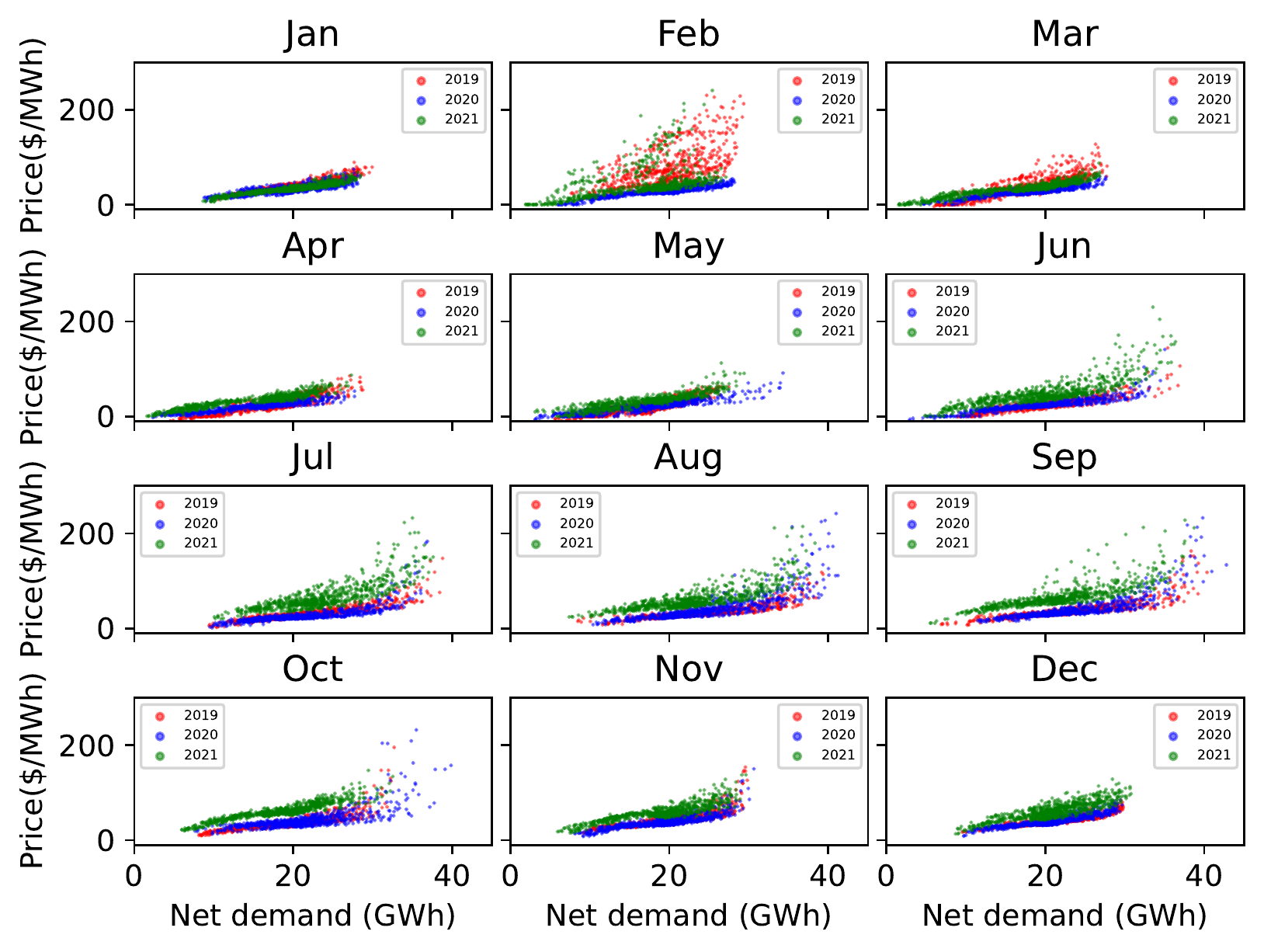}}}
	\hspace{-1ex}
	\subfigure[]{
		\raisebox{-2mm}{\includegraphics[width=3in]{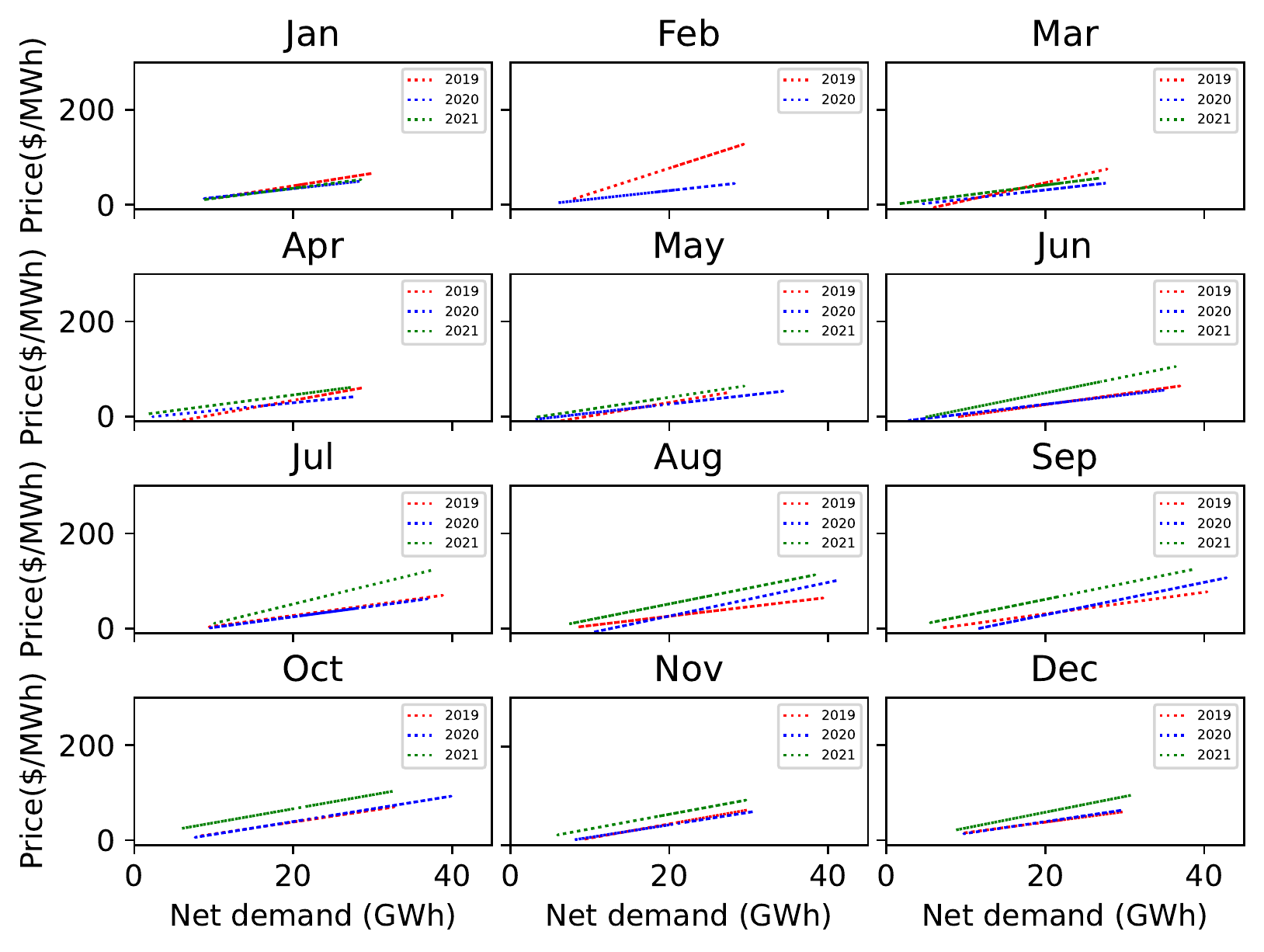}}}
	\vspace{-2mm}
	\caption{(a) \small Price as a function of net demand for all the hours in each month of 2019, 2020, and 2021; (b) Linear approximation between the price and net demand for each month of 2019, 2020, and 2021.}
	\label{fig:caiso_price_data}
	\vspace{-3ex}
\end{figure*}

To obtain $a^\omega[t]$, in Fig.\ref{fig:caiso_price_data}(a), we plot the data of day-ahead prices $\pi_o^\omega[t]$ and net demand  $D_0^\omega[t]-R_0^\omega[t]$ of all the hours with prices below 250\$/MWh  in each month from January to December in the years 2019 (red points), 2020 (blue points), and 2021 (green points), respectively.\footnote{We consider the prices below 250\$/MWh only due to the following reasons.  
  Prices over 250\$/MWh only occur in 0.33\% of all the hours. Some extremely high prices (e.g., over  600\$/MWh) occur due to some rare events like the Feb 2021 Texas outage, which have a limited impact on the market revenues in the long run. Second, one limitation of our linear model is that it cannot accurately model very high prices. A piece-wise linear price function can be more accurate to model the high prices but it also significantly increases the complexity of the analysis and the computational effort to find the Nash equilibrium. We leave this extension as future work.} Fig.\ref{fig:caiso_price_data} (a) indicates  that the linear model provides a reasonable approximation for the relationship between price and net demand.  In Fig.\ref{fig:caiso_price_data}(b), we use linear regression to characterize the slope $a^\omega[t]$ of supply functions for each month from January to December in the year 2019 (red curves), 2020 (blue curves), and 2021 (green curves), respectively.\footnote{There were abnormal prices in February 2021 due to the Texas outage. We exclude this month's data since the linear model does not have a good approximation.}  Note that we cluster the scenarios based on each month of each year and have a uniform $a^\omega[t]$ for each cluster. For the linear coefficient $b^\omega[t]$, it  can be different for each hour of each scenario according to \eqref{eq:con_cost_characterization}.

\subsubsection{Technology parameters} We consider three types of LERs: solar energy, wind energy, and energy storage, and focus on cost parameters for 2020. Specifically, we consider PV solar energy and onshore wind energy with capacity costs of 885\$/kW and 1355 \$/kW, respectively, and a lifetime of 25 years \cite{costsirena}. For storage, we use Li-ion batteries (LFP), with energy-capacity cost 385 \$/kWh, power-capacity cost 85 \$/kWh,  roundtrip efficiency 0.88, and a lifetime of 10 years \cite{mongird2020}. In the later sensitivity analysis, we reduce the capital cost below the 2020 baseline for illustration. We set the VOLL at 3500\$/MWh and vary it in some cases. Furthermore, we present all the results of profits and revenues in the values scaled to one day.

	\vspace{-1ex}
\subsection{P-mechanism: Imperfect competition}
With the penalty payment only, we show that imperfect competition can lead to the equilibrium system cost much higher than the social-optimum results because the investors withhold investment. We also find that the increased market competition due to more investors can change the capacity share of different resources in the market.

\begin{figure}[t]
	\centering
	\hspace{-2ex}
	\subfigure[]{
		\raisebox{-2mm}{\includegraphics[width=1.7in]{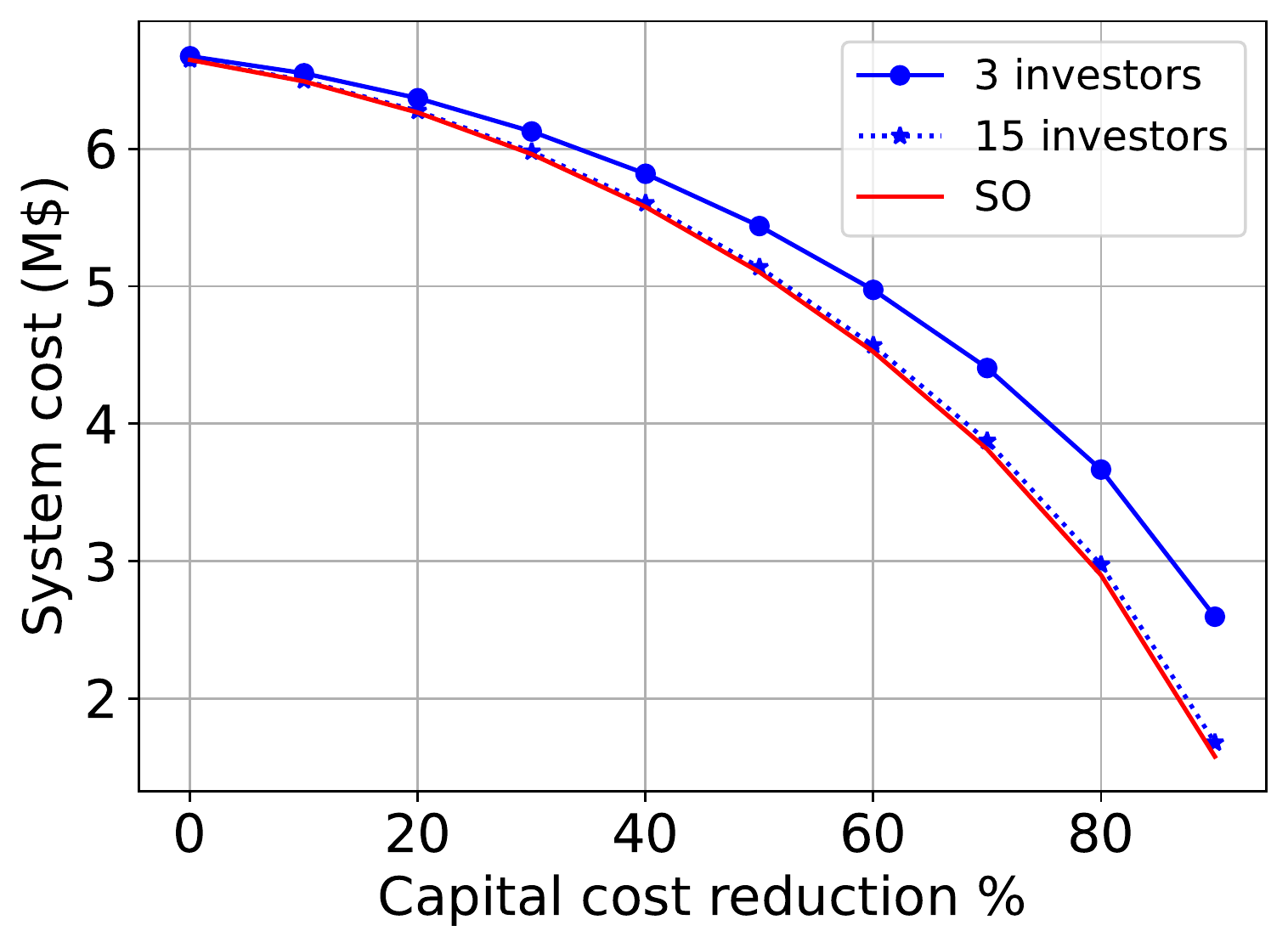}}}
	\hspace{-1.5ex}
	\subfigure[]{
		\raisebox{-2mm}{\includegraphics[width=1.7in]{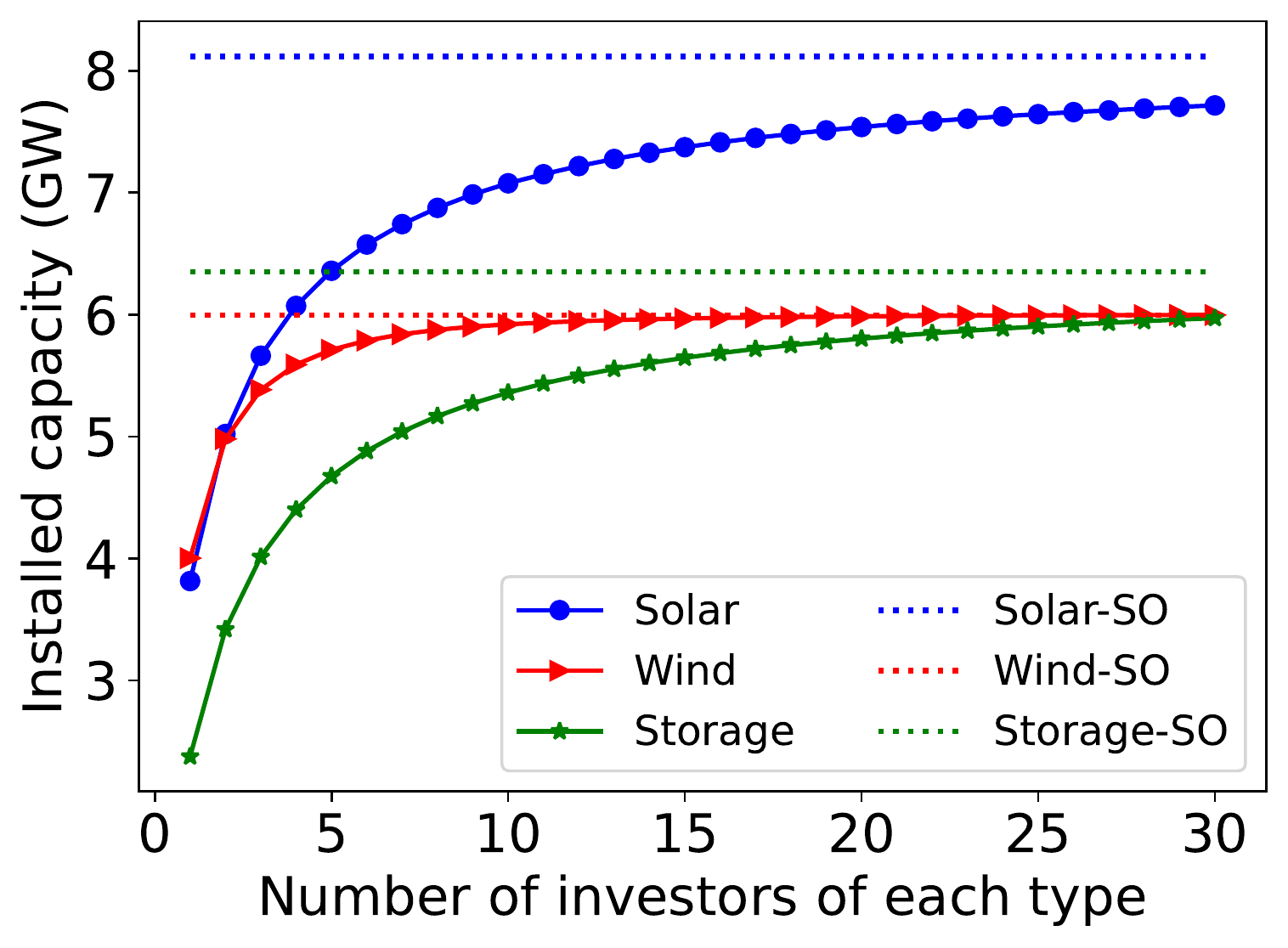}}}
	\vspace{-2mm}
	\caption{P-mechanism: (a) \small System cost with the capital cost reduction of new investments;  {(b) Total invested capacity of solar, wind, and storage with the number of investors of each resource.}}
	\label{fig:num}
		\vspace{-3ex}
\end{figure} 

In Fig.\ref{fig:num},  we focus on the P-mechanism. In Fig.\ref{fig:num}(a), we present the system costs as the capital cost of the LERs decreases. The blue solid curve shows the result under the competition of three investors, i.e., one solar-energy investor, one wind-energy investor, and one storage investor, while the blue dotted curve shows the competition among fifteen investors, i.e., five investors of each LER type. The red curve presents the social-optimum result.  Fig.\ref{fig:num}(b) shows the total installed power capacity of each LER type in the market when the investors' number of each type increases: the blue solid curve for solar energy, red for wind energy, and green for energy storage. The dotted curves are the social-optimum (SO) results. To better illustrate the impact of investors' number, in Fig.\ref{fig:num}(b), we reduce the capital cost by 80\%, 40\%, and 20\% for energy storage, solar energy, and wind energy, respectively.

In Fig.\ref{fig:num}(a),  the system cost under the 15-investor case is always close to the social-optimum result. However, the system cost under the 3-investor case can be much higher than the social-optimum result, especially if the capital cost is significantly reduced (by over 60\%). This reflects the ability of investors to withhold the invested capacity to support a higher market price, which is not beneficial to social welfare. In Fig.\ref{fig:num}(b), with more investors competing, the total capacity of each resource type will increase. The increasing number of investors will also change the market position of different resources. When there is only one investor for each type, the invested capacity of wind energy (red solid curve)  is larger than other types. However, as the number increases, the capacity of solar energy will surpass wind energy. As the number of each type goes to infinity, the results approach the social optimum, where wind-energy capacity will be the lowest (red dotted curve).

\vspace{-2ex}
\subsection{PI-mechanism: Impact of CER retirements}

Although the PI-mechanism can achieve social optimum, when the CERs retire,  the investors may face profit loss due to the decreasing market price and increasing demand for  LERs. We also find that in CAISO, a low CER retirement ratio results in more wind energy investments, while a high  ratio incentivizes more energy storage paired with solar energy.

\begin{figure}[t]
	\centering
	\hspace{-2ex}
	\subfigure[]{
		\raisebox{-2mm}{\includegraphics[width=1.7in]{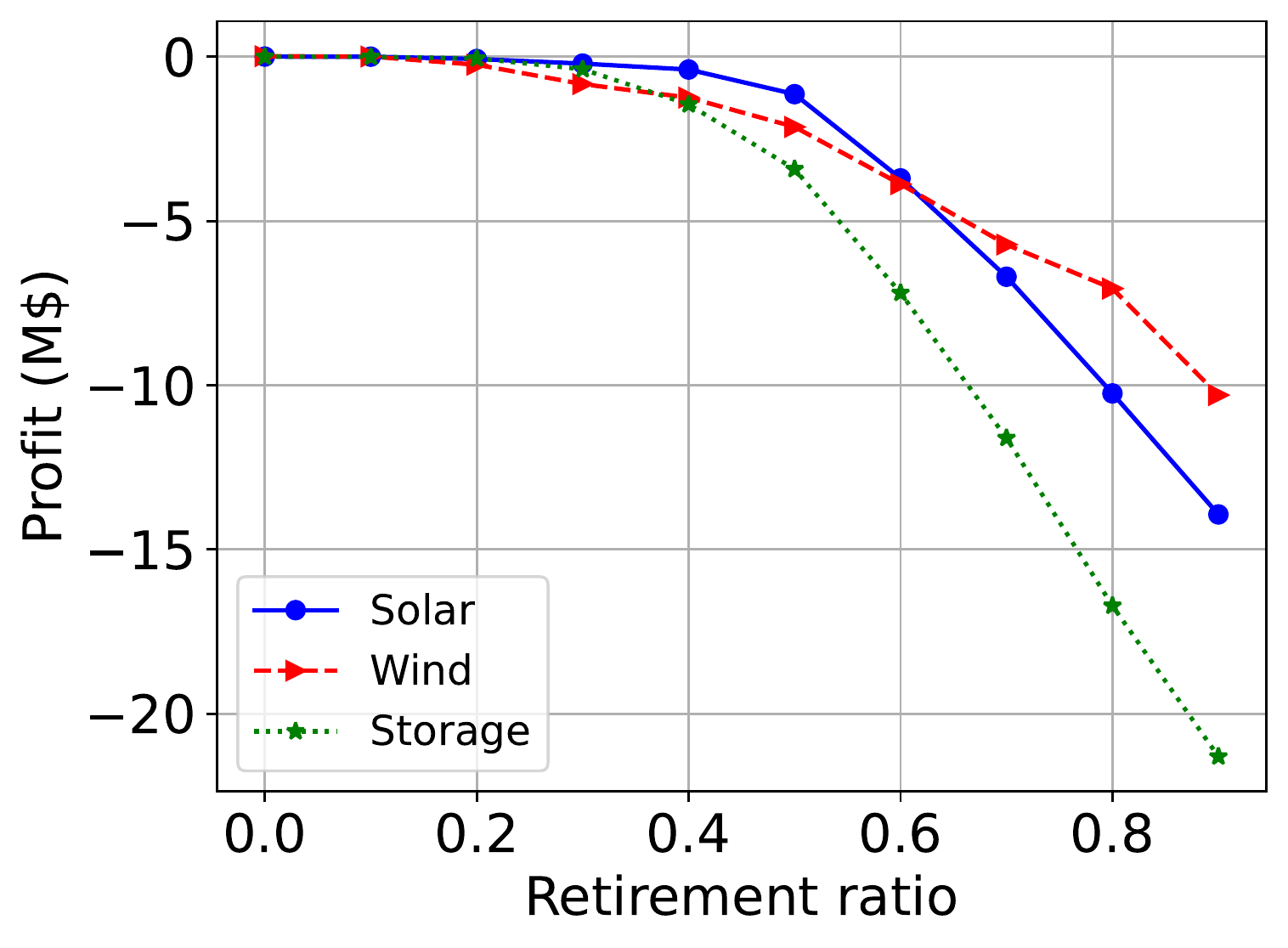}}}
	\hspace{-1.5ex}
	\subfigure[]{
		\raisebox{-2mm}{\includegraphics[width=1.7in]{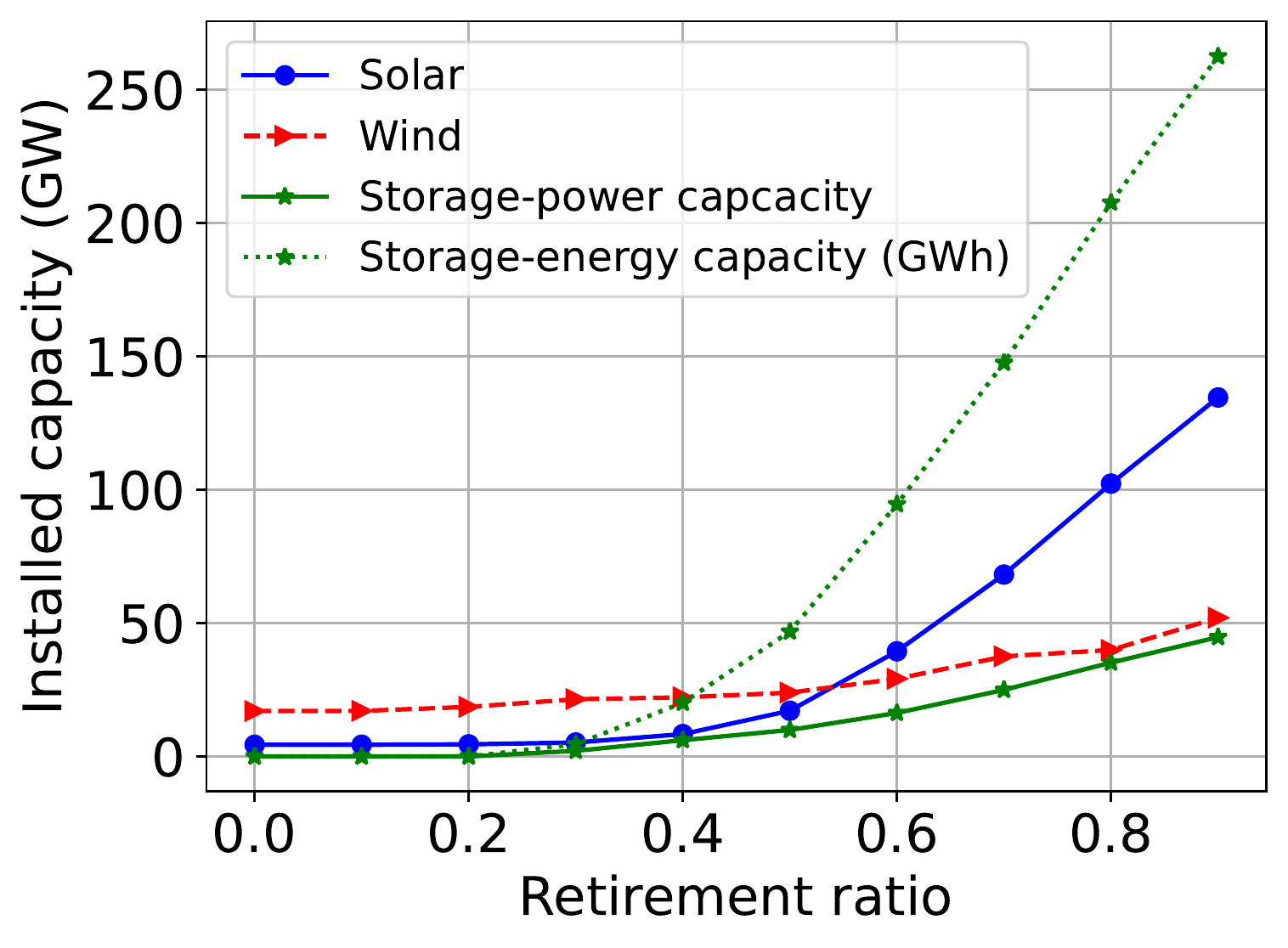}}}
	\vspace{-3mm}
	\caption{PI-mechanism: (a) \small Profit; (b) Capacity; As a function of CER retirement ratio under PI mechanism and 3 investors.}
	\label{fig:retire}
	\vspace{-2ex}
\end{figure}

In Fig.\ref{fig:retire},  we focus on the PI-mechanism where the investment results coincide with the social optimum. For illustration, we reduce the capacity costs of solar energy, wind energy, and storage by 50\%, respectively. We present the equilibrium profits of three investors in Fig.\ref{fig:retire}(a) and capacities in Fig.\ref{fig:retire}(b) as the CERs retire. We show the blue curve for solar energy, red curve for wind energy, and green curve for storage  (both power and energy capacity). 

In Fig.\ref{fig:retire}(a), when the retirement ratio of CERs is low, e.g., smaller than 10\%, all the investors have non-negative profits (zero for storage but small positive values for wind and solar energy). However, when the retirement ratio increases, the profits of the three investors turn negative. Note that if investors do not face the penalty payment of the lost load, the lowest profits are always zero since investors can choose not to invest. However, the penalty payment may result in a negative profit for the investor. In Fig.\ref{fig:retire}(b), the capacities of  LERs increase significantly to fill the supply gap under the retirement of the CERs. The optimal amount of investment trades off the costs of the LER investments and system lost load. Furthermore, as shown in Fig.\ref{fig:retire}(b), a lower or medium retirement ratio of the CERs, e.g., smaller than 40\%,  results in more wind energy investments (in red) than solar energy  (in blue) and storage (in green), while a high retirement ratio gives more solar energy paired with storage than wind energy.

\vspace{-1ex}
\subsection{PIU mechanism: Profit and system cost with price uplift}

The price uplift can flexibly adjust the investors' profits and prevent profit loss. Although the price uplift will lead to a higher system cost than the social optimum, a higher penetration level of LERs can mitigate this deviation.

\begin{figure}[t]
	\centering
	\hspace{-2ex}
	\subfigure[]{
		\raisebox{-2mm}{\includegraphics[width=1.7in]{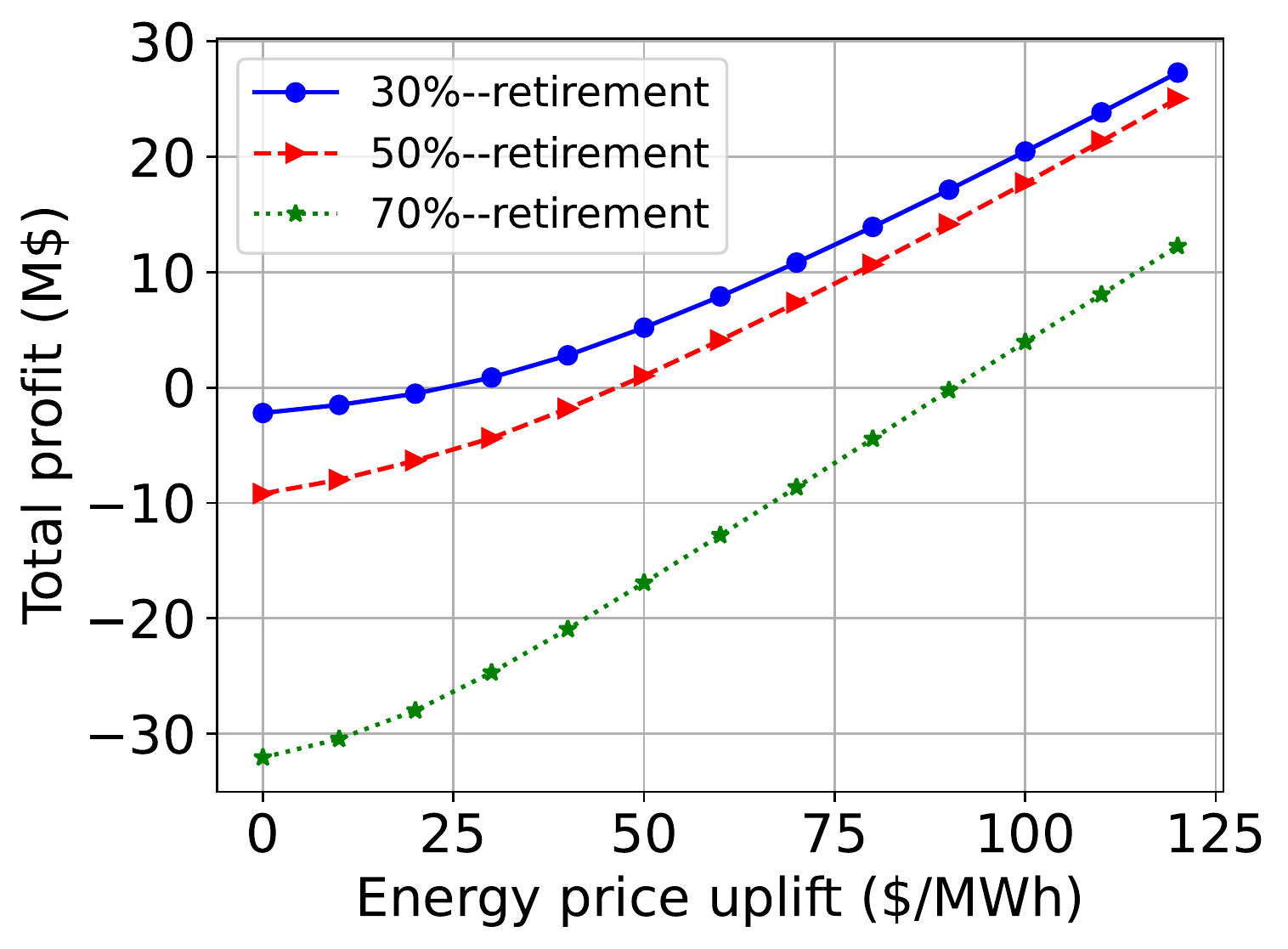}}}
	\hspace{-1.5ex}
	\subfigure[]{
		\raisebox{-2mm}{\includegraphics[width=1.7in]{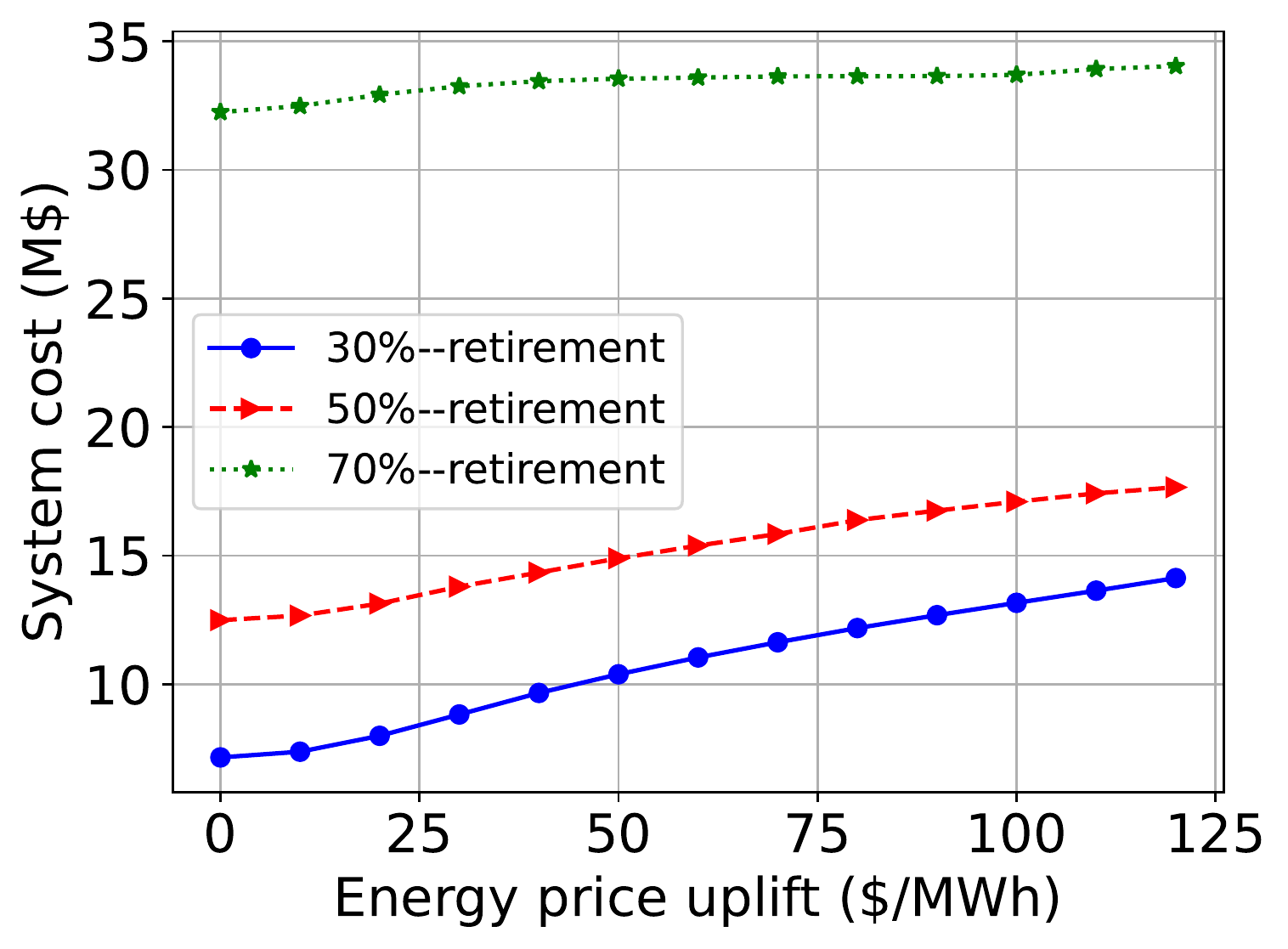}}}
	\vspace{-3mm}
	\caption{PIU-mechanism: (a) \small Total profit; (b) System cost; Both are with the energy price uplift under the different retirement ratios.}
	\label{fig:uplift}
		\vspace{-3ex}
\end{figure}

 In Fig.\ref{fig:uplift},  we focus on the PIU mechanism. As the price uplift (assumed uniform for  hours and scenarios) increases, we present the total profit of three investors in Fig.\ref{fig:uplift}(a) and the system cost in Fig.\ref{fig:uplift}(b), where the blue, red, green curves refer to retirement ratios of 30\%, 50\%, and 70\%, respectively. Considering future cost reduction, we reduce the capacity costs of solar energy, wind energy, and storage by 30\%, respectively.

In Fig.\ref{fig:uplift}(a), when the price uplift is zero, the total profit is negative. As the uplift increases, the total profit will turn positive. The break-even price uplift increases with the retirement ratio of the CERs, e.g.,  about 24\$/MWh under 30\% retirement (in blue) and about 90 \$/ MWh under 70\% retirement (in green). In Fig.\ref{fig:uplift}(b), when the  uplift is zero, the system cost is equal to the social optimum.  As the uplift increases, the system cost increases and deviates from the social optimum. However, such deviation is lessened with more retirement of the CERs. Under 70\% retirement, the system cost rises only slightly with the uplift, while the system cost increases significantly under 30\% retirement. The reason is that the uplift is equivalent to an increase in the operating cost of the CERs. This impact dampens when there are fewer CERs in the system. 

\vspace{-1ex}
\subsection{Economic surplus comparisons between the PIU mechanism and traditional MCP mechanism}

We show that the PIU mechanism can achieve a lower consumer cost than the  MCP mechanism, by reducing the excessive surplus of CERs. We summarize a surplus comparison table for all the  market participants in Appendix.I\cite{marketappendix}.

\subsubsection{Impact of energy price uplift} 
\begin{figure}[t]
	\centering
	\hspace{-2ex}
	\subfigure[]{
		\raisebox{-2mm}{\includegraphics[width=1.7in]{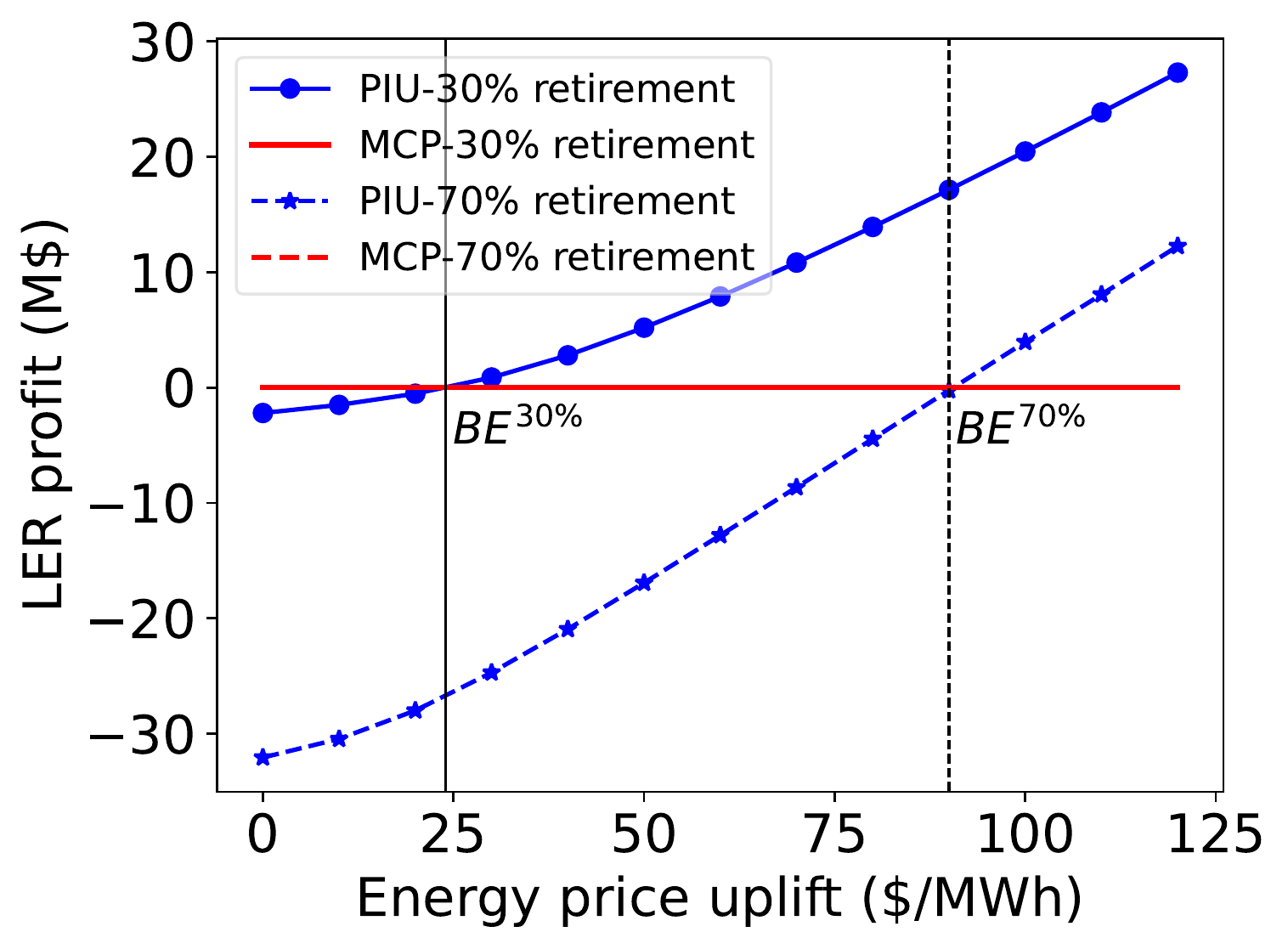}}}
	\hspace{-2ex}
	\subfigure[]{
		\raisebox{-2mm}{\includegraphics[width=1.7in]{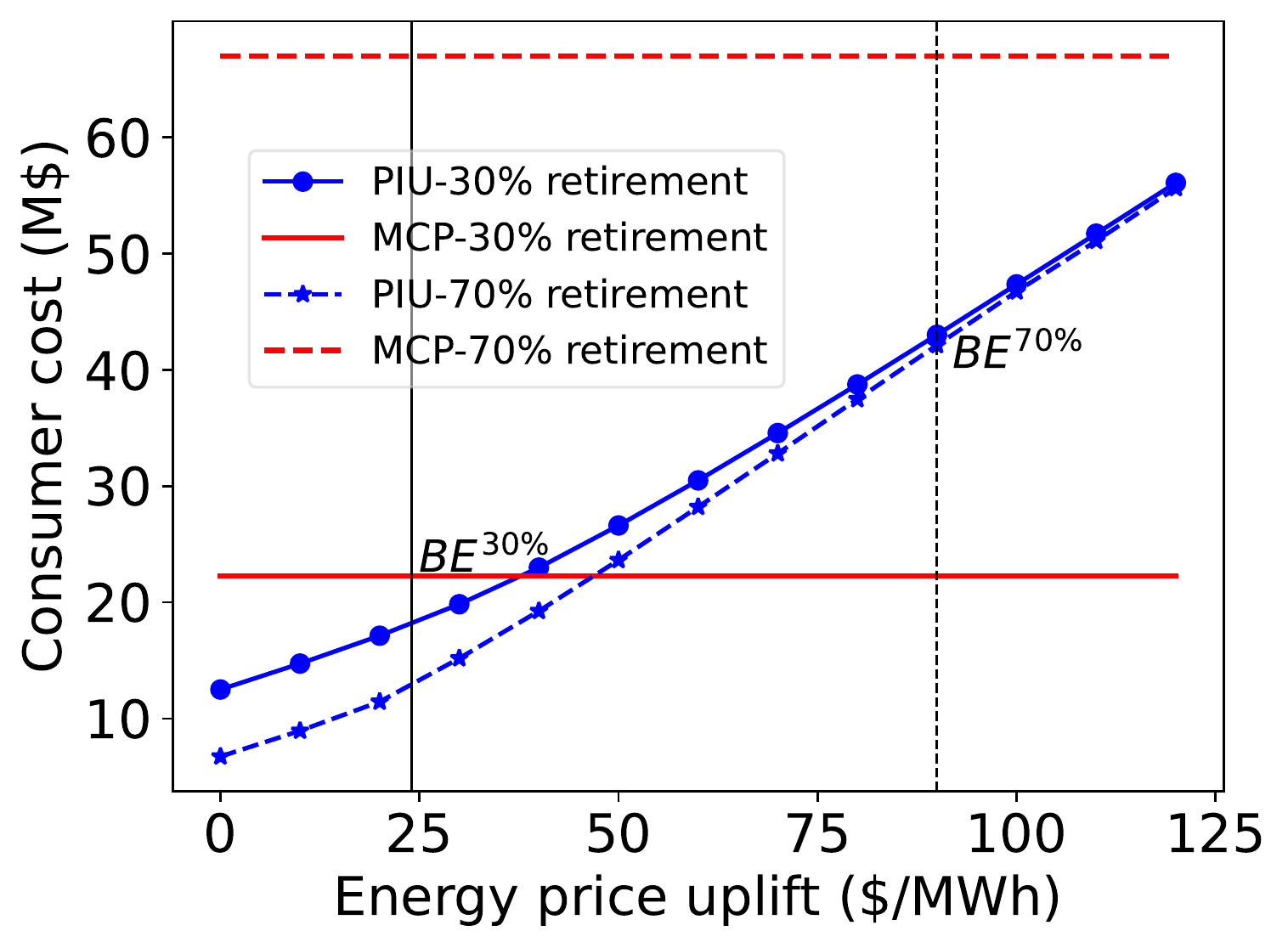}}}
	\hspace{-2ex}
	\subfigure[]{
		\raisebox{-2mm}{\includegraphics[width=1.7in]{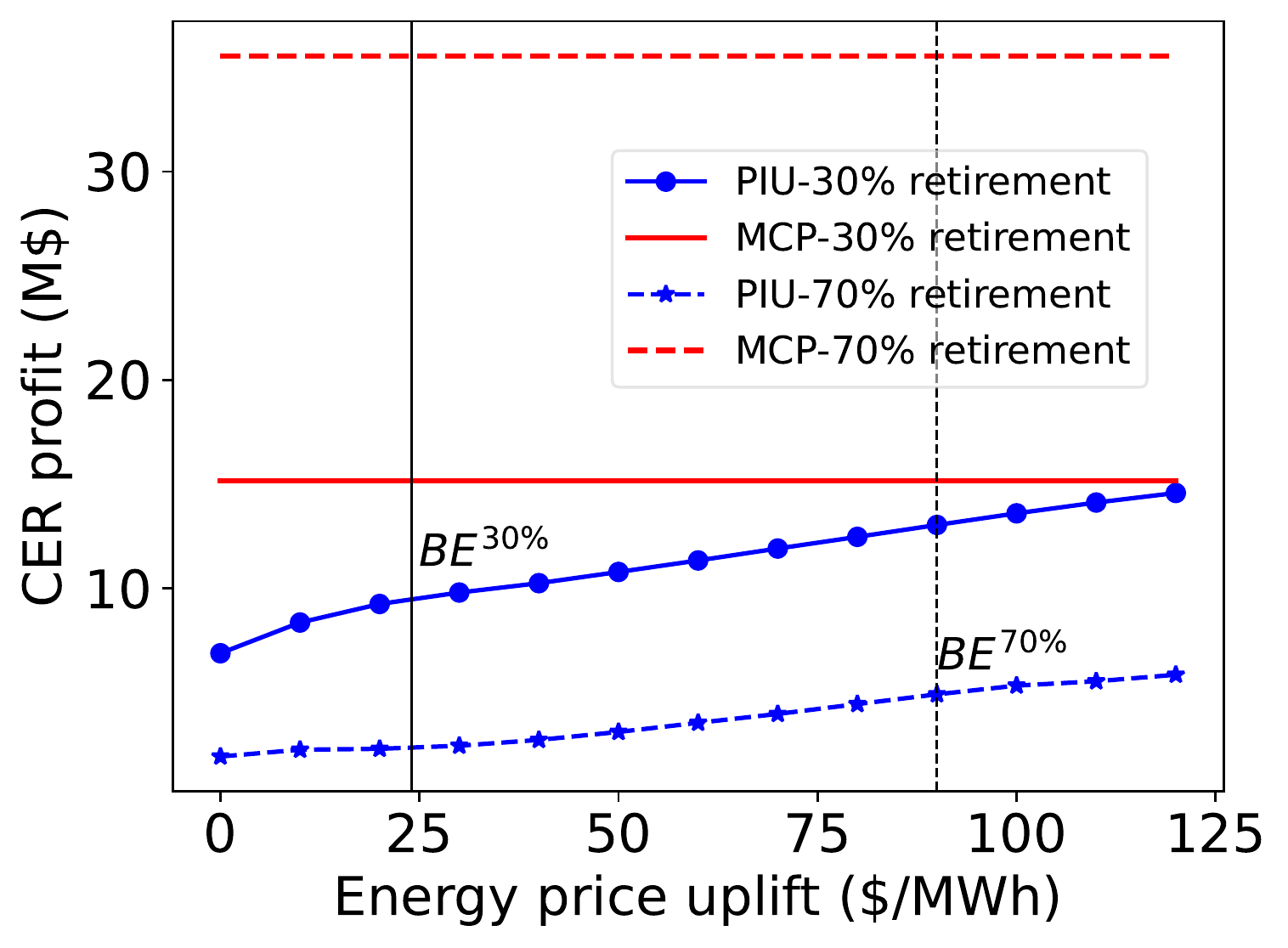}}}
			\hspace{-2ex}
	\subfigure[]{
		\raisebox{-2mm}{\includegraphics[width=1.7in]{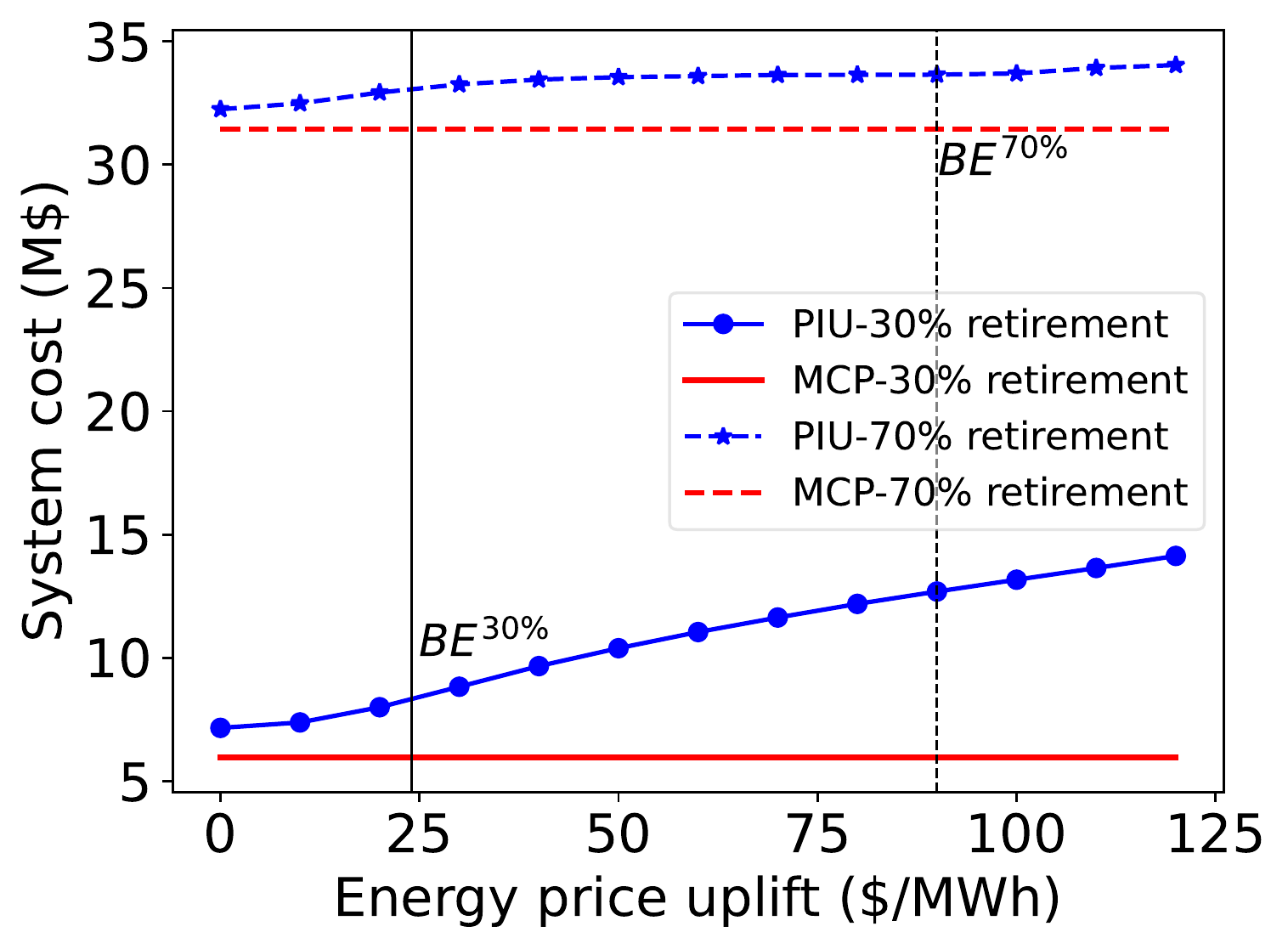}}}
	\vspace{-2mm}
	\caption{PIU vs MCP: (a) LER profit; (b) Consumer cost; (c)  CER profit; (d) System cost, as functions of price uplift.}
	\label{fig:surplus_uplift}
	\vspace{-2ex}
\end{figure}

We show how the energy price uplift will impact the surplus of participants under the PIU and  MCP mechanisms.  In Fig.\ref{fig:surplus_uplift}, we compare the profits and costs of different entities under the PIU and MCP mechanisms. In Proposition \ref{prop:shadowimperfect}, we have shown that the  MCP mechanism may cause manipulated scarcity prices under imperfect competition. We now focus on a comparison under perfect competition. In Fig.\ref{fig:surplus_uplift}(a)-(d), we present the LER profit, consumer cost, CER profit, and  system cost, respectively, as the price uplift varies. The blue and red curves show the PIU and MCP mechanisms, respectively.  The solid curves show 30\% retirement of CERs and the dashed curves show 70\% retirement. Considering future scenarios, we reduce the capacity cost of solar energy, wind energy, and storage by 30\% in this example.

In  Fig.\ref{fig:surplus_uplift}(a), the LER profit under the MCP mechanism is just zero. The profit under the PIU mechanism increases with the price uplift, and we denote the break-even points as ${BE}^{30\%}$ and ${BE}^{70\%}$ for the 30\% and 70\% CER retirement, respectively. In  Fig.\ref{fig:surplus_uplift}(b), as the uplift increases, the consumer cost under the PIU mechanism increases, but it is over 30\% lower than the MCP mechanism at the break-even point ${BE}^{70\%}$. The reason is that the CERs have a much higher profit under the MCP mechanism as shown in Fig.\ref{fig:surplus_uplift}(c), which increases the consumer cost. Comparing the red curves of Fig.\ref{fig:surplus_uplift}(c), the  CER profit under 30\% retirement is much lower than that under 70\% retirement. The high ratio of CER retirement creates more frequent scarcity prices and lets the remaining low-cost CERs get excessive profits under MCP mechanism, which is absent in the PIU mechanism. In Fig.\ref{fig:surplus_uplift}(d), the system cost under the PIU mechanism is higher than the social-optimum under the MCP mechanism, but the gap at the break-even point ${BE}^{70\%}$  is less than 7\%, when the CER-retirement ratio is high at 70\%.

\subsubsection{Impact of VOLL}  
\begin{figure}[t]
	\centering
 \hspace{-1ex}
	\subfigure[]{
		\raisebox{-2mm}
  {\includegraphics[width=1.6in]{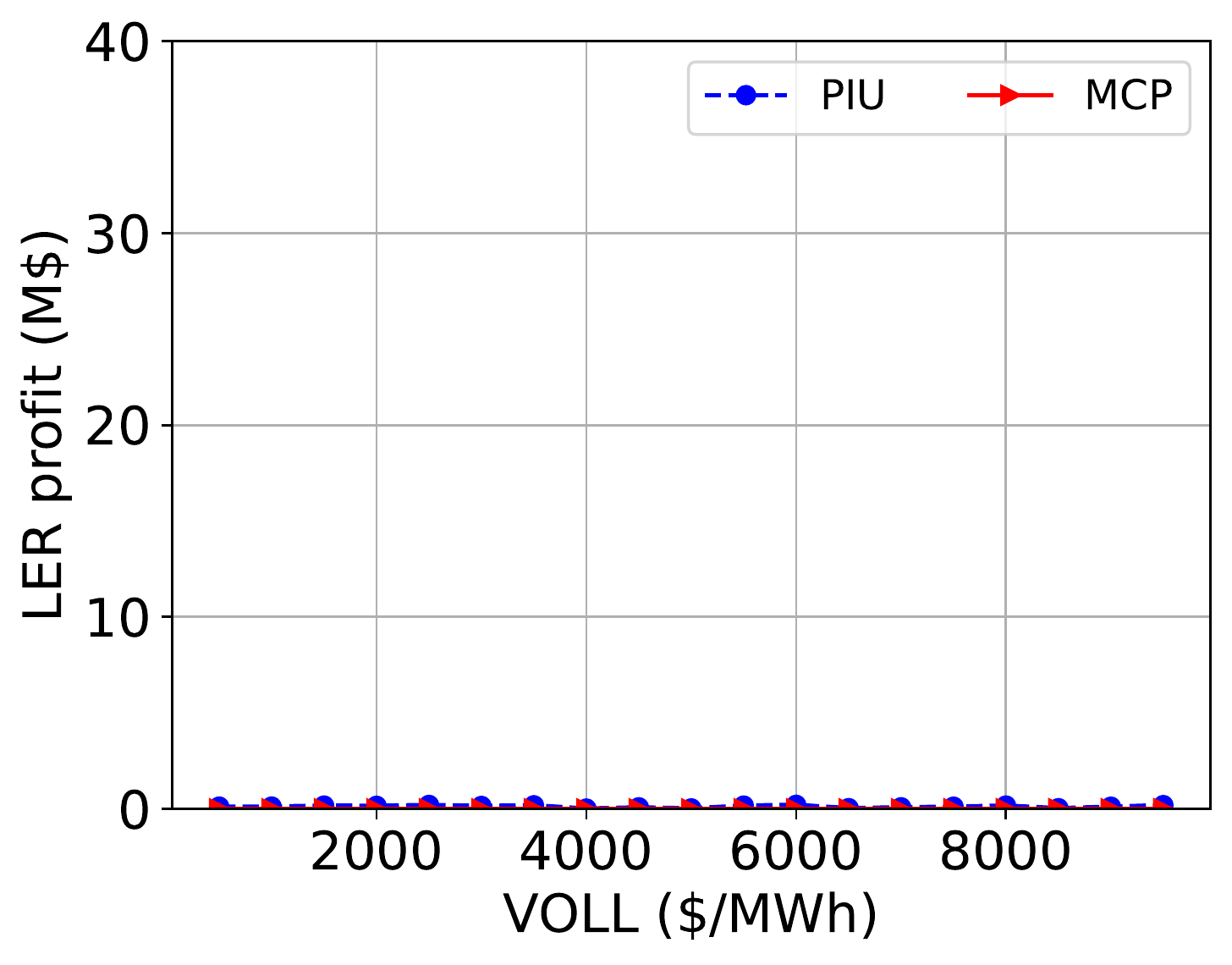}}}
	\hspace{-1ex}
	\subfigure[]{
		\raisebox{-2mm}
  {\includegraphics[width=1.6in]{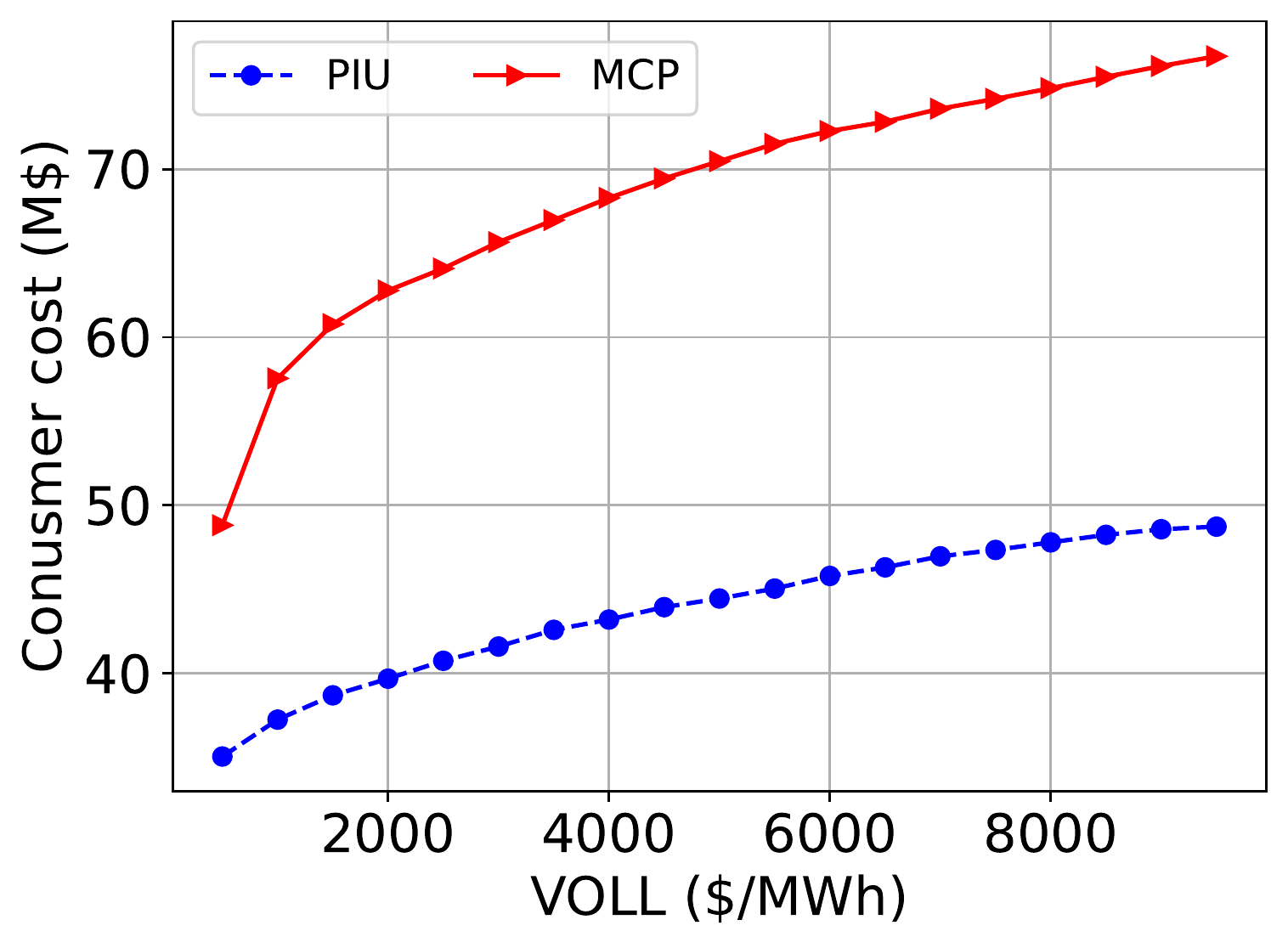}}}
	\hspace{-1ex}
	\subfigure[]{
		\raisebox{-2mm}{\includegraphics[width=1.6in]{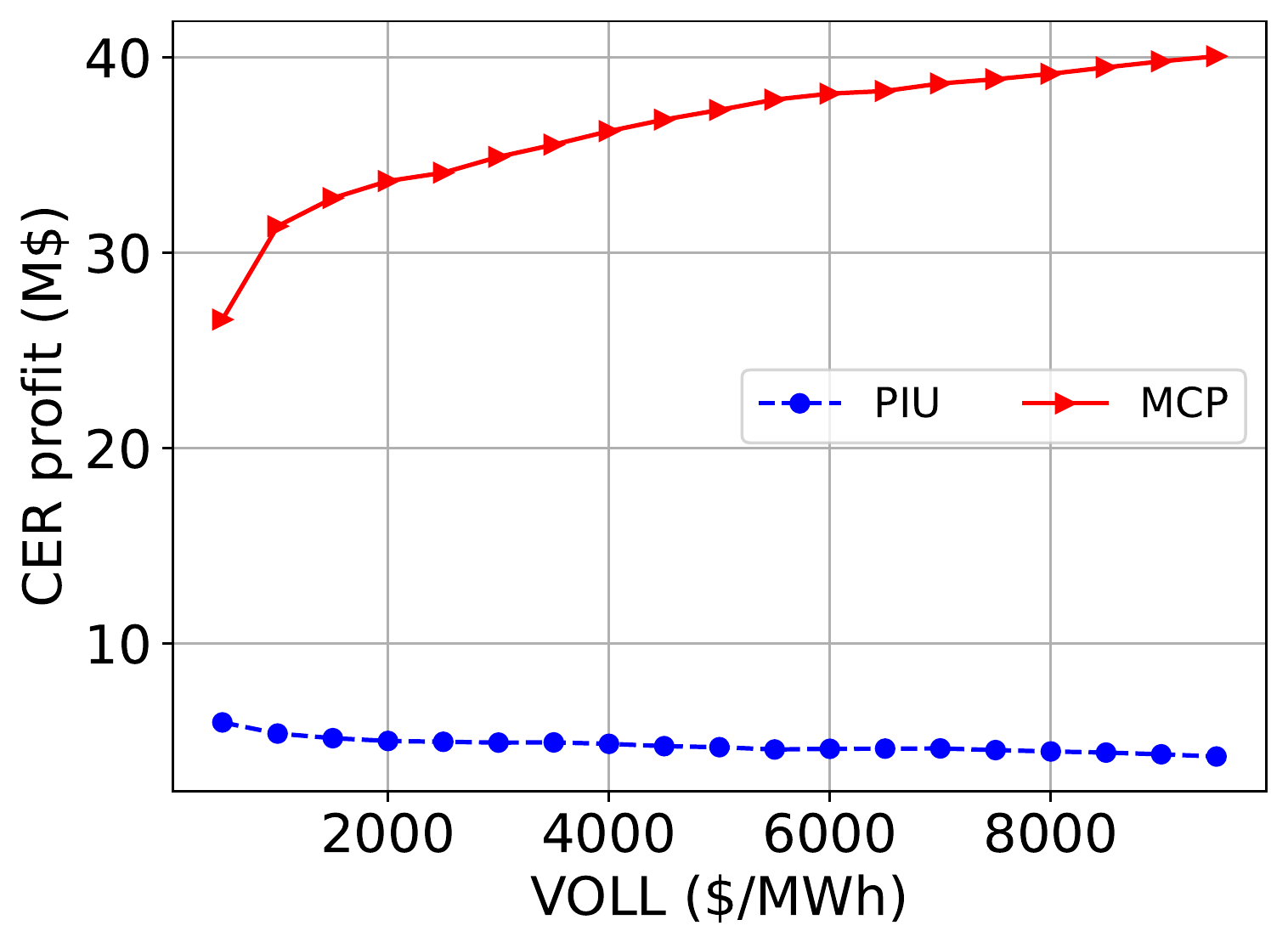}}}
  	\hspace{-1ex}
	\subfigure[]{
		\raisebox{-2mm}{\includegraphics[width=1.6in]{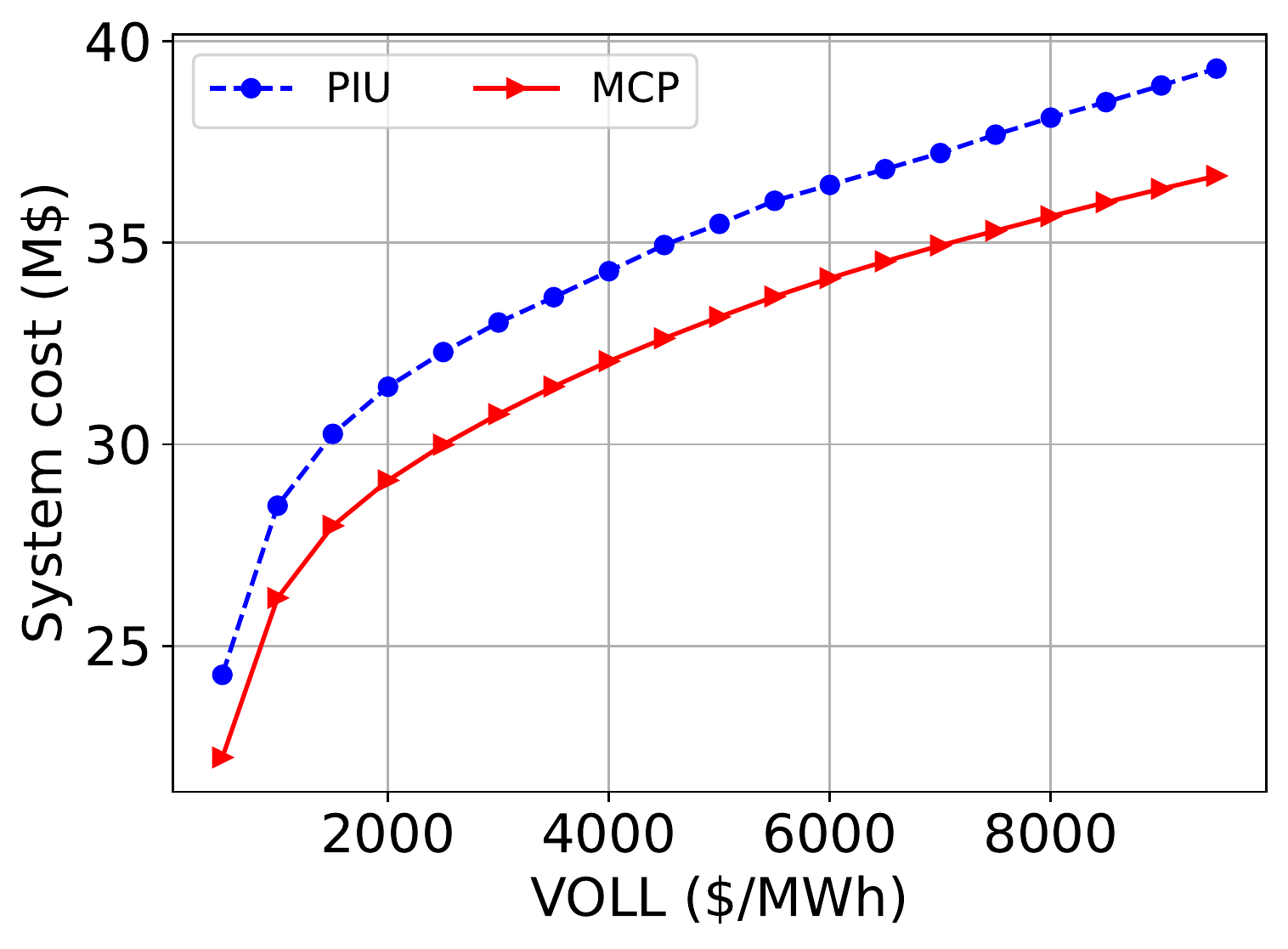}}}
	\vspace{-2mm}
	\caption{ {PIU vs MCP: (a) \small Consumer cost; (b) CER profit; (d) System cost, as functions of VOLL.}}
	\label{fig:voll}
		\vspace{-3ex}
\end{figure}

{We show how the parameter VOLL will impact the surplus of participants under the PIU and  MCP mechanisms.  In Fig.\ref{fig:voll},  we set a 70\% retirement ratio of CERs. We vary the values of VOLL from 500\$/MWh to 9500\$MWh. For the PIU mechanism, we choose the energy-price uplift that achieves break-even profit for LER investors as shown in Fig.\ref{fig:voll}a.  Recall that under the MCP mechanism, the equilibrium profit of LER investors is always zero.}
    
 {As the value of VOLL increases, we compare consumer costs, CER profits, and system costs under the break-even price uplift in Fig.\ref{fig:voll}b-\ref{fig:voll}d. As shown in Fig.\ref{fig:voll}b,  the consumer cost increases with VOLL. Under the PIU mechanism, the consumer cost is always much lower than the MCP mechanism, even under an unrealistic low VOLL of 500\$/MWh. Also, the PIU mechanism can reduce the excessive profits for CERs compared with MCP as shown in Fig.\ref{fig:voll}c. Although the system cost under the PIU mechanism is higher than the MCP mechanism in Fig.\ref{fig:voll}d, the gap is not significant compared with the reduction of the consumer cost.}

\vspace{-1ex}
\section{Conclusion}\label{sec:con}
In this paper, we establish a game-theoretical framework for the competition among LER investors, to investigate strategic market interactions, incentive mechanisms, and efficiency in future electricity markets. We present a PIU mechanism for LERs investors, which involves a penalty payment for lost load, a supply incentive, and an energy price uplift. With a quadratic supply cost function of CERs, the penalty payment leads to the resulting Nash equilibrium coinciding with the social optimum under perfect competition. The supply incentive achieves the social optimum under imperfect competition by encouraging investors to supply more energy, while the price uplift aims at ensuring adequate revenues for investors. The traditional MCP mechanism can  achieve the social optimum under perfect competition. However, under imperfect competition, investors can withhold investment to trigger more frequent scarcity prices, which will not happen under the PIU mechanism. The PIU mechanism can also reduce consumers' cost, by reducing the excess-surplus of low-cost CER suppliers under the MCP mechanism.

 {Although the proposed PIU mechanism shows nice properties, it still exhibits limitations that require attention in future research. First, it cannot ensure social optimum and revenue adequacy simultaneously. The energy-price uplift can ensure revenue adequacy for investors but make the system cost deviate from the social optimum. However, this deviation can be mitigated when more CERs retire in the market. Second, the analysis of Nash equilibrium  in this work relies on the assumption of a quadratic supply curve, which selects one equilibrium among potential multiple solutions. The impact of different supply curves and multiple Nash equilibria should be further investigated.  In future work, we will include  demand response in the market and study how consumers may respond to alternative mechanism designs. We will also study the practicality of implementing elements of the PIU mechanism in future electricity markets, e.g., through long-term insurance contracts where premiums can work as the energy price uplift and reimbursement can serve as the penalty payment.}

\vspace{-1ex}

\bibliographystyle{IEEEtran}
\bibliography{storage.bib}

%
\begin{IEEEbiography}[{\includegraphics[width=1in,height=1.25in,clip,keepaspectratio]{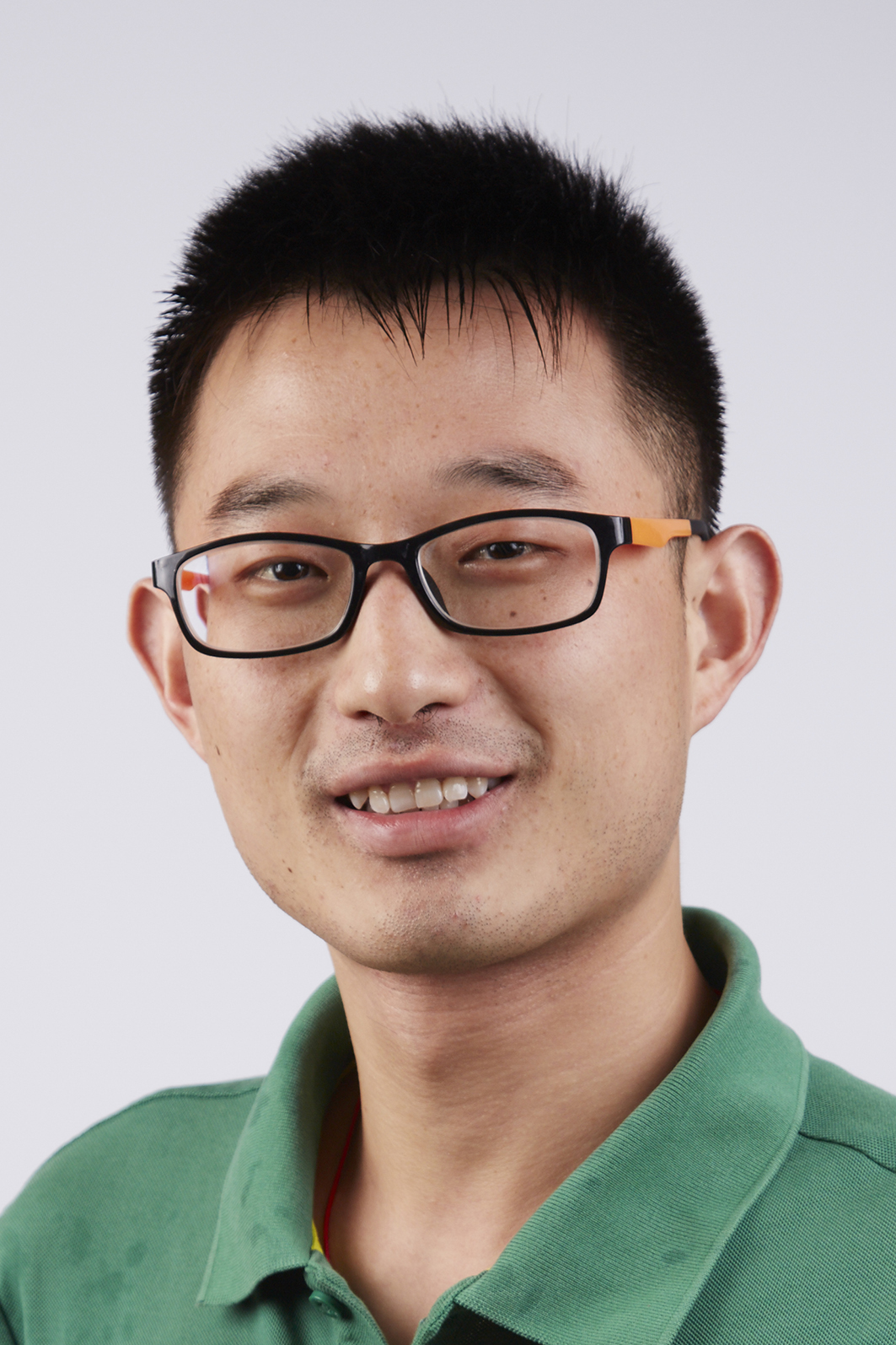}}]{Dongwei Zhao} (M'21) received the B.S. degree from Zhejiang University  in 2015, and  the  Ph.D.
degree from The Chinese University of Hong Kong in 2021. He is currently a postdoctoral associate at MIT Energy Initiative and Laboratory for Information \& Decision Systems, Massachusetts Institute of Technology.
His main research interests are in the optimization, game theory, and mechanism design of power and energy systems. More information at https://sites.google.com/view/joris-zhao
\end{IEEEbiography}

\begin{IEEEbiography}[{\includegraphics[width=1in,height=1.25in,clip,keepaspectratio]{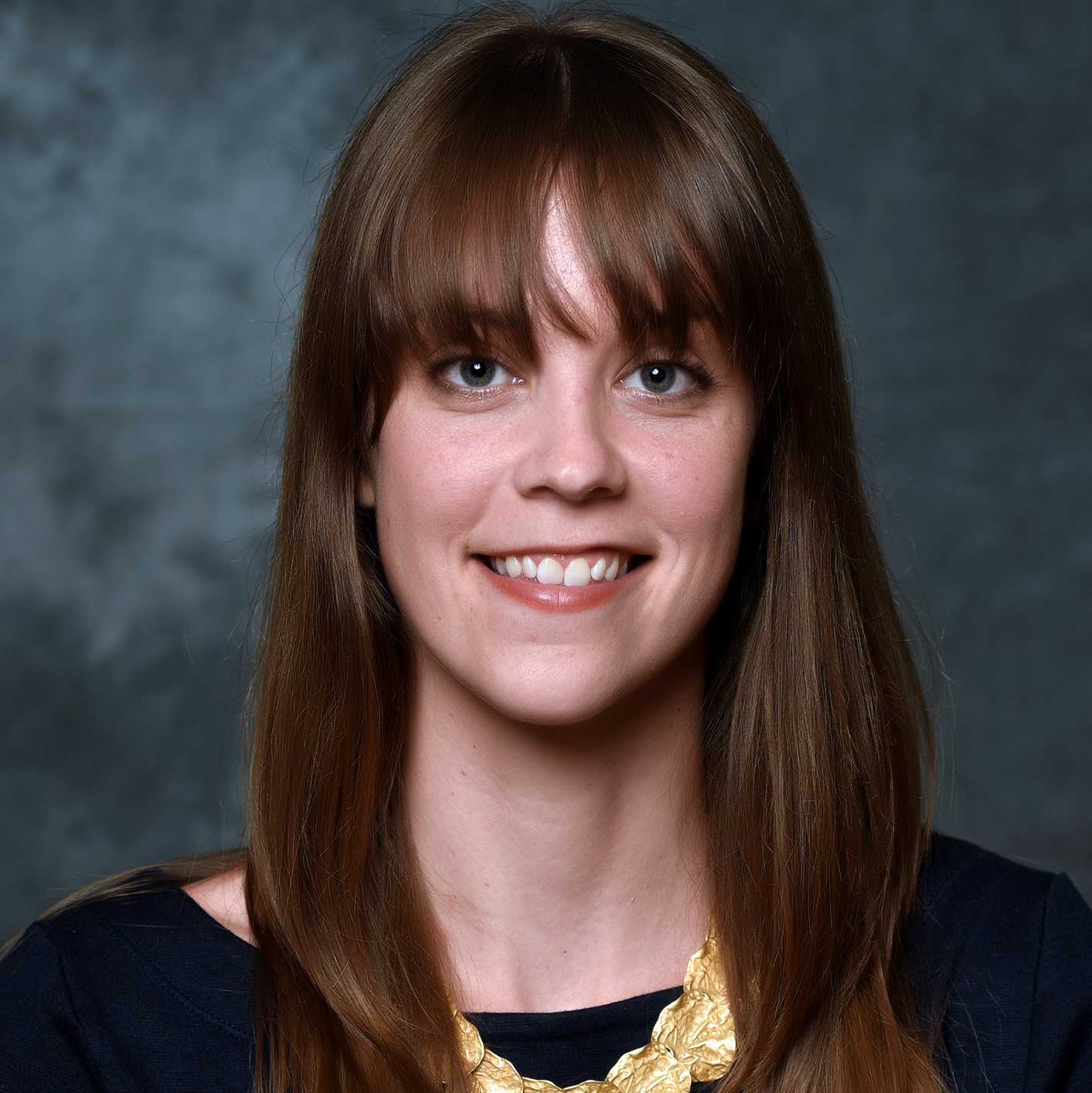}}]{Sarah Coyle} received a B.S. in Geological Sciences from the University of Texas  at Austin in 2012, a M.S. in Geophysics from the University of Texas at Austin in 2014, and a M.S. in System Design and Management from Massachusetts Institute of Technology in 2022. She is currently working as the Product Manager for Asset Development and Portfolio Optimization at Chevron.
\end{IEEEbiography}

\begin{IEEEbiography}[{\includegraphics[width=1in,height=1.25in,clip,keepaspectratio]{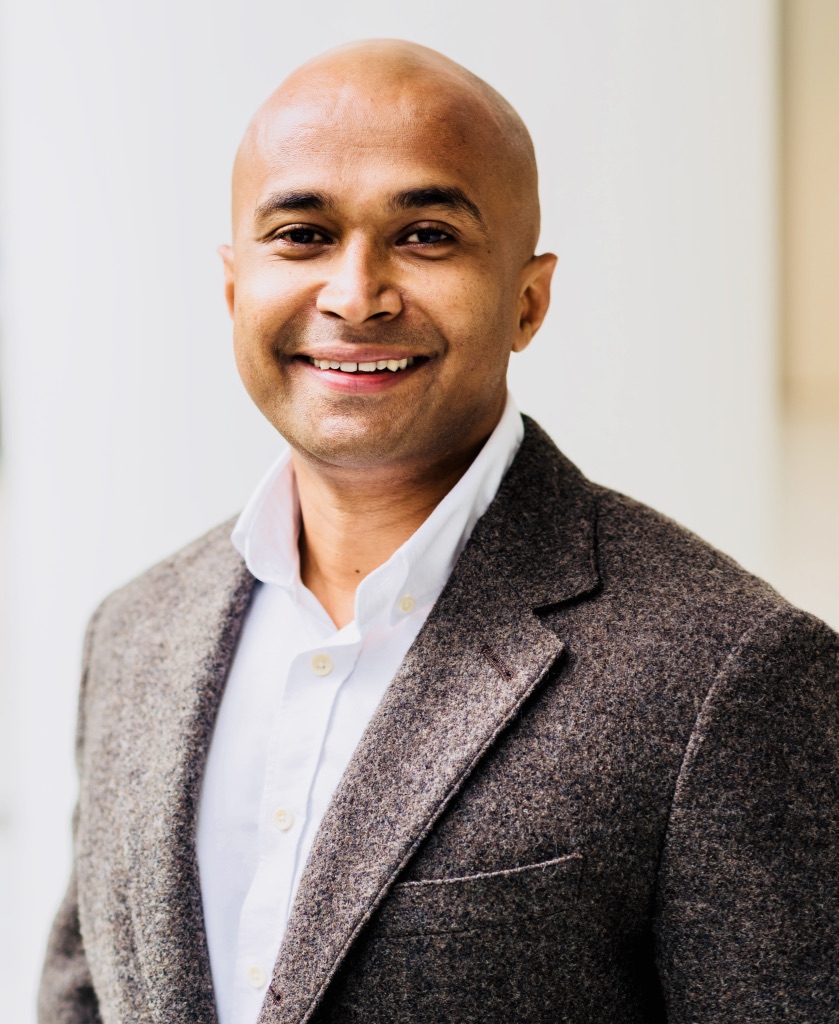}}]{Apurba Sakti} received his Ph.D. in Engineering and Public Policy from Carnegie Mellon University in 2013. He then pursued a postdoctoral position with the MIT Energy Initiative before being promoted to a Research Scientist. At MIT, Sakti's research focused on different techno-economic aspects of energy storage systems, which included the development of improved analytics for low-carbon energy systems to help understand evolving market dynamics. Since 2022, Sakti is an advisor with the data science team at Albemarle Corporation. 
\end{IEEEbiography}

\begin{IEEEbiography}[{\includegraphics[width=1in,height=1.25in,clip,keepaspectratio]{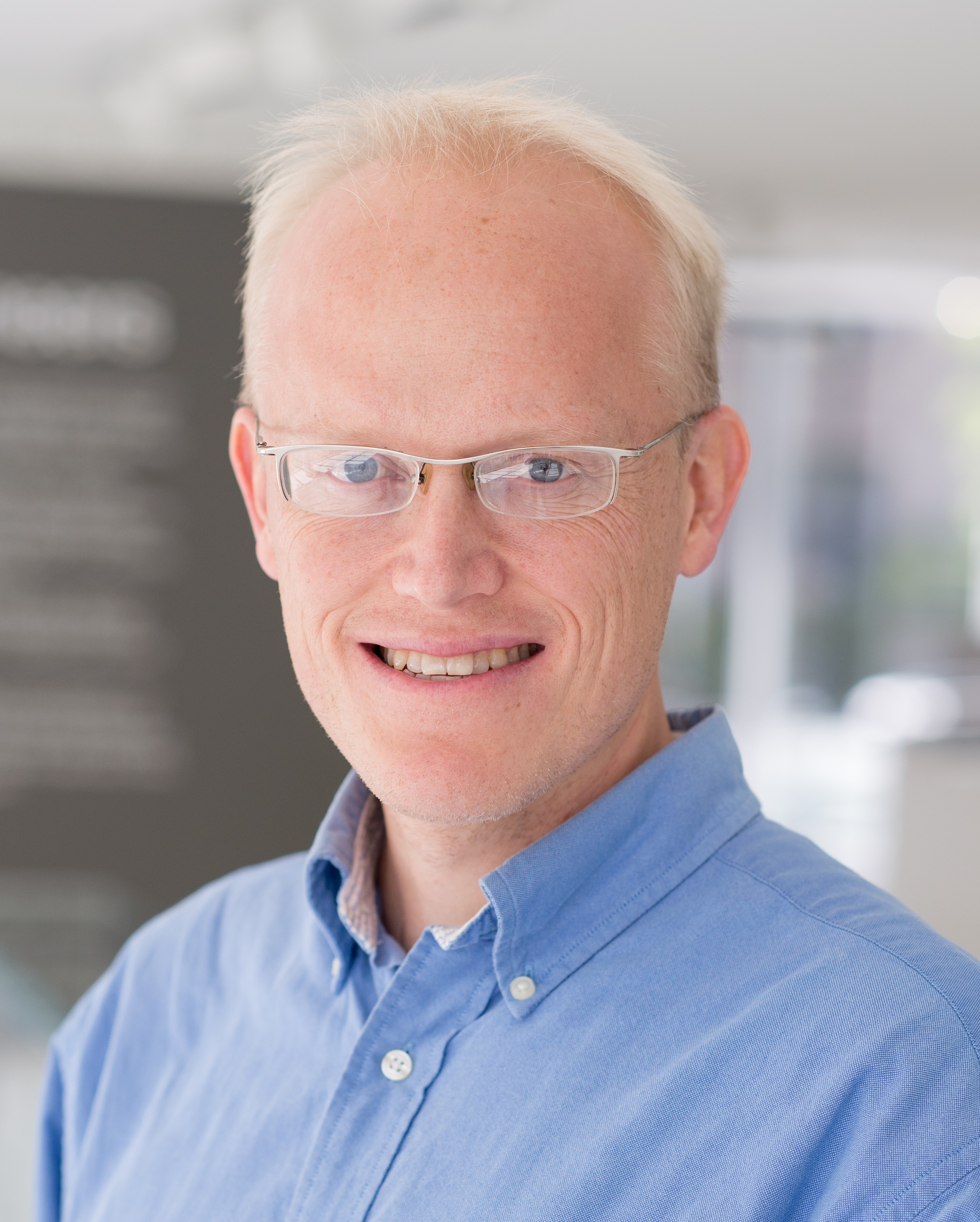}}]{Audun Botterud} Audun Botterud
is a Principal Research Scientist in Laboratory for Information and Decision
Systems (LIDS) at MIT, where he leads the Energy Analytics Group. He has a co-appointment in the Energy Systems and Infrastructure Analysis Division at Argonne National Laboratory. His research interests include power systems, electricity markets, renewable energy, and energy storage. Audun holds
a M.Sc. (Industrial Engineering) and a Ph.D. (Electrical Power Engineering), both from the Norwegian University of Science and Technology. He was previously with SINTEF Energy Research in Trondheim, Norway.
\end{IEEEbiography}

\clearpage
\newpage
\appendix

\begin{table*}[t]
	\caption{Comparison of market participants' surplus}
	\vspace{-2ex}
	\begin{center}
		\begin{tabularx}{\textwidth}{|p{3.5cm}|C|C|}
			\hline	
		 & \textbf{PIU mechanism}
		 & \textbf{MCP mechanism}\\
			\hline
			\textit{LERs}&Regulated by the uplift &0 \\
			\hline
			\textit{CERs}&Lower (by simulation) & Higher (by simulation)\\
			
			\hline
			\textit{Consumers}&Higher (by simulation)& Lower (by simulation)\\
			
			\hline
			\textit{System operator}& Non-negative under perfect competition &0 \\
			\hline
			\vspace{2ex}\textit{Social welfare}&
			\vspace{-1ex}
			\begin{itemize}
			    \item 	$\Delta \bm{\pi}=\bm{0}$: Social optimum; 
			\item $\Delta \bm{\pi}>\bm{0}$: It deviates from social optimum and incentivizes more LER investment but the gap is reduced as CERs retire.
			\end{itemize}
			& Social optimum \\
			\hline
		\end{tabularx}
		\label{tab:comparison}
	\end{center}
\end{table*}

\section*{Appendix.I: Surplus comparison between the PIU mechanism and MCP mechanism}\label{section:surpluscomparison}

We have demonstrated the shortcomings of the MCP mechanism in terms of the high scarce prices, and supply withhold under imperfection competition. However, the MCP mechanism can achieve the social optimum under perfect competition, while the PIU mechanism cannot. In this appendix, we will further compare the surplus of various market participants between the PIU mechanism and MCP mechanism under perfect competition.

The surplus of the LER investors is the profit that we have derived. The surplus of the CERs is the market revenue deducting the operation cost, i.e., 
        \begin{align}
        \mathbb{E}_\omega \sum_{t\in \mathcal{T}}\pi^{\omega}[t](\bm{z})\cdot p^{cv,\omega}[t]-G^{cv}(\bm{p}^{cv}).
    \end{align}
Consumers' surplus is the value of system served demand minus the energy payment
    \begin{align}
        \mathbb{E}_\omega \sum_{t\in \mathcal{T}}\left(\text{VOLL}-\pi^\omega[t](\bm{z})\right) \cdot  (D^\omega[t]-p^{sh}[t]).
    \end{align}
Here we use consumers' surplus instead of cost to be consistent with other participants' surplus. The system operator's surplus under the MCP mechanism is zero since the money flow is only between the consumers and suppliers. In the PIU mechanism,  the surplus is the total penalty payment charged from the investors minus the supply incentive, i.e.,
         \begin{align}
      \mathbb{E}_\omega \sum_{t\in \mathcal{T}}\sum_{i\in \mathcal{I}}\left(\Big(\text{VOLL}\hspace{-0.5mm}-\hspace{-0.5mm}\pi^\omega[t](\bm{z})\Big) \hspace{-0.5mm}\cdot \hspace{-0.5mm} p_i^{sh}[t]\hspace{-0.5mm}-\hspace{-0.5mm}\frac{1}{2}a^\omega[t](\tilde{A}_i^\omega[t])^2\right).
    \end{align}
Under perfect competition, i.e.,  a sufficiently large number of investors, the total supply incentive goes to zero because no supply incentive is needed. In this case, the operator's surplus is non-negative and can be distributed to consumers. 

We summarize the surplus comparison in Table \ref{tab:comparison} under perfect competition.

\section*{Appendix.II: Proof of Proposition \ref{prop:shadowperfect} }

\textbf{Proof:} We prove that the social-optimum solution is one Nash equilibrium by KKT conditions in Part 1. We show the zero profit of investors under the social optimum by the Lagrange dual problem in Part 2.

\subsection*{Part 1}
Given the decisions of the other investors and the system operator, any investor $i$ maximizes its own profit. 
\begin{align}
\max~	&f_i(\bm{z}_i, \bm{z}_{-i})=R_i(\bm{z}_i, \bm{z}_{-i})-C_i^{inv}(\bm{z}_i)-C_i^{opr}(\bm{z}_i)\notag\\
\text{s.t.} ~&\bm{z}_i\in  \mathcal{Z}_i^o\bigcup \mathcal{Z}_i^c(\bm{z}_{-i}),\notag\\
\text{var:} ~&\bm{z}_i.\notag
\end{align}
The market price is set at $\lambda^\omega[t]/\rho^\omega$, which is not affected by any investor's decision under perfect competition. We have the following equivalent problem given $\bm{z}_{-i}$, $\bm{p}^{cv}$ and $\bm{p}^{sh}$.
\begin{align}
\min~	&-\sum_{t\in \mathcal{T}} {A}_i^{\omega}[t] \cdot \lambda^{b,\omega}[t]+C_i^{inv}(\bm{z}_i)+C_i^{opr}(\bm{z}_i)\notag\\
\text{s.t.} &~\bm{z}_i\in  \mathcal{Z}_i^o\bigcup \mathcal{Z}_i^c(\bm{z}_{-i}).\notag\\
\text{var:} &~\bm{z}_i.
\end{align}

Since the above problem is convex, the optimal profit solution is equivalent to the solutions derived by the  KKT conditions. It is straightforward that the KKT conditions of the above problem are the necessary condition for the KKT conditions of Problem \textbf{SO}. Note that we can additionally add the constraint \eqref{eq:const_balancemodify} (which models the set $\mathcal{Z}_i^c(\bm{z}_{-i})$) to Problem \textbf{SO}. Therefore, under the social-optimum solution to Problem \textbf{SO}, any investor achieves the optimal profit and will not deviate unilaterally. \qed

\subsection*{Part 2}

We prove the zero profit of investors at the social optimum based on the Lagrange dual problem of Problem \textbf{SO}, which is formulated as follows.
\begin{align*}
\max_{\bm{\lambda}^{b},\bm{\mu}}~~\min_{\bm{z},\bm{p}^{cv},\bm{p}^{sh}}~	&\sum_{i\in \mathcal{I}} -f_i(\bm{\lambda}^b, \bm{z_i})\\
&\hspace{-14ex}+\sum_{t\in \mathcal{T}}\sum_{\omega\in \Omega}\Big(\big(-\lambda^{b,\omega}[t]-\underline{\mu}^{cv,\omega}[t]+\overline{\mu}^{cv,\omega}[t]\big)\cdot p^{cv,\omega}[t]\\
&\hspace{18ex}-\overline{\mu}^{cv,\omega}[t] \cdot \bar{p}^{cv}\Big)\\
&\hspace{-14ex}+\sum_{t\in \mathcal{T}}\sum_{\omega\in \Omega}\Big(-\lambda^{b,\omega}[t]-\underline{\mu}^{sh,\omega}[t] \Big) \cdot p^{sh,\omega}[t]\\
&\hspace{-14ex}+\sum_{t\in \mathcal{T}}\sum_{\omega\in \Omega} \lambda^{b,\omega}[t]\cdot\hspace{-0.5mm}D^\omega[t]+ G^{cv}(\bm{p}^{cv})+G^{sh}(\bm{p}^{sh})\\
&\hspace{-8ex} \text{s.t.}~ \eqref{eq:const_ren_a}-\eqref{eq:const_ren_b}, \eqref{eq:ch}-\eqref{eq:duration}, \forall i \in \mathcal{I},\\
\end{align*}
where we let
\begin{align}
-f_i(\bm{\lambda}^b, \bm{z_i})=C_i^{inv}(\bm{X}_i)\hspace{-0.5mm}+\hspace{-0.5mm} C_i^{op}(\bm{p}_i)\hspace{-0.5mm}-\hspace{-0.5mm}\sum_{t\in \mathcal{T}}\sum_{\omega\in \Omega} \lambda^{b,\omega}[t]\cdot A_i^{\omega}[t].\notag
\end{align}
Note that $\sum_{t\in \mathcal{T}}\sum_{\omega\in \Omega} \lambda^{b,\omega}[t] A_i^{\omega}[t]$ is equivalent to $\sum_{t\in \mathcal{T}}\mathbb{E}_{\omega} \left[\pi^{\lambda,\omega}[t] A_i^{\omega}[t]\right]$ based on \eqref{eq:shadowprice}, which means that  $f_i(\bm{\lambda}^b, \bm{z_i})$ is the profit of investor $i$. 
We denote the optimal solution above by $\bm{\lambda}^{b*},\bm{\mu}^*, \bm{z}^*, \bm{p}^{cv*}$, and $\bm{p}^{sh*}$. Given $\bm{\lambda}^{b*}$ and $\bm{\mu}^*$, the solution $\bm{z}^*, \bm{p}^{cv*}$, and $\bm{p}^{sh*}$ minimizes the objective function and the solution is equivalent to the social optimum in Problem \textbf{SO} due to strong duality.

We now prove the zero profit of investors at the social optimum by contradiction. (i) Suppose $- f_i(\bm{\lambda}^{b*}, \bm{z}_i^*)>0$ at the social optimum. Given $\bm{\lambda}^{b*}$ and $\bm{\mu}^*$, we can always have a  new solution  $\bm{z}_i'=\bm{0}$ satisfying the constraints such that $- f_i(\bm{\lambda}^{b*}, \bm{z}_i')=0$, which will reduce the objective value and contradicts the optimal solution. (ii) Suppose that we have $-\sum_{i\in \mathcal{I}} f_i(\bm{\lambda}^{b*}, \bm{z}_i^*)<0$ at the social optimum. Given $\bm{\lambda}^{b*}$ and $\bm{\mu}^*$, we can always have a  new solution  $\bm{z}_i'=\alpha\cdot \bm{z}_i^*,\alpha>1$ satisfying the affine constraints such that $-\sum_{i\in \mathcal{I}} f_i(\bm{\lambda}^{b*}, \bm{z}_i')<-\sum_{i\in \mathcal{I}} f_i(\bm{\lambda}^{b*}, \bm{z}_i^*)$, which will reduce the objective value and contradicts the optimal solution. Thus, we have $ f_i(\bm{\lambda}^{b*}, \bm{z}_i^*)=0$ at the social optimum. \qed

\section*{Appendix.III: Proof of Proposition \ref{prop:shadowimperfect} }
\textbf{Proof:} We prove Proposition \ref{prop:shadowimperfect} based on the symmetric solutions for the homogeneous VRE investors and near Nash equilibrium.

To begin with, We define $\alpha-$Nash equilibrium as follows.

\begin{defi}[$\alpha-$Nash equilibrium]\label{def:nearequilibrium}
	In the competition game $G =\langle \mathcal{I},(\mathcal{Z}_i),(f_i)\rangle$, the strategy profile $\bm{z}^*\in \Pi_{i\in \mathcal{I}} \mathcal{Z}_i(\bm{z}_{-i}^*)$ is  an $\alpha-$Nash equilibrium if for every investor $i\in \mathcal{I}$, $f_i(\bm{z}_i^*,\bm{z}_{-i}^*)\geq f_i(\bm{z}_i,\bm{z}_{-i}^*)-\alpha$ for any $\bm{z}_i\in \mathcal{Z}_i(\bm{z}_{-i}^*)$.
\end{defi}

Furthermore, the optimization problem in Proposition \ref{prop:shadowimperfect} is concave, so we consider the symmetric solutions among investors. We denote each investor's solution as $\bm{z}_i^*=\bm{z}_u^*$. Next, we will prove the Nash equilibrium results by discussing  $\sum_{i\in \mathcal{I}}A_i^{\omega}[t]$ in the ranges  $\sum_{i\in \mathcal{I}}A_i^{\omega}[t]\leq D^\omega[t]-\gamma \cdot \bar{p}^{cv}-\epsilon^\omega[t]$, $D^\omega[t]-\gamma \cdot \bar{p}^{cv}-\epsilon^\omega[t]\leq \sum_{i\in \mathcal{I}}A_i^{\omega}[t]\leq D^\omega[t]-\gamma \cdot \bar{p}^{cv}$, and  $D^\omega[t]-\gamma \cdot \bar{p}^{cv}< \sum_{i\in \mathcal{I}}A_i^{\omega}[t]\leq D^\omega[t]$, respectively.

\subsection*{Step 1}
We define a constraint  $\mathcal{M}_1^\omega[t]$ for $A_i^\omega[t]$ as $A_i^\omega[t]+\sum_{j\in \mathcal{I}\setminus i}A_j^{\omega*}[t]\leq D^\omega[t]-\gamma \cdot \bar{p}^{cv}-\epsilon^\omega[t]$ for some time $t$ of scenario $\omega$. 
We show that  any investor $i$ will not change its strategy unilaterally to any $\bm{z}_i$ that satisfies $ \mathcal{M}_1^\omega[t] $ for some time $t$ of scenario $\omega$.

Note that the market price is fixed at $\pi=\text{VOLL}$ in this constraint range, any investor $i$ who changes his strategy unilaterally within this constraint will not impact other investors' profits. Meanwhile, the optimization problem in Proposition \ref{prop:shadowimperfect} maximizes all the investors' total profit, which implies that any investor $i$'s unilateral strategy change will not increase its profit within this constraint $\mathcal{M}_1^\omega[t]$, i.e., 
\begin{align}
f_i(\bm{z}_i^*, \bm{z}_{-i}^*)\geq f_i(\bm{z}_i, \bm{z}_{-i}^*), \label{eq:begin}
\end{align}
for any $\bm{z}_i$ that satisfies $ \mathcal{M}_1^\omega[t]$.

\subsection*{Step 2}
We define a constraint $\mathcal{M}_2^\omega[t]$ for $A_i^\omega[t]$ as $D^\omega[t]-\gamma \cdot \bar{p}^{cv}-\epsilon^\omega[t]\leq A_i^{\omega}[t]+\sum_{j\in \mathcal{I}\setminus i}A_j^{\omega*}[t]\leq D^\omega[t]-\gamma \cdot \bar{p}^{cv}$.
We will show that when any investor $i$ changes its strategy unilaterally from $\bm{z}^*$  to any $\bm{z}_i^\dagger$ such that $\mathcal{M}_1^{\omega_a}[t_a]$ holds for some time $t_a$ of scenario $\omega_a$ and $\mathcal{M}_2^{\omega_b}[t_b]$ holds for some time $t_b$ of scenario $\omega_b$, this investor's profit increase will be bounded above by a function of ${\epsilon}^\omega[t]$.

Specifically, we consider a benchmark strategy $\bm{z}_i'$ for $\bm{z}_i^\dagger$, i.e., investor $i$ changes its strategy unilaterally from $\bm{z}_i^*$ to $\bm{z}_i'$ such that  $D^{\omega_b}[t_b]-\gamma \cdot \bar{p}^{cv}-\epsilon^{\omega_b}[t_b]= A_i^{\omega_b'}[t_b]+\sum_{j\in \mathcal{I}\setminus i}A_j^{\omega_b*}[t_b]$ for time $t_b$ of scenario $\omega_b$. Meanwhile we let $\bm{z}_i' =\bm{z}_i^\dagger$, except at time $t_b$ of scenario $\omega_b$. Since $\bm{z}_i^\dagger$ satisfies  $\mathcal{M}_1^{\omega_a}[t_a]$ and $\mathcal{M}_2^{\omega_b}[t_b]$, we easily have
\begin{align}
    R_i^\omega(\bm{z}_i', \bm{z}_{-i}^*)[t]\geq R_i^\omega(\bm{z}_i^\dagger, \bm{z}_{-i}^*)[t]-\epsilon^\omega[t]\cdot \text{VOLL}, \label{eq:mi}
\end{align}
i.e., the  expected daily revenue
\begin{align}
    R_i(\bm{z}_i', \bm{z}_{-i}^*)\geq R_i(\bm{z}_i^\dagger, \bm{z}_{-i}^*)-\mathbb{E}_\omega\sum_{t\in\mathcal{T}}\epsilon^\omega[t]\cdot \text{VOLL}, \label{eq:miexp}
\end{align}
 Note that when $A_i^{\omega}+\sum_{j\in \mathcal{I}\setminus i }A_j^{\omega*}[t]=D^\omega[t]-\gamma \cdot \bar{p}^{cv}$, \eqref{eq:mi} is always satisfied for any market price $\pi^\omega[t]\leq \text{VOLL}$.

Since $A_i^{\omega_b'}[t_b]\leq A_i^{\omega_b\dagger}[t_b]$, the corresponding invested capacity $X_i'$ of investor $i$ under $\bm{z}_i'$ is no greater than the capacity $X_i^\dagger$  under $\bm{z}_i^\dagger$, which makes the expected daily profit satisfy
\begin{align}
    f_i(\bm{z}_i', \bm{z}_{-i}^*)\geq f_i(\bm{z}_i^\dagger, \bm{z}_{-i}^*)-\mathbb{E}_\omega\sum_{t\in\mathcal{T}}\epsilon^\omega[t]\cdot \text{VOLL}, 
\end{align}
 which leads to 
\begin{align}
    f_i(\bm{z}_i^*, \bm{z}_{-i}^*)\geq f_i(\bm{z}_i^\dagger, \bm{z}_{-i}^*)-\mathbb{E}_\omega\sum_{t\in\mathcal{T}}\epsilon^\omega[t]\cdot \text{VOLL}\label{eq:step2}
\end{align}
based on \eqref{eq:begin}. Thus, we have $\bm{z}^*$ is $\big( \mathbb{E}_\omega\sum_{t\in\mathcal{T}}\epsilon^\omega[t]\big)-$near Nash equilibrium under the constraint set $\sum_{i\in \mathcal{I}}A_i^{\omega}[t]\leq D^\omega[t]-\gamma \cdot \bar{p}^{cv}$.

\subsection*{Step 3}
We define a constraint $\mathcal{M}_3^{\omega}[t]$ for $A_i^{\omega}[t]$ as $D^\omega[t]-\gamma \cdot \bar{p}^{cv}< A_i^{\omega}[t]+\sum_{j\in \mathcal{I}\setminus i}A_j^{\omega*}[t]\leq D^\omega[t]$. 
We will characterize a sufficient condition such  that $\bm{z}^*$ is  $\big(\mathbb{E}_\omega\sum_{t\in\mathcal{T}}\epsilon^\omega[t]\big)-$near Nash equilibrium globally.

To characterize such a sufficient  condition,  we suppose that
any investor $i$ changes its strategy unilaterally  to any $\bm{z}_i^\ddagger$ such that $\mathcal{M}_1^{\omega_a}[t_a]$ holds for some time $t_a$ of scenario $\omega_a$, $\mathcal{M}_2^{\omega_b}[t_b]$ holds for some time $t_b$ of scenario $\omega_b$, and $\mathcal{M}_3^{\omega_c}[t_c]$ holds for some time $t_c$ of scenario $\omega_c$. Then, we compare it with a benchmark case: investor $i$ changes its strategy unilaterally to $z_i''$ that satisfies  $A_i^{\omega_c''}[t_c]+\sum_{j\in \mathcal{I}\setminus i }A_j^{\omega_c*}[t_c]=D^{\omega_c}[t_c]-\gamma \cdot \bar{p}^{cv}$, $\mathcal{M}_2^{\omega_b}[t_b]$, and $\mathcal{M}_1^{\omega_a}[t_a]$. We let  $\pi^{\omega_c}(\bm{z}_i'', z_{-i}^*)[t_c]= \text{VOLL}$ 
because \eqref{eq:mi} is always satisfied for any market price $\pi^\omega[t]\leq \text{VOLL}$.  We let $\bm{z}_i''=\bm{z}_i^\ddagger$, except at time $t_c$ of $\omega_c$.

For $\bm{z}_i''$, at time $t_c$ of scenario $\omega_c$, investor $i$'s revenue satisfies
\begin{align}
    R_i^{\omega_c}(\bm{z}_i'', \bm{z}_{-i}^*)[t_c]&=\text{VOLL}\cdot A_i^{\omega_c''}[t_c]\notag\\
    &\hspace{-15ex}=\text{VOLL}\cdot (D^{\omega_c}[t_c]-\gamma \cdot \bar{p}^{cv}-(N-1)A_u^{\omega_c*}[t_c]).
\end{align}
For $\bm{z}_i^\ddagger$, at time $t_c$ of scenario $\omega_c$,  investor $i$'s revenue satisfies 
\begin{align}
    &R_i^{\omega_c}(\bm{z}_i^\ddagger, \bm{z}_{-i}^*)[t_c]\notag\\
    &~\leq \big( D^{\omega_c} [t_c]\hspace{-0.5mm} -\hspace{-0.5mm}(N\hspace{-0.5mm}-\hspace{-0.5mm}1)A_u^{\omega_c*}[t_c] \big) \cdot \frac{d g^{cv,\omega_c}[t_c]}{d p^{cv,\omega_c}[t_c]}\Big|_{\gamma \cdot \bar{p}^{cv}}.
\end{align}
To let  $R_i^{\omega_c}(\bm{z}_i'', \bm{z}_{-i}^*)[t_c]\geq R_i^{\omega_c}(\bm{z}_i^\ddagger, \bm{z}_{-i}^*)[t_c]$, it suffices to show
\begin{align}
    &\text{VOLL}\cdot (D^\omega[t]-\gamma \cdot \bar{p}^{cv}-(N-1)A_u^{\omega*})\notag\\
    &~~~~~~\geq \big( D^\omega [t] -(N-1)A_u^{\omega*} \big) \cdot \frac{d g^{cv,\omega}[t]}{d p^{cv,\omega}[t]}\Big|_{\gamma \cdot \bar{p}^{cv}},\label{eq:mid2}
\end{align}
for all the time $t$ of scenario $\omega$. Note that $\text{VOLL}\geq  \frac{d g^{cv,\omega}[t]}{d p^{cv,\omega}[t]}\Big|_{\gamma \cdot \bar{p}^{cv}}$ and $A_u^{\omega*}[t]< \frac{D^\omega[t]-\gamma \cdot \bar{p}^{cv}}{N}$. To satisfy \eqref{eq:mid2}, it suffices to show 
\begin{align}
    &\text{VOLL}\cdot (D^\omega[t]-\gamma \cdot \bar{p}^{cv}-(N-1)\frac{D^\omega[t]-\gamma \cdot \bar{p}^{cv}}{N})\notag\\
    &\geq \Big( D^\omega [t]\hspace{-0.5mm}-\hspace{-0.5mm}(N\hspace{-0.5mm}-\hspace{-0.5mm}1)\frac{D^\omega[t]\hspace{-0.5mm}-\hspace{-0.5mm}\gamma \cdot \bar{p}^{cv}}{N} \Big) \cdot \frac{d g^{cv,\omega}[t]}{d p^{cv,\omega}[t]}\Big|_{\gamma \cdot \bar{p}^{cv}},\label{eq:mid5}
\end{align}
i.e., \begin{align}
\text{VOLL} \geq \left(1+ \frac{N\cdot \gamma \bar{p}}{D^\omega[t]-\gamma \bar{p}}\right)\cdot \frac{d g^{cv,\omega}[t]}{d p^{cv,\omega}[t]}\Big|_{\gamma \cdot \bar{p}^{cv}}. \label{eq:mid3}
\end{align}

Therefore, as long as \eqref{eq:mid3} is satisfied, we have
$R_i^{\omega_c}(\bm{z}_i'', \bm{z}_{-i}^*)[t_c]\geq R_i^{\omega_c}(\bm{z}_i^\ddagger, \bm{z}_{-i}^*)[t_c]$ ,  which easily makes the  the expected daily revenue $R$ satisfy 
\begin{align}
    R_i(\bm{z}_i'', \bm{z}_{-i}^*)\geq R_i(\bm{z}_i^\ddagger, \bm{z}_{-i}^*).
\end{align}
Since $A_i^{\omega_c''}[t_c]\leq A_i^{\omega_c\ddagger}[t_c]$, the corresponding invested capacity $X_i''$ of investor $i$ under $\bm{z}_i''$ is no greater than the capacity $X_i^\ddagger$  under $\bm{z}_i^\ddagger$, which makes the expected daily profit also satisfy
\begin{align}
    f_i(\bm{z}_i'', \bm{z}_{-i}^*)\geq f_i(\bm{z}_i^\ddagger, \bm{z}_{-i}^*).
\end{align}

Note that here $\bm{z}_i''$ is one special case of $\bm{z}_i^\dagger$ in \textit{Step 2}. Based on \eqref{eq:step2}, we have 
\begin{align}
    f_i(\bm{z}_i^*, \bm{z}_{-i}^*)\geq f_i(\bm{z}_i^\ddagger, \bm{z}_{-i}^*)-\mathbb{E}_\omega \sum_{t\in \mathcal{T}} \epsilon^\omega[t]\cdot \text{VOLL}, 
\end{align}
for any $\bm{z}_i^\ddagger$ such that  $\mathcal{M}_1(A_i^{\omega_a}[t_a])$ holds for some time $t_a$ of scenario $\omega_a$, $\mathcal{M}_2(A_i^{\omega_b}[t_b])$ holds for some time $t_b$ of scenario $\omega_b$, and $\mathcal{M}_3(A_i^{\omega_c}[t_c])$ holds for some time $t_c$ of scenario $\omega_c$. 

Overall, we have $\bm{z}^*$ is $\big( \mathbb{E}_\omega\sum_{t\in\mathcal{T}}\epsilon^\omega[t]\big)-$near Nash equilibrium globally, which   approaches a pure-strategy Nash equilibrium as $\epsilon^\omega[t]\rightarrow 0$. \qed

\section*{Appendix.IV: Proof of Proposition \ref{prop:penaltyfinite}} 

\subsection*{ {A.Proof based on potential game theory}} 
 {We can prove Proposition \ref{prop:penaltyfinite} from the perspective of the potential game theory, which we define as follows.  Let $H =\langle \mathcal{N},(\mathcal{Z}_i),(h_i)\rangle$ be any strategic game and  $\mathcal{Z} =\times_{i\in \mathcal{N}} \mathcal{Z}_i$ be the collection
 of all players' strategy profiles. We have the following result for potential games \cite{monderer1996potential} \cite{la2016potential}.}

 \begin{defi}[Potential game] \label{defi:potentialgame}
 	 {A function $\Phi: \mathcal{Z} \rightarrow \mathbb{R}$  is called an (exact) potential function for the game $H$ if for each $i \in \mathcal{N}$ and all $\bm{z}_{-i} \in  \mathcal{Z}_{-i}$, $$\Phi(\bm{z}'_i,\bm{z}_{-i})-\Phi(\bm{z}_i,\bm{z}_{-i})=h_i(\bm{z}'_i,\bm{z}_{-i})-h_i(\bm{z}_i,\bm{z}_{-i}),$$ for all $\bm{z}'_i, \bm{z}_i\in \mathcal{Z}_i$. The game	$H$ is called an (exact) potential game if it admits a potential function.}
 \end{defi}

\begin{prop}[Pure strategy Nash equilibrium]\label{prop:existenceofpotential}
 {If the potential function $\Phi$ has a maximum point $\bm{z}^*$ in $\mathcal{Z}$, then the game $H$ has a pure-strategy Nash equilibrium that coincides
with $\bm{z}^*$.}
\end{prop}

 {The above proposition shows that if we can find a potential function for this game, the optimum of the potential function is one Nash equilibrium.
Our game-theoretical model is in nature a Cournot competition model. Based on the linear supply functions, we can characterize a potential function, i.e.,  the objective \eqref{eq:nashobj} in Proposition \ref{prop:penaltyfinite}. A stylized Cournot model with a characterized potential function is given in  \cite{monderer1996potential}. Our previous work \cite{zhao2022storagecaiso} extends it to the storage competition problem \cite{zhao2022storagecaiso}, and this work extends to the mechanism design for the joint investment of renewable energy and energy storage.}

\subsection*{B.Proof based on KKT conditions}
At the Nash equilibrium of the game-theoretical model, each investor's decision is the best response given other investors' decisions. We show that the KKT conditions of the optimization problem in Proposition \ref{prop:penaltyfinite} suffice to satisfy the investors' KKT condition set when each investor optimizes its own profit given other investors' decisions.

For each investor $i$, at the Nash equilibrium, each investor decides its decision $\bm{z}_i$ to maximize the profit given other investors' decisions $\bm{z}_{-i}$, i.e.,
\begin{align*}
 \max~ &\tilde{f}_i(\tilde{\bm{z}})\\
    \text{s.t.}~ &\tilde{\bm{z}}_i\in \tilde{\mathcal{Z}}_i^o\bigcup \tilde{\mathcal{Z}}_i^c(\tilde{\bm{z}}_{-i})
\end{align*}
 Based on the equation \eqref{eq:const_balancenew}, we have 
$p^{cv,\omega}[t]=D^\omega[t]-\sum_{i\in \mathcal{I}}\tilde{A}_i^{\omega}[t],\forall t\in\mathcal{T},\forall \omega \in {\Omega}$, which leads to
\begin{align}
    \tilde{f}_i(\tilde{\bm{z}})&=\sum_{t\in \mathcal{T}}\mathbb{E}_\omega\left[\frac{d g^{cv,\omega}[t]}{d p^{cv,\omega}[t]}\cdot\tilde{A}_i^{\omega}[t]\right]-C_i^{inv}(\tilde{\bm{z}}_i)-C_i^{opr}(\tilde{\bm{z}}_i)\notag\\
    &\hspace{-6ex}=\sum_{t\in \mathcal{T}}\mathbb{E}_\omega\left[a^\omega[t](D^\omega[t]-\sum_{i\in \mathcal{I}}\tilde{A}_i^{\omega}[t])\cdot \tilde{A}_i^{\omega}[t]+b^\omega[t] \cdot\tilde{A}_i^{\omega}[t]\right]\notag\\
    &\hspace{18ex}-C_i^{inv}(\tilde{\bm{z}}_i)-C_i^{opr}(\tilde{\bm{z}}_i).\label{eq:medc}
\end{align}
Investors' profits are coupled with $\tilde{\bm{A}}^{\omega}[t]$ directly so we examine  $\frac{d   \tilde{f}_i(\tilde{\bm{z}})}{d \tilde{A}_i^{\omega}[t]}$: 
\begin{align}
 \frac{d   \tilde{f}_i(\tilde{\bm{z}})}{d \tilde{A}_i^{\omega}[t]} 
    &=a^\omega[t]\Big(D^\omega[t]-\sum_{i\in \mathcal{I}}\tilde{A}_i^{\omega}[t]\Big)-a^\omega[t]\cdot \tilde{A}_i^{\omega}[t] +b^\omega[t]\notag\\
&\hspace{19ex}-\frac{C_i^{inv}(\bm{z}_i)+C_i^{opr}(\bm{z}_i)}{d \tilde{A}_i^{\omega}[t]}.
\end{align}

Considering the objective function of the optimization problem in Proposition \ref{prop:penaltyfinite}, we easily verify
\begin{align}
     &\frac{d  \left(\sum_i  \tilde{f}_i(\tilde{\bm{z}})+\mathbb{E}_{\omega \in \Omega}\Big[\sum_{t\in \mathcal{T}}\sum_{1\leq i<j\leq I} a^\omega[t]\cdot \Phi^\omega(\tilde{\bm{z}_i},\tilde{\bm{z}}_j)\Big]\right)}{d \tilde{A}_i^{\omega}[t]}\\
     &= \frac{d   \tilde{f}_i(\tilde{\bm{z}})}{d \tilde{A}_i^{\omega}[t]}. 
\end{align}
Note that the joint constraints of all the investors in the game-theoretical model coincide with the constraints of the optimization problem in Proposition \ref{prop:penaltyfinite}. It is straightforward that the KKT conditions of the optimization problem in Proposition \ref{prop:penaltyfinite} suffice to satisfy the investors' KKT condition set when each investor optimizes its own profit given other investors' decisions in the game-theoretical model. \qed

\section*{Appendix.V: Proof of Proposition \ref{prop:penaltyinf}}
\textbf{Proof:} We prove it based on the Nash equilibrium derived in Proposition \ref{prop:penaltyfinite} and symmetric solutions for the homogeneous investors. 

First, based on \eqref{eq:medc}, the objective function of the optimization problems in  Proposition \ref{prop:penaltyfinite} can be written into
\begin{align}
   \hspace{2ex} &F(\tilde{\bm{z}})\triangleq \sum_{i\in\mathcal{I}}  \tilde{f}_i(\tilde{\bm{z}})+\mathbb{E}_{\omega \in \Omega}\Big[\sum_{t\in \mathcal{T}}\sum_{1\leq i<j\leq I} a^\omega[t]\hspace{-0.5mm}\cdot\hspace{-0.5mm} \Phi^\omega(\tilde{\bm{z}_i},\tilde{\bm{z}}_j)\Big]\notag \\
    &=\sum_{t\in \mathcal{T}}\mathbb{E}_\omega\Big[a^\omega[t] \cdot D^\omega[t]\cdot \sum_{i\in\mathcal{I}}\tilde{A}_i^{\omega}[t]+b^\omega[t] \cdot\sum_{i\in\mathcal{I}}\tilde{A}_i^{\omega}[t]\Big]\notag\\
    &-\sum_{t\in \mathcal{T}}\mathbb{E}_\omega\Big[\frac{1}{2}a^\omega[t]\Big( (\sum_{i\in \mathcal{I}}\tilde{A}_i^{\omega}[t])^2+\sum_{i\in\mathcal{I}}(\tilde{A}_i^{\omega}[t])^2\Big) \Big]\notag\\
    &- \mathbb{E}_\omega \hspace{-0.5mm}\sum_{t\in \mathcal{T}}\sum_{i\in \mathcal{I}} \text{VOLL} \hspace{-0.5mm}\cdot\hspace{-0.5mm} p_i^{sh,\omega}[t]\hspace{-0.7mm}-\hspace{-0.7mm}\sum_{i\in\mathcal{I}}C_i^{inv}(\tilde{\bm{z}}_i)\hspace{-0.7mm}-\hspace{-0.7mm}\sum_{i\in\mathcal{I}}C_i^{opr}(\tilde{\bm{z}}_i).\notag
\end{align}

Then, based on the convexity of the optimization problem in Proposition \ref{prop:penaltyfinite}, we assume symmetric solutions for investors in the same type, i.e., any investor $i$ in type $k$ has $\tilde{A}_i^{\omega}[t]=\tilde{A}_k^{\omega}[t]$, which leads to
 \begin{align}
  \sum_{i\in\mathcal{I}}\tilde{A}_i^{\omega}[t]=\sum_{k\in\mathcal{K}} N^k \tilde{A}_k^{\omega}[t].
 \end{align}
We let $N^k \tilde{A}_k^{\omega}[t] \triangleq  \tilde{A}_k^{\omega'}[t]$ and $ p^{sh,\omega}[t] \triangleq \sum_{i\in\mathcal{I}} p_i^{sh,\omega}[t]$ and have
\begin{align}
   &~~~~F(\tilde{\bm{z}}) \notag\\
    &=\sum_{t\in \mathcal{T}}\mathbb{E}_\omega\Big[a^\omega[t] \cdot D^\omega[t]\cdot \sum_{k\in\mathcal{K}}\tilde{A}_k^{\omega'}[t]+b^\omega[t] \cdot\sum_{k\in\mathcal{K}}\tilde{A}_k^{\omega'}[t]\Big]\notag\\
    &-\sum_{t\in \mathcal{T}}\mathbb{E}_\omega\Big[\frac{1}{2}a^\omega[t]\Big( (\sum_{k\in \mathcal{K}}\tilde{A}_k^{\omega'}[t])^2+\sum_{k\in\mathcal{K}}\frac{(\tilde{A}_k^{\omega'}[t])^2}{N^k}\Big) \Big]\notag\\
    &- \mathbb{E}_\omega \sum_{t\in \mathcal{T}} \text{VOLL} \cdot p^{sh,\omega}[t]-\sum_{i\in \mathcal{I}}C_i^{inv}(\tilde{\bm{z}}_i)-\sum_{i\in \mathcal{I}}C_i^{opr}(\tilde{\bm{z}}_i).\notag
\end{align}

Now, we consider the case when $N^k\rightarrow +\infty$. Note that the optimal value $F(\tilde{\bm{z}})$ is lower bounded by the special solution of no investment, which implies that the solutions of the invested capacity of any resource will be bounded above. Thus,  $\tilde{A}_k^{\omega'}[t]$ is bounded above.  As $N^k\rightarrow +\infty$ for any $k\in \mathcal{K}$, we have  $\sum_{k\in\mathcal{K}}\frac{(\tilde{A}_k^{\omega'}[t])^2}{N^k}\rightarrow 0$, which leads to 
\begin{align}
   &~~~~F(\tilde{\bm{z}}) \label{eq:xfinite}\\
    &=\sum_{t\in \mathcal{T}}\mathbb{E}_\omega\Big[a^\omega[t] \cdot D^\omega[t]\cdot \sum_{k\in\mathcal{K}}\tilde{A}_k^{\omega'}[t]+b^\omega[t] \cdot\sum_{k\in\mathcal{K}}\tilde{A}_k^{\omega'}[t]\Big]\notag\\
    &-\sum_{t\in \mathcal{T}}\mathbb{E}_\omega\Big[\frac{1}{2}a^\omega[t]\Big(\sum_{k\in \mathcal{K}}\tilde{A}_k^{\omega'}[t]\Big)^2 \Big]\notag\\
    &- \mathbb{E}_\omega \sum_{t\in \mathcal{T}} \text{VOLL} \cdot p^{sh,\omega}[t]-\sum_{i\in \mathcal{I}}C_i^{inv}(\tilde{\bm{z}}_i)-\sum_{i\in \mathcal{I}}C_i^{opr}(\tilde{\bm{z}}_i).\notag
\end{align}
Since $\sum_{k\in\mathcal{K}} \tilde{A}_k^{\omega'}[t]=D^\omega[t]-p^{cv,\omega}[t]$, we have
\begin{align}
    &~~~~F(\tilde{\bm{z}}) \notag \\
    &=\sum_{t\in \mathcal{T}}\mathbb{E}_\omega\Big[ \frac{1}{2} a^\omega[t] \cdot D^\omega[t]\cdot D^\omega[t]+b^\omega[t] \cdot D^\omega[t] \Big]\notag\\
    &-\sum_{t\in \mathcal{T}}\mathbb{E}_\omega\Big[\frac{1}{2}a^\omega[t](p^{cv,\omega}[t])^2 + b^\omega[t] \cdot p^{cv,\omega}[t]\Big]\notag\\
    &- \mathbb{E}_\omega \sum_{t\in \mathcal{T}} \text{VOLL} \cdot p^{sh,\omega}[t]-\sum_iC_i^{inv}(\tilde{\bm{z}}_i)-\sum_iC_i^{opr}(\tilde{\bm{z}}_i).\notag
\end{align}
Thus, $\max F(\tilde{\bm{z}})$ is equivalent to 
\begin{align}
   \min 
    &\sum_{t\in \mathcal{T}}\mathbb{E}_\omega\Big[\frac{1}{2}a^\omega[t](p^{cv,\omega}[t])^2 + b^\omega[t] \cdot p^{cv,\omega}[t]\Big]\notag\\
    &+ \mathbb{E}_\omega \sum_{t\in \mathcal{T}} \text{VOLL} \cdot p^{sh,\omega}[t]+\sum_iC_i^{inv}(\tilde{\bm{z}}_i)+\sum_iC_i^{opr}(\tilde{\bm{z}}_i)\notag,
\end{align}
which is the system cost in Problem \textbf{SO}. The Nash equilibrium derived in Proposition approaches the social optimum as $N^k\rightarrow +\infty$ for any $k$.\qed

\section*{Appendix.VI: Proof of Proposition \ref{prop:incentive}}
\textbf{Proof:} As implied in \eqref{eq:xfinite} in Appendix.V, if the  part $-\sum_{t\in \mathcal{T}}\mathbb{E}_\omega\Big[\frac{1}{2}a^\omega[t]\sum_{i\in\mathcal{I}}(\tilde{A}_i^{\omega}[t])^2 \Big]$ in  $F(\tilde{\bm{z}})$ is eliminated, the Nash equilibrium derived in Proposition \ref{prop:penaltyfinite} will coincide with the social optimum.  Thus, we design the additional supply incentive $\frac{1}{2}a^\omega[t](\tilde{A}_i^{\omega}[t])^2$ for each time $t\in \mathcal{T}$ and scenario $\omega \in \Omega$
added to investor $i$'s profit.  \qed

\section*{Appendix.VII: Proof of Proposition \ref{prop:uplift}}
\textbf{Proof: }The market price is set by the marginal cost of CERs: 
\begin{align}
\frac{d g^{cv,\omega}[t]}{d p^{cv,\omega}[t]}=a^\omega[t]\cdot p^{cv,\omega}[t]+b^\omega[t]
\end{align}

Adding the price uplift is equivalent to changing $b^\omega[t]$ to $b^\omega[t]+\Delta \pi^\omega[t]$, which will be reflected in the corresponding social-optimum result. \qed

\section*{Appendix.VIII: Generalization of transmission network}

 {Our current work assumes no congestion in the network. We focus on the key insights from the PIU mechanism involving the penalty payment, supply incentive, and energy price uplift. However, we can partially generalize our model to incorporate networks with DC power flow, where some key results will still hold. Next, we will first present the system constraints based on the DC power flow model. Then, we show how to generalize the game-theoretical model and the PIU mechanism considering networks.}

 {For the transmission network, we denote by $\Lambda$ the set of transmission lines, indexed by $(n,m)$, where $n/m$ is sending/receiving end of a transmission line.  We denote by $\mathcal{N}$ the set of grid buses, indexed by $n$, where $(\cdot)_{n}$ denotes the mapping of $(\cdot)$ into the set of buses. The notation $\delta_{n}^\omega[t]$ denotes voltage angles and $x_{n,m}$ is transmission line $(n,m)$ reactance. Without loss of generality, we assume that each LER investor is located on only one bus with only one resource  in the network.}

\subsection{System constraints and costs}

 {We now update the system constraints and costs   based on the DC power flow model.}

 {For the system constraints, we have}
\begingroup
\allowdisplaybreaks
 {
	\begin{align}
	& \hspace{-2mm}\sum_{i\in \mathcal{I}_n}A_i^{\omega}[t]\hspace{-0.5mm}+\hspace{-0.5mm}p_n^{cv,\omega}[t]\hspace{-0.5mm}+\hspace{-0.5mm}p_n^{sh,\omega}[t]-p_n^{inj,\omega}[t]=\hspace{-0.5mm}D_n^\omega[t],\notag \\
 &\hspace{20ex}\forall n\in\mathcal{N},\forall t\in\mathcal{T},
 \forall \omega \in {\Omega},\label{eq:app_const_balance_net}\\
	& 0\leq p_n^{cv,\omega}[t]\leq \gamma \cdot \bar{p}_n^{cv},\forall n\in\mathcal{N},	\forall t\in\mathcal{T},\forall \omega \in {\Omega},\label{eq:app_const_convention_net}\\
	& 0\leq p_n^{sh,\omega}[t],	\forall t\in\mathcal{T},\forall \omega \in \Omega,\label{eq:app_const_loss_net}\\
	&-\overline{F}_{n,m}\leqslant p_n^{inj,\omega}[t] \leqslant \overline{F}_{n,m}, \forall n\in\mathcal{N}, \forall (n,m) \in \Lambda,\forall t \in \mathcal{T},\label{eq:app_daline} \\
 &p_n^{inj,\omega}[t]=\sum_{m:(n,m)\in\Lambda}\frac{\delta_{n}^\omega[t]-\delta_{m}^\omega[t]}{x_{n,m}},\forall n\in\mathcal{N}.\label{eq:app_trans}
	\end{align}}
 \endgroup
 {Compared with the initial model with no network, we have additional constraints  \eqref{eq:app_daline}-\eqref{eq:app_trans} for the transmission power flow. Note that the additional constraints  are linear.}

 {The system costs include the operation costs of lost load and CERs, as well as the investment and operation costs of  VRE and ES. We calculate the expected cost of lost load for all the buses based on  the value of lost load (VOLL): $G^{sh}(\bm{p}^{sh})=\mathbb{E}_\omega \sum_{t\in \mathcal{T}}\sum_{n\in \mathcal{N}} \text{VOLL} \cdot p_n^{sh,\omega}[t].$
For the operation cost of CERs, we assume a convex supply-cost function $g_n^{cv,\omega}[t](\cdot)$ for CERs at each bus $n\in \mathcal{N}$. The total expected cost is $G^{cv}(\bm{p}^{cv})=\mathbb{E}_\omega \sum_{t\in \mathcal{T}} \sum_{n\in \mathcal{N}} g_n^{cv,\omega}[t]({p}_n^{cv,\omega}[t]).$ The investment and operation costs of VRE and ES are calculated following the same way as the  initial no-network model.}
 
\subsection{Social-optimum benchmark}
 {We can easily generalize the social-optimum benchmark based on the system cost minimization problem. Compared with the case  of  no network,  we replace the initial system constraints (4)-(6) of the manuscript  with the new system constraints \eqref{eq:app_const_balance_net}-\eqref{eq:app_trans}. The system cost is updated accordingly.}

\subsection{Game-theoretical model}
 {We now generalize the game-theoretical model. We introduce the system operator as the player in the game. The models for LER investors maintain the same structure as the initial model of no network.}

 {Besides investors, we assume that the system operator is also a player in the game, which aims to minimize the system cost given LER investors' decisions. Compared with the initial model that does not consider networks, we need the system operator as an additional player to optimize transmission power flow. Specifically, the system operator decides the system variables, denoted as $\bm{z}_s$, which includes the CER supply $\bm{p}^{cv}$, the lost load $\bm{p}^{sh}$, the voltage angle $\bm{\delta}$, and transmission power flow $\bm{p}^{inj}$. The constraint set of the system operator is denoted as $\mathcal{Z}_s$ based on system constraints \eqref{eq:app_const_balance_net}-\eqref{eq:app_trans}. Note that  $\mathcal{Z}_s$ can be written as $\mathcal{Z}_s(\bm{z}_{-s})$ that is dependent on all the LER investors' decisions $\bm{z}_{-s}=(\bm{z}_1, \ldots,\bm{z}_I)$.}

 {We set up the game-theoretical model in the following.}
\begin{itemize}
 \item Players: Investors $i\in \mathcal{I}$ and system operator $s$
    \item Strategy:
    \begin{itemize}
        \item investor $i\in \mathcal{I}$: $\bm{z}_i \in \mathcal{Z}_i(\bm{z}_{-i})$
            \item system operator: $\bm{z}_s \in \mathcal{Z}_s(\bm{z}_{-s})$
    \end{itemize}
    \item Payoff: 
  \begin{itemize}
      \item Investor $i$: maximize its profit 
      
      $\max_{\bm{z}_i \in \mathcal{Z}_i(\bm{z}_{-i})} f_i(\bm{z}_i, \bm{z}_{-i})$
        \item System operator $i$:  minimize  the system cost (given investors' decisions)
        $$\min_{\bm{z}_s \in \mathcal{Z}_s(\bm{z}_{-s})} G^{cv}(\bm{p}^{cv})+G^{sh}(\bm{p}^{sh})$$
  \end{itemize}
\end{itemize}

 {Next, we generalize the PIU mechanism based on the established game-theoretical model and analyze the Nash equilibrium.}
\subsection{PIU mechanism}

 {We will first generalize the price function for each bus. Then, we generalize the penalty payment, supply incentive, and energy price uplift accordingly considering the transmission network.}

We have the price function for each bus. 
\begingroup
\allowdisplaybreaks
 {\begin{align}
\hat{\pi}_n^\omega[t] (\bm{z})=\left \{
\begin{aligned}
&\frac{d g_n^{cv,\omega}[t]}{d p_n^{cv,\omega}[t]}\Bigg|_{\gamma\cdot \bar{p}_n^{cv}},  ~\text{if}~ \sum_{i\in \mathcal{I}_n}A_i^\omega[t]< D_n^\omega[t]\notag\\
&\hspace{22ex}+p_n^{inj,\omega}[t]-\gamma\cdot \bar{p}_n^{cv};\\
&\frac{d g_n^{cv,\omega}[t]}{d p_n^{cv,\omega}[t]},~~~~~\text{if}~  D_n^\omega[t]+p_n^{inj,\omega}[t]-\gamma \cdot \bar{p}_n^{cv} \\
&\hspace{22ex}\leq \sum_{i\in \mathcal{I}_n}A_i^\omega[t]\leq D_n^\omega[t].
\end{aligned}
\right. \label{eq:app_ourprice}
\end{align}}
\endgroup

\subsubsection{Penalty payment} 
 {As we do for the case of no network, we let each investor bear a penalty on the allocated lost load $p_i^{sh,\omega}$. The lost load satisfies the  power balance as follows.}
 {	\begin{align}
	& \sum_{i\in \mathcal{I}_n}A_i^{\omega}[t]\hspace{-0.5mm}+\hspace{-0.5mm}p_n^{cv,\omega}[t]\hspace{-0.5mm}+\hspace{-0.5mm}\sum_{i\in \mathcal{I}_n}p_i^{sh,\omega}[t]-p_n^{inj,\omega}=\hspace{-0.5mm}D_n^\omega[t],
 \notag\\
   &\hspace{20ex}\forall n\in\mathcal{N},\forall t\in\mathcal{T},\forall \omega \in {\Omega}\\
	&p_i^{sh,\omega}[t]\geq 0, \forall i \in \mathcal{I},\forall t\in\mathcal{T},\forall \omega \in \Omega.
	\end{align}
We let $\tilde{A}_i^\omega[t]={A}_i^\omega[t]+p_i^{sh,\omega}[t]$ and  $\Tilde{\bm{z}}_i=\left(\bm{z}_i,\bm{p}_i^{sh}\right)$.
The profit of investor $i$ is modified to  $\tilde{f}_i$.
	\begin{align}
\tilde{f}_i(\tilde{\bm{z}}_i, \tilde{\bm{z}}_{-i})=f_i (\tilde{\bm{z}}_i, \tilde{\bm{z}}_{-i})- \mathbb{E}_\omega \sum_{t\in \mathcal{T}} \text{VOLL} \cdot p_i^{sh,\omega}[t]. \label{mech:penalty}
\end{align}}

 {We will characterize the Nash equilibrium based on Proposition \ref{prop:penaltyfinitex}, which is a modification of Proposition \ref{prop:penaltyfinite}.}

\begin{prop}\label{prop:penaltyfinitex}
 {One Nash equilibrium coincides with the optimal solution to the following problem.
\begin{align*}
\min	&\sum_{n \in \mathcal{N}}\sum_{i\in \mathcal{I}_n} \Big(C_i^{inv}(\bm{X}_i)+ C_i^{op} (\bm{p}_i)+G_i^{sh}(\bm{p}_i^{sh})\Big)
\\&+ \sum_{n \in \mathcal{N}} G_n^{cv}(\bm{p}_n^{cv})+\sum_{n\in \mathcal{N}}\sum_{i\in \mathcal{I}_n}\sum_{t\in \mathcal{T}}\frac{1}{2}\cdot a_n^\omega[t]\cdot (\tilde{A}_i^{\omega}[t])^2\\
 \text{s.t.}~& \tilde{\bm{z}}_i\in \tilde{\mathcal{Z}}_i^o, \\
 ~&\text{system constraints} ~\eqref{eq:app_const_balance_net}-\eqref{eq:app_trans}.
\end{align*}}
\end{prop}

 {Proposition \ref{prop:penaltyinf} in the main text still holds, i.e.,  the Nash equilibrium coincides with the social optimum under perfect competition when we have the penalty mechanism.}

\subsubsection{Supply incentive} 
 {As we do for the case of no network, the supply incentive still has the  structure $\frac{1}{2}a^\omega[t].(\tilde{A}_i^\omega[t])^2$. The profit of investor $i$ is 
      	\begin{align}
\tilde{\tilde{f}}_i (\tilde{\bm{z}}_i, \tilde{\bm{z}}_{-i})=\tilde{f}_i (\tilde{\bm{z}}_i, \tilde{\bm{z}}_{-i})+ \mathbb{E}_\omega \sum_{t\in \mathcal{T}} \frac{1}{2}a^\omega[t](\tilde{A}_i^\omega[t])^2. \label{mech:incentive}
\end{align}}

 {Proposition \ref{prop:incentive} still holds, i.e., under imperfect competition, one Nash equilibrium
coincides with the social-optimum solution when we have the penalty payment and supply incentive.}

\subsubsection{Energy price uplift} 
 {We give an uplift $\Delta \pi_n^\omega[t]$ for the price of each bus.
      \begin{align}
\tilde{\pi}_n^\omega[t] (\bm{z})=\hat{\pi}_n^\omega[t] (\bm{z}) +\Delta \pi_n^\omega[t]. \label{mech:uplift}
\end{align}}

 {Proposition 6 of the main text still holds, where we replace $b^\omega[t]$ and  $\pi^\omega[t]$ with $b_n^\omega[t]$ and  $\pi_n^\omega[t]$. Still, the proposition shows that the price uplift will equivalently
increase the cost of the CERs.}

 {Note that we only partially generalize our results for the transmission network. The propositions not mentioned here may not be extended.}

\end{document}